%% file: AGNPS.tex
\documentclass[usegraphicx,usenatbib]{mn2e}

\usepackage{graphicx}
\usepackage{natbib}
\usepackage{fixltx2e}
\usepackage{mathptmx}
\usepackage{amsmath}
\usepackage{wasysym}
\usepackage{url}
\usepackage{times}
\usepackage{captcont}
\usepackage{epsfig}
\usepackage{ulem}
\usepackage{deluxetable}

\input{defn}

\begin{document}

\title[AGN in brightest cluster galaxies]{Radiative efficiency, variability and Bondi accretion onto massive black holes: from mechanical to quasar feedback in brightest cluster galaxies}
\author[H.R. Russell et al.]  
    {\parbox[]{7.in}{H.~R. Russell$^1$\thanks{E-mail: 
          helen.russell@uwaterloo.ca}, B.~R. McNamara$^{1,2,3}$, A.~C. Edge$^{4}$, M.~T. Hogan$^{4}$, R.~A. Main$^1$, A.~N. Vantyghem$^1$\\
    \footnotesize 
    $^1$ Department of Physics and Astronomy, University of Waterloo, Waterloo, ON N2L 3G1, Canada\\ 
    $^2$ Perimeter Institute for Theoretical Physics, Waterloo, Canada\\ 
    $^3$ Harvard-Smithsonian Center for Astrophysics, 60 Garden Street, Cambridge, MA 02138, USA\\ 
    $^4$ Department of Physics, Durham University, Durham DH1 3LE\\
  }
}


\maketitle

\begin{abstract}
We examine unresolved nuclear X-ray sources in 57 brightest cluster galaxies to study the relationship between nuclear X-ray emission and accretion onto supermassive black holes.  The majority of the clusters in our sample have prominent X-ray cavities embedded in the surrounding hot atmospheres, which we use to estimate mean jet power and average accretion rate onto the supermassive black holes over the past several hundred Myr.  We find that roughly half of the sample have detectable nuclear X-ray emission.  The nuclear X-ray luminosity is correlated with average accretion rate determined using X-ray cavities, which is consistent with the hypothesis that nuclear X-ray emission traces ongoing accretion.  The results imply that jets in systems that have experienced recent AGN outbursts, in the last $\sim10^7\yr$, are `on' at least half of the time.  Nuclear X-ray sources become more luminous with respect to the mechanical jet power as the mean accretion rate rises.  We show that nuclear radiation exceeds the jet power when the mean accretion rate rises above a few percent of the Eddington rate, or a power output of $\sim10^{45}\ergps$, where the AGN apparently transitions to a quasar.  The nuclear X-ray emission from three objects (A2052, Hydra A, M84) varies by factors of $2-10$ on timescales of 6 months to 10 years. If variability at this level is a common phenomenon, it can account for much of the scatter in the relationship between mean accretion rate and nuclear X-ray luminosity. We find no significant change in the spectral energy distribution as a function of luminosity in the variable objects.  The relationship between accretion and nuclear X-ray luminosity is consistent with emission from either a jet, an ADAF, or a combination of the two, although other origins are possible.  We also consider the longstanding problem of whether jets are powered by the accretion of cold circumnuclear gas or nearly spherical inflows of hot keV gas.  For a subset of 13 nearby systems in our sample, we re-examine the relationship between the jet power and the Bondi accretion rate.  The results indicate weaker evidence for a trend between Bondi accretion and jet power, primarily due to the uncertainty in the cavity volumes.  We suggest that cold gas fuelling could be a likely source of accretion power in these objects however we cannot rule out Bondi accretion, which could play a significant role in low power jets.
\end{abstract}

\begin{keywords}
  X-rays: galaxies: clusters --- galaxies:active --- galaxies:jets --- accretion, accretion discs
\end{keywords}

\section{Introduction}
\label{sec:intro}

%

Energetic feedback from supermassive black holes (SMBHs) plays an
important role in the formation and evolution of galaxies.  With this
realisation, the exploration and understanding of active galactic
nuclei has taken a new emphasis.  Key lines of evidence include the
relationship between nuclear black hole mass and the mass of the host
galaxy (M-$\sigma$ relation; \citealt{Magorrian98}; \citealt{Kormendy01};
\citealt{Merritt01}), which was likely imprinted through the quasar
era (`quasar mode feedback'; \citealt{Silk98}; \citealt{Haehnelt98}; \citealt{DiMatteo05}), and the prevalence of radio bubbles in
the X-ray atmospheres of giant elliptical and brightest cluster
galaxies (`radio mode feedback'; \citealt{McNamara00}; \citealt{Churazov00};
\citealt{FabianPer00}; \citealt{McNamaraNulsen07}).  These forms of
feedback are ultimately powered by binding energy released by
accretion onto massive black holes.  However, energy is released
primarily in the form of radiation from quasars, while radio jets
release their energy in a mechanical form.

The reason why black holes release their binding energy in different
forms is poorly understood.  Clues have come from the so-called
fundamental plane of black holes (\citealt{Merloni03};
\citealt{Falcke04}).  This relationship between the radio and X-ray
power emerging from the vicinity of the black hole and the black hole
mass extends over nine orders of magnitude from stellar mass black
holes to supermassive black holes.  The continuity of this
relationship indicates that the emergent properties of black holes are
fundamentally the same regardless of their mass.  Furthermore,
accreting black hole binaries undergo changes in emission states that
seem to correlate with the accretion rate normalized to the black hole
mass (eg. \citealt{Fender99}; \citealt{Maccarone03};
\citealt{Remillard06}).  

Changes in the specific accretion rate apparently lead to structural
changes in the accretion disk which govern the release of binding
energy in the form of a jetted outflow or radiation.  During periods
of high accretion rate a geometrically thin and optically thick disk
forms that dissipates its energy primarily in the form of radiation
(\citealt{Shakura73}; \citealt{Novikov73}; \citealt{Frank02}).  During
periods of more modest accretion, a hot, radiatively inefficient
accretion flow (RIAF) or advection-dominated accretion flow (ADAF)
forms that releases its energy in the form of a jetted outflow or wind
(eg. \citealt{Narayan94}; \citealt{Abramowicz95};
\citealt{Narayan08}).  ADAFs are geometrically thick, optically thin
disks where the ion temperature substantially exceeds the electron
temperature.  In these disks, the inflow timescale is much shorter
than the cooling timescale.  Thus the accretion energy cannot be
radiated and is either advected inward with the flow or is released in
a wind or radio jet (ADIOS; \citealt{Blandford99,Blandford04}).  In
the context of SMBHs, a radiatively efficient disk is formed when the
accretion rate approaches the Eddington value giving rise to a Seyfert
nucleus or quasar.  When the accretion rate falls below a few percent
of the Eddington rate, the nucleus becomes faint and a radio galaxy is
formed (eg. \citealt{Churazov05}).

Galaxy formation models incorporating AGN feedback distinguish between
radiatively-dominated quasar feedback at early times and a
mechanically-dominated radio mode at late times
(eg. \citealt{Springel05}; \citealt{Croton06}; \citealt{Sijacki06}; \citealt{Hopkins06};
\citealt{Bower06}).  Quasar feedback operates through intense
radiation that is expected to couple to the gas and strong winds which drive
gas from the host galaxy, quenching star formation and regulating the
growth of the SMBH.  This will eventually starve the SMBH of fuel
and, as the accretion rate drops, a transition to
mechanically-dominated radio mode feedback is expected.  The SMBH
launches jets which regulate radiative cooling in the surrounding hot
atmosphere and the growth of the most massive galaxies.  The AGN
activity is closely correlated with the properties of the host halo
indicating that they form a feedback loop (eg. \citealt{Birzan04};
\citealt{Dunn04}; \citealt{Rafferty06}).  


Despite evidence for rapid accretion onto their SMBHs, giant
elliptical and brightest cluster galaxies in the nearby Universe
rarely harbour quasars.  The exceptions include H1821+643, 3C\,186 and
IRAS09104+4109 (\citealt{Crawfordqsos99}; \citealt{Belsole07};
\citealt{Russell10}; \citealt{Siemiginowska10}, Cavagnolo et
al. submitted).  Instead most harbour low luminosity AGN ie. ADAFs
(\citealt{Fabian95}; \citealt{Sambruna00}; \citealt{DiMatteo00},
\citealt{HlavacekLarrondo11}).  These low luminosities imply that
their host galaxies harbour massive black holes exceeding
$\sim10^{9}\Msun$.  Nuclear emission likely signals ongoing accretion.
However, any dependence of the amplitude and form of the emission
emerging from the nucleus on the accretion rate, as found in X-ray
binaries, requires an independent means of estimating the accretion
rate itself.  Therefore it is difficult to test nuclear emission
models that predict a strong dependence on the form of nuclear power
output with nuclear accretion rate (\citealt{Falcke04};
\citealt{Churazov05}).  Here we examine the nuclear emission
properties in a sample of over 50 BCGs using the energy demands of
X-ray cavities as a measure of mean nuclear accretion rate.


We assume $H_0=70\kmpspMpc$, $\Omega_m=0.3$ and $\Omega_\Lambda=0.7$.
All errors are $1\sigma$ unless otherwise noted.

\section{Data reduction and analysis}

\subsection{Sample selection}


We intend to explore the emergent properties of accreting black holes
at the centres of clusters.  Therefore we have selected objects with
large X-ray cavities which we use to estimate the mean accretion rate
of each object over the past $10^{7}-10^{8}\yr$.  Sources with a range
of cavity powers, and thus accretion rates, were selected.  The
sources were selected from cluster, group and elliptical galaxy
samples which show evidence of AGN activity in the form of cavities in
X-ray images (\citealt{Birzan04}; \citealt{Rafferty06};
\citealt{Allen06}; \citealt{Cavagnolo10}; \citealt{OSullivan11}).
These objects were supplemented with other recently discovered X-ray
point sources in cavity systems (RXCJ0352.9+1941, RXCJ1459.4-1811,
RXCJ1524.2-3154, RXCJ1558.3-1410, Zw\,348) and three noncavity systems
(Zw\,2089, A2667, A611) each with a bright point source.  

We have also included three quasars taken from the literature for
comparison, H1821+643, IRAS09104+4109 and 3C\,186
(\citealt{Russell10}; Cavagnolo et al. submitted;
\citealt{Siemiginowska05,Siemiginowska10}).  These sources have very
different spectral energy distributions (SEDs) from the low luminosity
AGN dominating this sample.  All three quasars have broad optical
emission lines and bolometric luminosities of $\sim10^{47}\ergps$, far
greater than the rest of the sample.  Both H1821+643 and 3C\,186 have
strong big blue bumps in the optical-UV band whereas IRAS09104+4109 is
a heavily obscurred quasar with most of its bolometric luminosity
emerging in the IR.  The big blue bump emission is usually interpreted
as thermal emission from an accretion disk around the SMBH, which has
then been re-radiated in the infrared in IRAS09104+4109.  This
component is absent in radiatively inefficient low luminosity AGN (eg. \citealt{Chiaberge05}).

This sample is neither complete nor unbiased and we are mindful of
this in our interpretation and analysis.  Our sample overlaps the 15
objects analysed by \citet{Merloni07} and we have quadrupled the
sample size, including a large number of upper limits.  In total, 57
sources were selected covering a redshift range from the nearest Virgo
ellipticals at a distance of only $17\Mpc$ to 3C186 at $z=1.06$ and a
mass range from single elliptical galaxies to rich clusters.  We
sample mean accretion rates from $2\times10^{-6}$ to
$0.6\dot{M}_{\mathrm{Edd}}$ for the first time.  The full list of
sources, excluding the three quasars, is shown in Table
\ref{tab:fluxes}.


\subsection{\textit{Chandra} data reduction}

For each object in this sample, we selected the deepest \textit{Chandra}
observation in the archive for analysis (Table \ref{tab:fluxes}).  Each observation
was reprocessed using CIAO 4.4 and CALDB 4.4.7 provided by the
\textit{Chandra} X-ray Center (CXC).  The level 1 event files were
reprocessed to apply the latest gain and charge transfer inefficiency
correction and then filtered to remove photons detected with bad
grades.  The improved background screening provided by VFAINT mode was
also applied where available.  Background light curves were extracted
from the level 2 event files of neighbouring chips for observations on
ACIS-I and from ACIS-S1 for observations on ACIS-S3.  The background
light curves were filtered using the \textsc{lc\_clean}
script\footnote{See http://cxc.harvard.edu/contrib/maxim/acisbg/}
provided by M. Markevitch to identify periods affected by flares.  The
final cleaned exposure times of each observation are detailed in Table
\ref{tab:fluxes}.  Standard blank-sky backgrounds were extracted for each
observation, processed identically to the events file and reprojected
to the corresponding sky position.  The blank-sky background was then
normalized to match the count rate in the $9.5-12\keV$ energy band in
the observed dataset.  This correction was less than 10 per cent for the
majority of the observations.  Each normalized blank sky background
was also checked against the observed background spectrum extracted
from a source-free region of each dataset to ensure it was a good
match.

\subsection{X-ray point source flux}
\label{sec:psflux}

\begin{figure*}
\begin{minipage}{\textwidth}
\centering
\includegraphics[width=0.4\columnwidth]{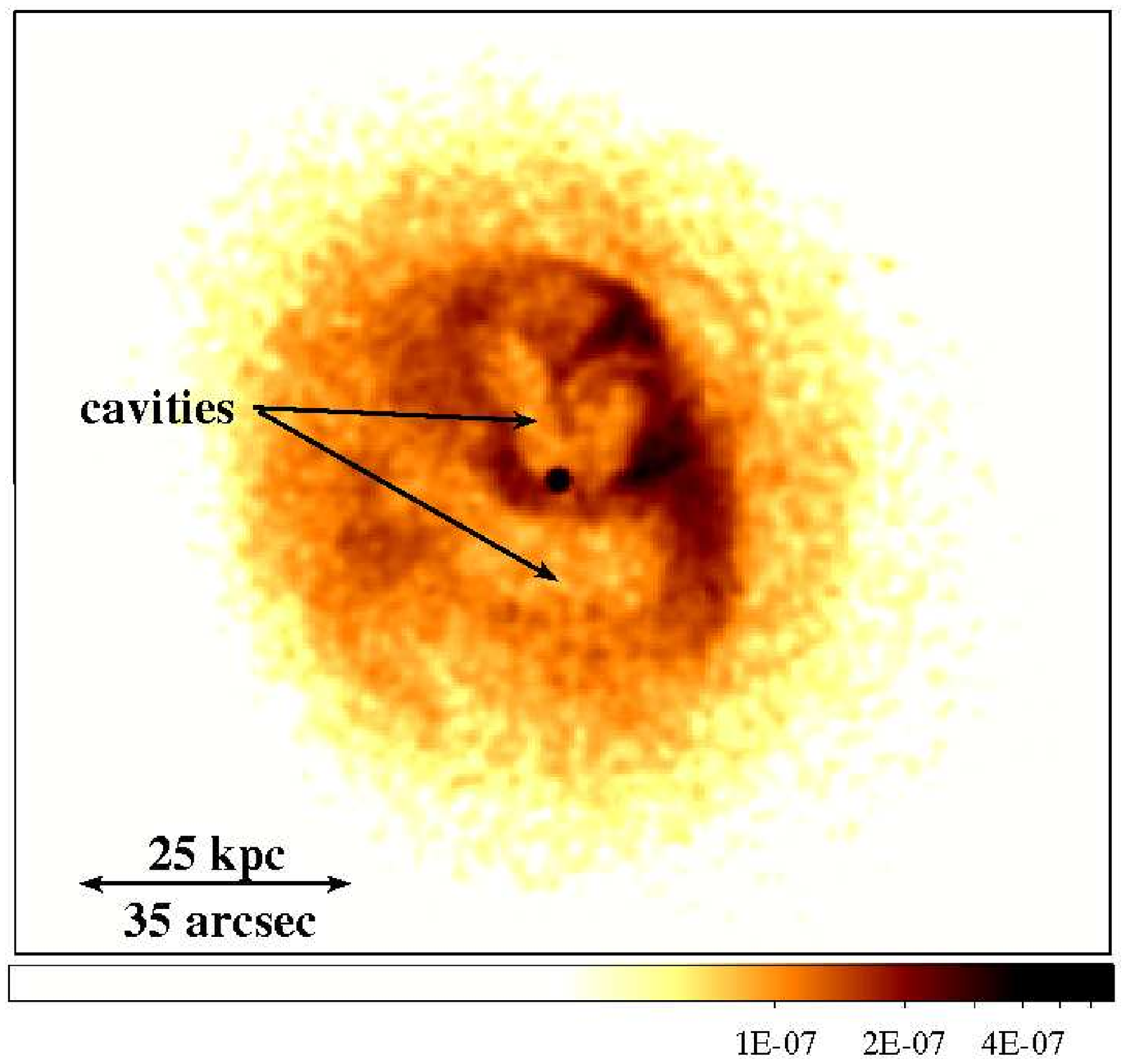}
\includegraphics[width=0.4\columnwidth]{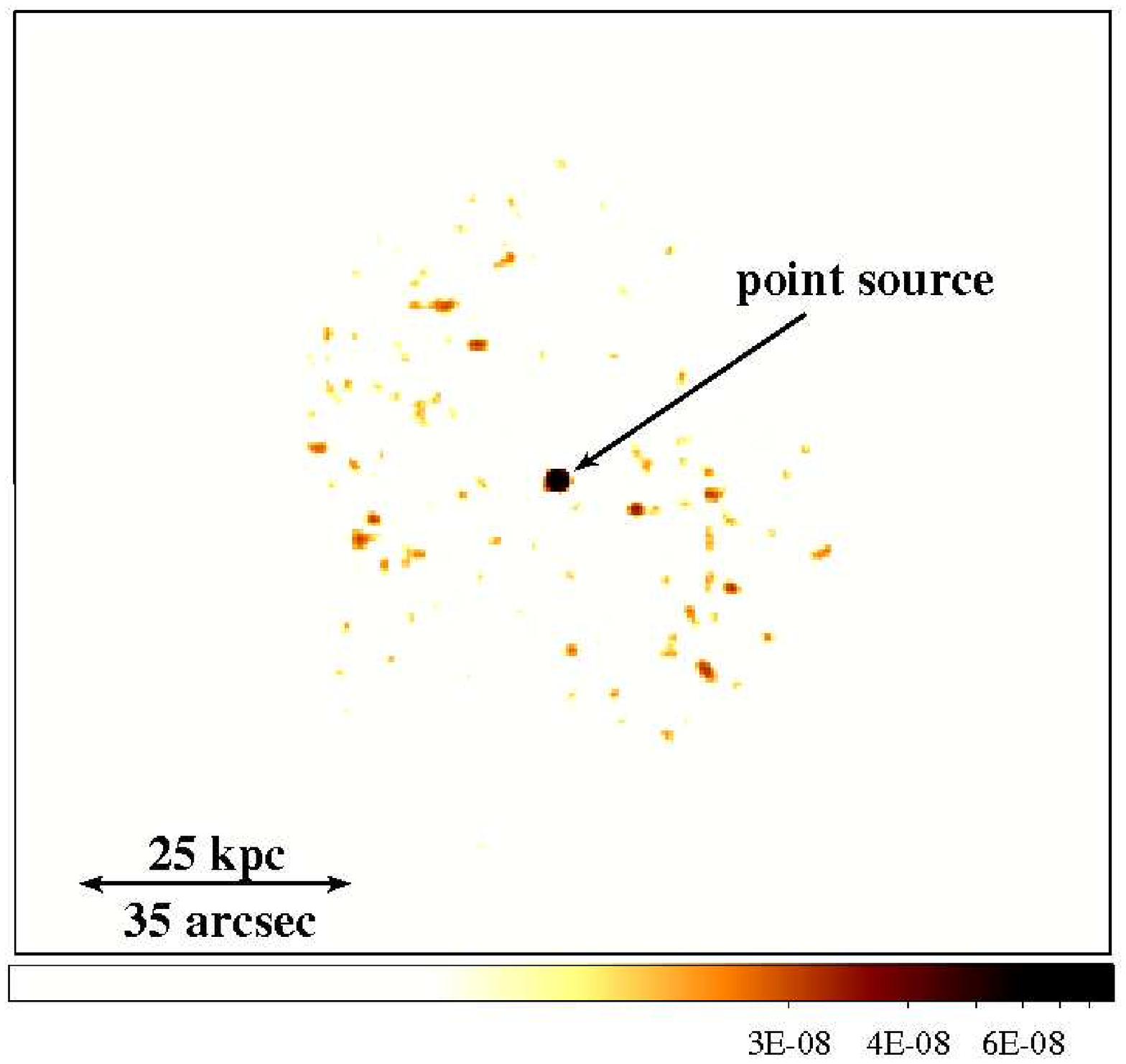}
\caption{Exposure-corrected \textit{Chandra} images covering the same field of Abell 2052 (see \citealt{Blanton03,Blanton09,Blanton11}).  The colour bar has units $\expmapcorr$.  Left: $0.5-7\keV$ energy band showing the X-ray cavities.  Right: $3-7\keV$ energy band showing the AGN point source detection.}
\label{fig:psfimgs}
\end{minipage}
\end{figure*}


Fig. \ref{fig:psfimgs} shows an example of a clear detection of surface brightness
depressions indicating cavities and an X-ray point source at the
centre of A2052 (\citealt{Blanton03,Blanton09,Blanton11}).  We required at least a 3$\sigma$ detection above the
background, where the background error considers only Poisson
statistics, in a hard $3-7\keV$ energy band image to confirm the
detection of an X-ray point source.  The majority of the sources were
not detected with sufficient counts above the cluster background to
generate a reasonable spectrum.  Therefore, we used two alternative
methods, based on those described by \citet{HlavacekLarrondo11}, to
calculate the X-ray flux of the confirmed sources and $1\sigma$ upper
limits on the non-detections.  

The photometric method sums the point source flux in a $1\asec$ radius
region using an exposure-corrected image, which was weighted by a
spectral model for the point source.  This region was centred on the
peak in the X-ray cluster emission if no point source was detected.
The only exception was the Centaurus cluster where we used the
position of the radio source as the centre (\citealt{Taylor06}).  For
the majority of the sample, we used an absorbed powerlaw model
\textsc{phabs(zphabs(powerlaw))} with no intrinsic absorption, a
photon index $\Gamma=1.9$ (eg. \citealt{Gilli07}) and Galactic
absorption from \citet{Kalberla05}.  For sources detected with several
hundred source counts or more, this model was fitted to an extracted
spectrum using \textsc{xspec} version 12 (\citealt{Arnaud96}) with the
intrinsic absorption and the photon index left free.  These model
parameters are detailed in Table \ref{tab:fluxes}.  We used the
modified version of the C-statistic available in \textsc{xspec} to
determine the best-fit parameters for spectra with a low number of
counts (\citealt{Cash79}; \citealt{Wachter79}).  Note that the
intrinsic absorption was set to zero for sources with only upper
limits and the photon index was set to 1.9 where this was consistent
with the best-fit value within the error.  The cluster background was
subtracted using an annulus around the point source from
$1.5-2.5\asec$ and an exposure-corrected image that was uniformly
weighted at the average peak in the cluster spectrum around $1.5\keV$.
The point source flux was also corrected for the fraction of the point
spread function (PSF) falling in the $1\asec$ region ($~90$ per
cent\footnote{http://cxc.harvard.edu/proposer/POG}).  A larger region
encompassing more of the PSF also included a greater fraction of
cluster background, which increased the measurement uncertainty.


However, for galaxy clusters and groups with steep central surface
brightness peaks this method is likely to significantly undersubtract
the background cluster emission.  We therefore also employed a
spectroscopic method where we fit a spectrum extracted from the
$1\asec$ radius point source region with a model for both the point
source and cluster emission.  The parameters for the cluster model
were determined by extrapolating profiles of the projected cluster
properties in to the point source region.  These profiles were
generated by extracting spectra from a series of circular annuli
centred on the point source and fitting them with a suitable spectral
model to determine the temperature, metallicity and normalization.  We
required a minimum of $\sim2000$ source counts per region to ensure
good constraints on the cluster parameters.  However, fewer counts per region
were allowed for the low temperature sources in the sample where the
Fe L line emission improves temperature diagnostics.  Point sources
were identified using the CIAO algorithm \textsc{wavdetect}, visually
confirmed and excluded from the analysis (\citealt{Freeman02}).
Regions of nonthermal jet emission were also excluded.  The
cluster spectra were grouped to contain a minimum of 20 counts per
spectral channel, restricted to the energy range $0.5-7\keV$ and fit
in \textsc{xspec} with appropriate responses, ancillary responses and
backgrounds.


For the majority of the clusters, an absorbed single temperature
\textsc{phabs(mekal)} model (\citealt{Balucinska92}; \citealt{Mewe85};
\citeyear{Mewe86}; \citealt{Kaastra92}; \citealt{Liedahl95}) provided
a good fit to the cluster emission in each annulus.  This model was
insufficient for some of the bright, nearby clusters with multiphase
gas signatures, such as Centaurus, M87 and A2052.  In these cases, a
second \textsc{mekal} component or a \textsc{mkcflow} component was
added to the spectral model.  The cluster redshift and Galactic column
density were fixed to the values given in Table \ref{tab:fluxes}.
Abundances were measured assuming the abundance ratios of
\citet{AndersGrevesse89}.  For the multi-temperature models, the
metallicity was tied between the two components and, for the
\textsc{mkcflow} model, the lower temperature was fixed to $0.1\keV$
and the higher temperature was tied to that of \textsc{mekal}
component.  We produced radial profiles for the best fit temperature,
metallicity and normalization parameters and used powerlaw fits to the
inner points to extrapolate these properties into the point source
region.  The cluster spectra were also used to determine the
Galactic absorption for sources located on regions of the sky where
the absorption is highly variable.  The spectral fits were repeated
for annuli at large radii, excluding cooler gas components in the
cluster core, leaving the $n_{\mathrm{H}}$ parameter free (Table
\ref{tab:fluxes}).

The spectrum extracted from the point source region was then fitted
with a combined model for both the point source and cluster emission.
The similarity between the powerlaw and thermal model components,
particularly for higher temperature clusters, can make it difficult to
distinguish them with \textit{Chandra}'s spectral resolution.  Most of
the model parameters were therefore constrained for the fit.  The
parameters for the point source model were set to those detailed in
Table \ref{tab:fluxes} with only the normalization left free.  The
cluster model temperature was fixed to the value determined by
extrapolating from neighbouring annuli.  The metallicity was fixed to
the value in the neighbouring cluster annulus as it was generally not
found to vary significantly at these small radii.  The normalization
of the cluster component was problematic because it was strongly
affected by cavity substructure in the core, which produced large
variation.  We therefore constrained the cluster normalization in the
point source region to be no less than the normalization from the
neighbouring annulus, scaled by the ratio of their respective areas.
The \textsc{xspec} \textsc{cflux} model was used to determine the flux
or an upper limit for the unabsorbed point source component (Table
\ref{tab:fluxes}).  


The main source of error in both the photometric and spectroscopic
measurements of the point source flux is the subtraction of the
cluster emission.  The photometric method is likely to underestimate
the background cluster emission and should therefore be treated as an
upper limit on the point source flux.  The spectroscopic method
improves on this by allowing for the increase in cluster surface
brightness towards the cluster centre but may significantly
overestimate the cluster background because it is indistinguishable
from the powerlaw component.  For strongly obscured point sources and
low temperature cluster emission, the spectroscopic method is likely
to be a significant improvement over the photometric method.  For
higher temperature clusters, the photometric method may be more
accurate.  We have therefore listed both the photometric and
spectroscopic fluxes in Table \ref{tab:fluxes} but used the generally
more accurate spectroscopic flux in our analysis.  The possible bias
in this measurement for higher temperature clusters with strong
$3-7\keV$ emission is discussed in section \ref{sec:seleff}.


Several of the brighter point sources in the sample were significantly
piled up in the longest \textit{Chandra} exposures initially selected
for analysis.  Pile up occurs whenever two or more photons, arriving
in the same detector region and within a single ACIS frame integration
time, are detected as a single event (\citealt{Davis01}).  For M87 and
Cygnus A, there were alternative observations available in the archive
with shorter $0.4\s$ frame times for which the point source was not
piled up.  These short frame time observations were used to calculate
the point source flux and the cluster background was analysed using
the deeper exposures.  All the archival observations of the point
source in Perseus, where the cluster centre is not positioned far off
axis distorting the PSF, were found to be significantly piled up.
Perseus was therefore excluded from this sample.

\begin{figure}
\centering
\includegraphics[width=0.98\columnwidth]{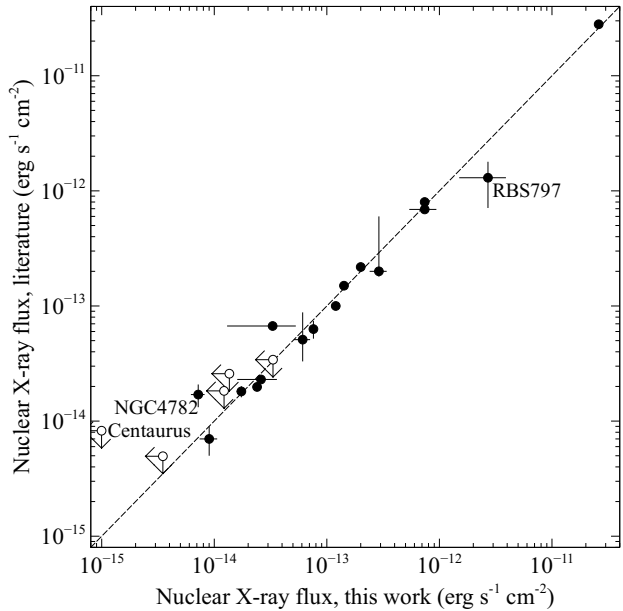}
\caption{Comparison of the spectroscopic point source fluxes with values from the literature (see Table \ref{tab:fluxes}).  Note that the point source fluxes shown here were evaluated in the same energy band as those given in the literature and may differ significantly from those in Table \ref{tab:fluxes}.}
\label{fig:fluxlitcomp}
\end{figure}

Fig. \ref{fig:fluxlitcomp} compares our spectroscopic point source
fluxes with measurements for the same sources available in the
literature.  The majority of the fluxes are consistent within the
errors.  Small variations are expected due to differences in
background subtraction and the position selected for the upper limits
but there are three sources, Centaurus, NGC4782 and RBS797, with
significantly different values which we have considered in detail.
There is only a modest discrepancy for RBS797 given the large errors
and this is likely due to the additional model components used for the
\citet{Cavagnolo11} result. The differences for Centaurus and NGC4782
are due to our use of a two temperature rather than a single
temperature cluster model.  The best-fit single temperature falls
midway between the preferred higher and lower temperature values of
the two component model and therefore significantly underestimates the
cluster surface brightness in the $2-10\keV$ band used to determine
the point source flux.  Our two component model therefore finds a
higher cluster background and a significantly lower nuclear point
source upper limit.

\subsection{Cavity power}
\label{sec:cavpow}

The cavities observed in the X-ray images of this sample allow a
direct measurement of the mechanical output from the AGN
(\citealt{Churazov00}; \citealt{Dunn04}; \citealt{Birzan04};
\citealt{McNamaraNulsen07}).  For a bubble filled with relativistic
plasma, the energy required to inflate it is given by $E=4PV$, where
the bubble is assumed to be in pressure equilibrium with the
surrounding ICM.  The bubble energy is then divided by the sound speed
timescale or the buoyant rise time to estimate the power input to the
ICM (see eg. \citealt{Birzan04}).  Cavity powers for the majority of
the targets in this sample were available in the literature and
otherwise estimated following \citet{Birzan04} (see Table
\ref{tab:fluxes}).  These values should be treated generally as lower
limits on the total mechanical energy input by the central AGN.  The
datasets available in the \textit{Chandra} archive for this cluster
sample vary from snapshot $10\ks$ exposures to almost complete orbit
exposures over $100\ks$.  Even for the nearest and brightest galaxy
clusters, deeper exposures continue to discover new cavities
(eg. \citealt{Fabian11}).  Some cavities will therefore have been
missed from these estimates of the total power.  In addition, weak
shocks and sound waves have also been found to contribute
significantly to the power output of the central AGN
(eg. \citealt{FabianPer03,FabianPer06}; \citealt{FormanM8705}).

Estimates of the cavity power can be complicated by the presence of
significant X-ray emission from jets.  For example, the synchrotron
self-Compton emission in the centre of 3C\,295 makes it difficult to
determine the extent of the inner cavities and therefore the total
power is significantly underestimated for this system
(\citealt{Harris00}; \citealt{Allen01}).  The cavity power estimate
for 3C\,295 is therefore shown as a lower limit.  Three sources in our
sample, Zw\,2089, A611 and A2667, have point sources but no detected
cavities in the X-ray images.  Ideally, these sources would be
included in the analysis with upper limits on the cavity power.
However, it is difficult to place meaningful constraints on the
possible size of non-detected cavities, particularly at large radius
where the noise increases and the X-ray surface brightness drops off
rapidly (see Birzan et al. 2012).  These sources have therefore been
included with only illustrative upper limits, not quantitative.


By comparing the radiative output from the X-ray point source with the
cavity power, we can calculate the radiative efficiency of the AGN.
The total power output from the AGN can also be used to infer a mean
accretion rate, given an assumption of the accretion efficiency.  We
refer throughout to a mean rather than instantaneous accretion rate as
this method estimates the average accretion requirements over the
$\sim10^7\yr$ age of each cavity.  We have also calculated the
theoretical Eddington and Bondi accretion rates for comparison with
the inferred accretion rates for the sample.

\subsection{Eddington luminosity}
\label{sec:eddlumin}

The Eddington luminosity indicates the limiting luminosity of the SMBH
when the outward pressure of radiation prevents the gravitational
infall of accreting material.  For a fully ionized plasma, the
Eddington luminosity can be expressed as

\begin{equation}
\frac{L_{\mathrm{Edd}}}{\ergps}=1.26\times10^{47}\left(\frac{M_{\mathrm{BH}}}{10^{9}\Msun}\right),
\end{equation}


\noindent where $M_{\mathrm{BH}}$ is the SMBH mass.  Dynamical
estimates of the black hole mass were used where these were available
(Cygnus A, \citealt{Tadhunter03}; \citealt{Rafferty06}; M84, \citealt{Walsh10}; M87,
\citealt{Gebhardt11}; NGC4261,
\citealt{Ferrarese96}).  For the majority of
the sample without dynamical masses, we relied on well-studied
relations between the black hole mass and the properties of the host
galaxy.  Following \citet{Graham07}, the apparent K-band magnitude of
the host galaxy from the 2MASS
catalogue\footnote{http://www.ipac.caltech.edu/2mass}
(\citealt{Skrutskie06}) was converted to an absolute magnitude,
corrected for Galactic extinction (\citealt{Schlegel98}), redshift and
evolution (\citealt{Poggianti97}), and used to estimate the black
hole mass.  3C\,401, RBS797 and Zw\,348 have detected X-ray point
sources but no available apparent K-band magnitudes for this analysis
and are not included in plots with Eddington accretion rate or
luminosity scaling.  \citet{DallaBonta09} find an upper limit on the
central black hole mass in A2052 from dynamical measurements,
which is consistent with the value calculated from the apparent K-band
magnitude.

There are however potential problems with the use of K-band magnitudes
to estimate black hole masses.  \citet{Lauer07} suggest that the
apparent magnitudes from 2MASS are not deep enough to capture the
extent of the BCG envelope and therefore will underestimate the total
luminosity and black hole mass.  \citet{Batcheldor07} instead suggest
that the BCG extended envelope, formed of debris from tidal stripping,
is unlikely to be closely associated with the central galaxy dynamics.
We have therefore considered these black hole masses to be estimates
and, where possible, present the results with and without scaling by the Eddington
accretion rate.

\subsection{Bondi accretion rate}

The primary aim of this paper is to examine the relationship between
accretion and power output in SMBHs. One of the outstanding problems
is whether the jets are powered by cold accretion from circumnuclear
accretion disks of atomic and molecular gas or whether they are
fuelled by a nearly spherical inflow of hot keV gas.  It is impossible
to prove either case because we lack imaging on the scale of and below
the Bondi radius for almost all objects in our sample.  However this
has been done in nearby ellipticals where \textit{Chandra}'s
resolution is close to the size of the Bondi sphere
(eg. \citealt{DiMatteo01,DiMatteo03}; \citealt{Churazov02};
\citealt{Pellegrini03}).  This problem has been addressed by
\citet{Allen06} in a sample of nearby ellipticals and by
\citet{Rafferty06} in a more distant sample with properties similar to
our own and these studies arrived at somewhat different conclusions.
\citet{Allen06} found a strong trend between jet power and an estimate
of the Bondi accretion rate based on circumnuclear X-ray properties
whereas the sample analysed by \citet{Rafferty06} required greater
extrapolation in to the Bondi sphere making their conclusions on Bondi
accretion highly uncertain.  Nevertheless, using energetic arguments,
they found Bondi power accretion was unable to fuel the most powerful
jets unless their black holes were much larger than implied by the
M-$\sigma$ relation.  This question is crucial for understanding the
relationship between jet power and nuclear emission therefore here we
re-analyse data for a subsample of 13 systems where we have resolution
close to the Bondi sphere.  This subsample includes the 9 galaxies
targeted by \citet{Allen06} and 4 additional objects with new, deeper
\textit{Chandra} observations (Table \ref{tab:bondi}).

Assuming spherical symmetry and negligible angular momentum, the Bondi
rate, $\dot{M}_{\mathrm{B}}$, is the accretion rate for a black hole
embedded in an atmosphere of temperature, $T$, and density, $n_e$,
(\citealt{Bondi52}) and can be expressed as

\begin{equation}
\frac{\dot{M}_{\mathrm{B}}}{\Msunpyr}=0.012\left(\frac{k_{B}T}{\keV}\right)^{-3/2}\left(\frac{n_e}{\pcmcu}\right)\left(\frac{M_{\mathrm{BH}}}{10^9\Msun}\right)^2
\label{eq:bondiacc}
\end{equation}

\noindent for an adiabatic index $\gamma=5/3$.  This accretion occurs
within the Bondi radius, $r_{\mathrm{B}}$, where the gravitational
potential of the black hole dominates over the thermal energy of the
surrounding gas,

\begin{equation}
\frac{r_{\mathrm{B}}}{\kpc}=0.031\left(\frac{k_BT}{\keV}\right)^{-1}\left(\frac{M_{\mathrm{BH}}}{10^9\Msun}\right).
\label{eq:bondirad}
\end{equation}

\noindent The Bondi accretion rate is therefore an estimate of the rate of
accretion from the hot ICM directly onto the black hole and depends on
the temperature and density of the cluster atmosphere at the Bondi
radius.  

The projected cluster spectra, extracted from a series of annuli as
described in section \ref{sec:psflux}, were deprojected using the
model-independent spectral deprojection routine \textsc{dsdeproj}
(\citealt{SandersFabian07}; \citealt{Russell08}).  Assuming only
spherical symmetry, \textsc{dsdeproj} starts from the
background-subtracted spectra and uses a geometric method
(\citealt{Fabian81}; \citealt{Kriss83}) to subtract the projected
emission off the spectrum from each successive annulus.  The resulting
deprojected spectra were each fitted in \textsc{xspec} with an
absorbed single temperature \textsc{mekal} model to determine the
temperature and density of the gas, as described in section
\ref{sec:psflux}.  Several of the selected clusters have clear
evidence for multiple temperature components in the inner regions
(eg. M87, \citealt{FormanM8705}; Centaurus, \citealt{Fabian05}) and it
is likely that deeper exposures of other clusters in the subsample
will also produce robust detections of multi-phase gas.  However, given the
available range in exposure depth, a uniform method with a single
temperature model was used to determine the emission-weighted average
temperature of the ICM in the cluster centre.  We also generated
deprojected electron density profiles with finer radial binning from
the surface brightness profiles and incorporating the temperature and
metallicity variations (eg. \citealt{Cavagnolo09}).


These profiles trace the cluster parameters to radii within an order
of magnitude of the Bondi radius and therefore some extrapolation is
required.  Note that although the Bondi radius in M87 is resolved by
\textit{Chandra}, a significant region is affected by pileup and the
PSF from the jet knot HST-1 and must be excluded.  The temperature
profiles were generally found to flatten in the central regions of
these systems.  We have therefore assumed that the temperature at the
accretion radius has only decreased by a further factor of 2 from the
innermost temperature bin.  Using equation \ref{eq:bondirad}, the
Bondi radius was calculated from this innermost temperature value and
the black hole mass.  Dynamical black hole mass estimates were used
where available.  For comparison with \citet{Allen06}, stellar
velocity dispersions from the HyperLeda
Database\footnote{http://leda.univ-lyon1.fr/} were then used to
calculate black hole masses for the rest of the subsample
(\citealt{Tremaine02}).  Only HCG62 did not have a dynamical or
velocity mass estimate and the K band magnitude was used (section \ref{sec:eddlumin}).
\citet{Tremaine02} find an intrinsic dispersion of $0.25-0.3\dex$ in
$\mathrm{log}(M_{\mathrm{BH}})$ in the $M_{\mathrm{BH}}-\sigma$
relation, which dominates over the error in the measurement of the
velocity dispersion.  

The deprojected electron density continues to increase in the cluster
centre and was therefore extrapolated to the Bondi radius using
several different model profiles.  Three different models were
considered: a powerlaw model continuing a steep density gradient to
$r_{\mathrm{B}}$, a $\beta$-model flattening to a constant and a
shallowing S\'ersic profile with $n=4$.  These models were fitted to
the density profile and used to calculate the gas density at
$r_{\mathrm{B}}$.  The density at $r_{\mathrm{B}}$ is shown as a range
of likely values from the powerlaw model upper limit to the
$\beta$-model or S\'ersic model lower limit.  Using equation
\ref{eq:bondiacc}, the Bondi accretion rate was calculated from the
density and temperature at $r_{\mathrm{B}}$ and the black hole mass.

\subsection{Radio point source flux}


We also compared the nuclear X-ray flux with the radio core flux for a
subset of the sample to try to determine the origin of the nuclear
X-ray emission.  Radio observations were available for 22 sources in
our sample which allowed us to reliably distinguish ongoing core
activity.  If the sole source of the X-ray flux is in the base of a
jet, a direct relationship between the radio and X-ray core flux would
be expected.

Nine of the sources in this subsample were observed simultaneously at
C and X bands with the ATCA (project C1958, PI Edge).  All but one
of the remaining sources were observed at C-band with the VLA-C array
(various projects, PI Edge), with the last source (A478)
having been observed simultaneously at L and X band with the VLA-A
array (project AE117).  For each BCG, the radio-SED was further
populated with data from the major radio catalogs (including but not
limited to: AT20G at $20\GHz$, NVSS/FIRST at $1.4\GHz$, SUMSS at $0.843\GHz$,
WENSS/WISH at $0.325\GHz$, VLSS at $0.074\GHz$).  Additional fluxes were
found by searches around the radio-peak coordinates in both the NED
and HEASARC online databases.  All literature fluxes were individually
scrutinised to ensure reliable matches. Where the synthesised beam
size was considered limiting, leading to source confusion, data were
discarded.  Four of the sources have VLBA observations at C-band.

Core flux contributions were calculated by considering both the
morphology and SEDs of each of the sources. The VLBA C-band
observations provided direct measurements of the core flux.  For the
remainder, the SEDs were decomposed into two major components; a
flatter spectrum, active component attributed to ongoing activity
within the core of the AGN and a steeper spectrum component, most
dominant at lower frequencies attributed to either past AGN activity
or alternate acceleration mechanisms (e.g. radio lobes, mini-haloes,
etc.).

Where a clearly resolved core was present in the observations, two
component SEDs were fitted directly.  For sources which were
resolution limited at at C-band, spectral breakdown of the SEDs was
performed on a case-by-case basis.  Consideration was given to extent
seen at other wavelengths, spectral shape and variability, with the
proviso that variable sources are more likely contain a strong
currently active core.  Simple mathematical models were fit to the
SEDs using IDL routines where a strong case could be made for
believing distinct components were present.  For sources where past
and current activity could not be reliably distinguished, limits were
placed on the core contribution.  Full details of the SED analysis
will be presented in Hogan et al. (in prep).

\section{Results}

\begin{figure}
\centering
\includegraphics[width=0.98\columnwidth]{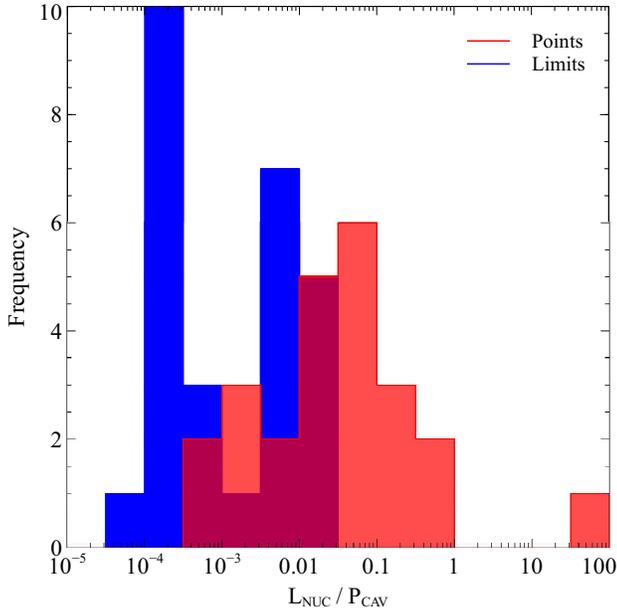}
\caption{Histogram showing the distribution of the ratio of X-ray
  point source luminosity to cavity power for the low luminosity AGN
  in the sample.  Three point sources and one upper limit have no detected
  cavities and are therefore not included.}
\label{fig:histocorejet}
\end{figure}

In total, 27 out of 54 BCGs in this sample were found to have X-ray
central point sources detected above $3\sigma$ in the $3-7\keV$ energy
band.  Although this is not a complete sample of objects, central
X-ray point sources appear to be common in BCGs with detected X-ray
cavities.  Fig. \ref{fig:histocorejet} shows the distribution of the
ratio of X-ray nuclear luminosity in the $2-10\keV$ energy range to
cavity power, $L_{\mathrm{X}}/P_{\mathrm{cav}}$, for the cluster
sample.  The detected point sources in this sample cover a broad range
in radiative efficiency from $L_{\mathrm{X}}/P_{\mathrm{cav}}>0.1$
(eg. 3C\,295) to $L_{\mathrm{X}}/P_{\mathrm{cav}}<10^{-3}$
(eg. A2199).  This distribution presumably reflects a broad range in average
accretion rate with the majority accreting in a radiatively inefficient mode
ie. they are ADAFs (see section \ref{sec:intro}).

\subsection{Radiative and cavity power output}
\label{sec:radcavpowout}

\begin{figure*}
\begin{minipage}{\textwidth}
\centering
\includegraphics[width=0.48\columnwidth]{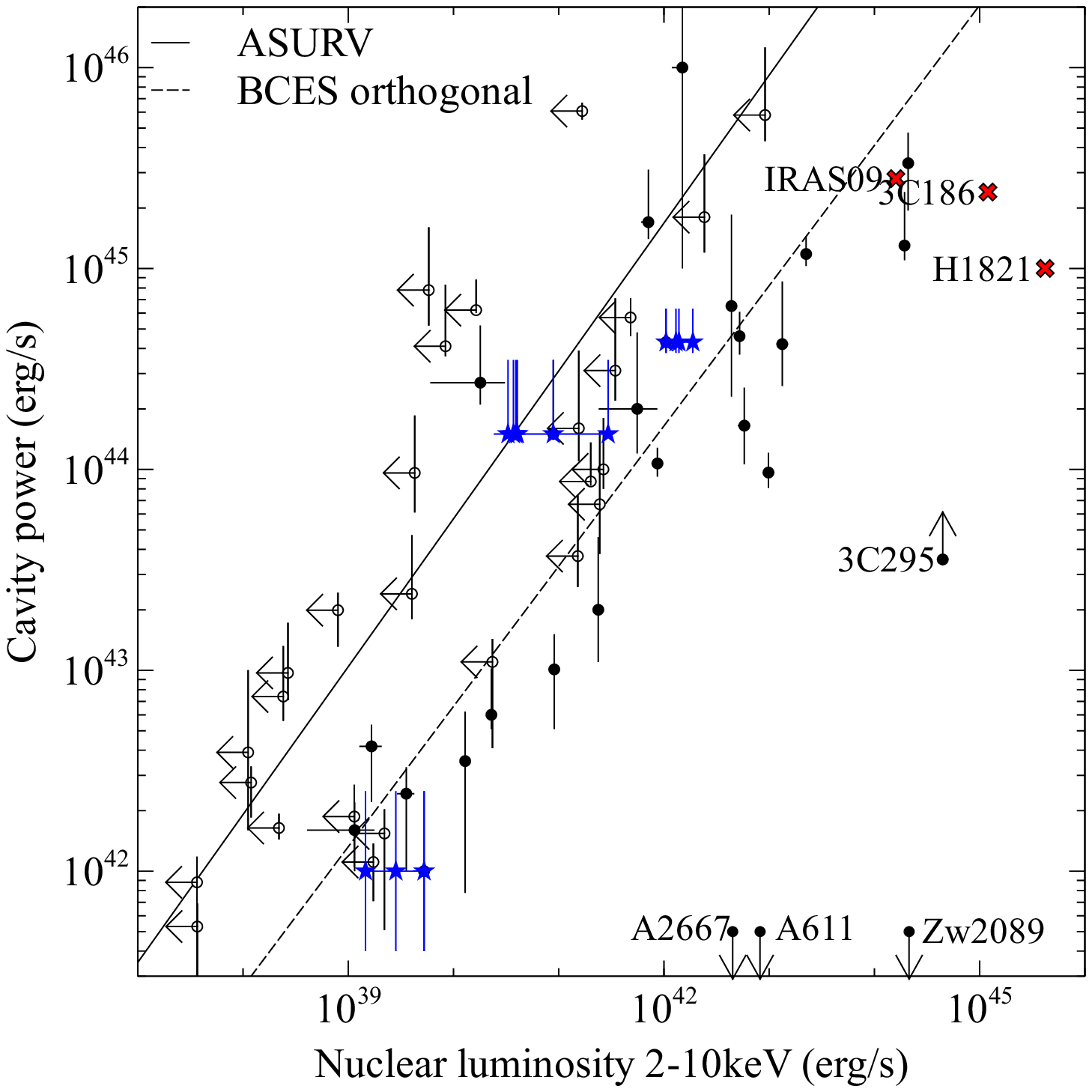}
\includegraphics[width=0.48\columnwidth]{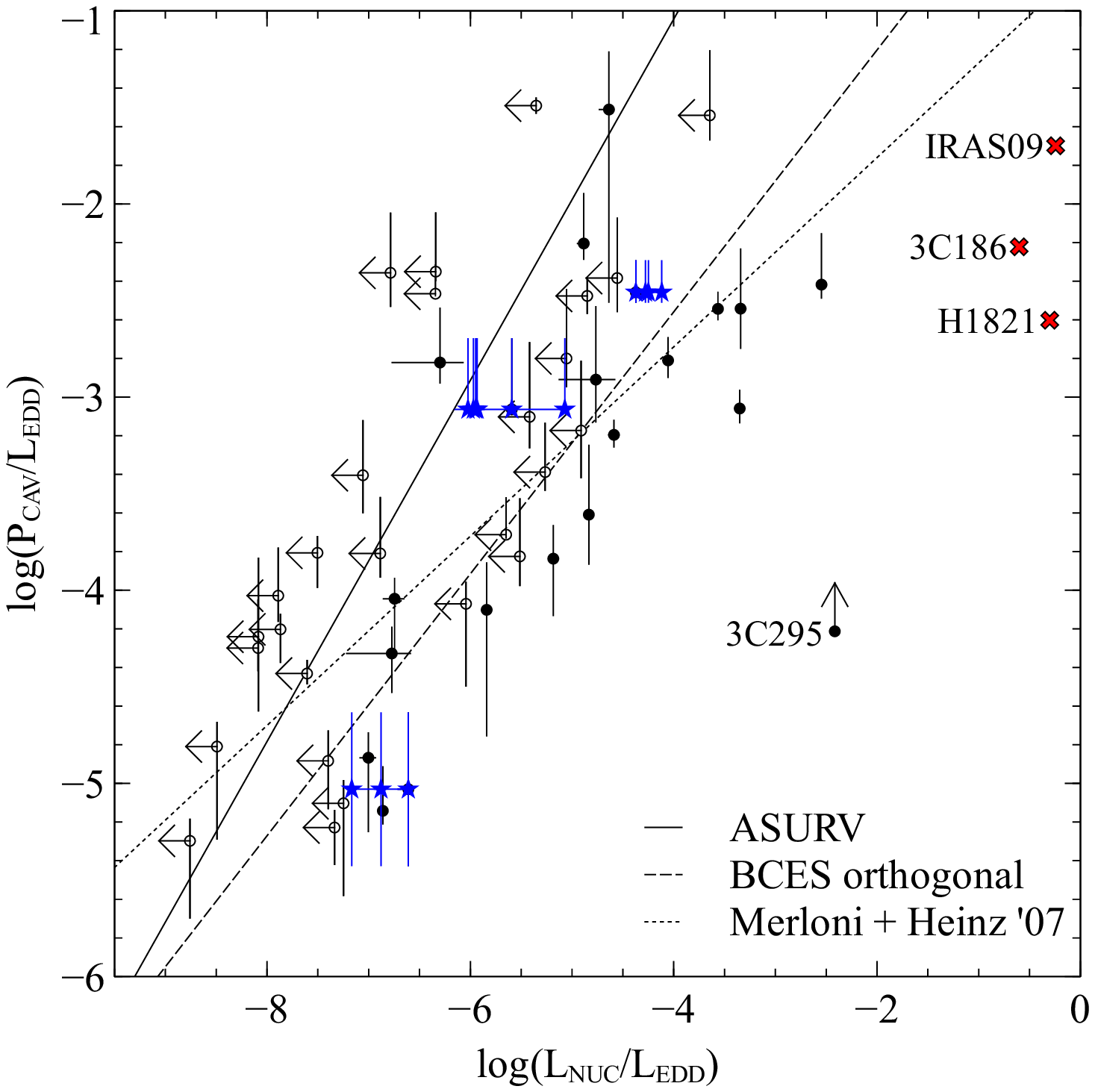}
\caption{Left: Nuclear X-ray luminosity in the energy range $2-10\keV$
  calculated using the spectroscopic method versus cavity power from
  the literature.  Right: bolometric nuclear luminosity versus cavity
  power where both quantities are scaled by the Eddington luminosity.
  Clusters with confirmed point source detections are shown by the
  filled circles and upper limits are shown by the open circles.
  Three quasar sources from the literature are included for comparison
  (red crosses).  The variable sources are shown as blue stars (see
  section \ref{sec:variab}).  The best-fit to both points and upper
  limits using the \textsc{asurv} \citet{Buckley79} estimator for
  censored data is shown by the solid line.  The BCES orthogonal fit
  to the data points only is shown by the dashed line.  The best-fit
  from \citet{Merloni07} is shown as a dotted line and is consistent
  with the BCES fit to the data points within the errors.}
\label{fig:cavpow}
\end{minipage}
\end{figure*}


Fig. \ref{fig:cavpow} (left) shows a correlation between the nuclear
point source $2-10\keV$ luminosity and the cavity power injected into
the surrounding ICM.  This correlation was first found by \citet{Merloni07} for a
sample of 15 AGN with measured cavity powers, 13 of which also had
X-ray point source detections.  The scatter in this correlation covers three
orders of magnitude therefore the generalised Kendall's $\tau$ rank
correlation coefficient for censored data (\citealt{Brown74}) from the
survival analysis package \textsc{asurv} Rev 1.3 (\citealt{Isobe86};
\citealt{Isobe90}; \citealt{Lavalley92}) was used to evaluate its
significance.  We find a probability of accepting the null hypothesis
that there is no correlation between the point source luminosity and
the cavity power of $P_{\mathrm{null}}=4\times10^{-4}$.  Given the
substantial difference in the timescales, ie. six orders of magnitude,
it is surprising to observe a trend at all between these properties
and suggests that AGN feedback is persistant.

It is not clear if the systems with upper limits on the point source
luminosity form part of this trend, and are only just too faint to
detect, or are currently `off'.  We have therefore considered these
two scenarios separately using the BCES estimators
(\citealt{Akritas96}) for a linear regression fit to the data points
only and the \textsc{asurv} non-parametric Buckley-James linear
regression fit to both the points and upper limits
(\citealt{Buckley79}).  Note that this analysis excluded the point
sources with no detected cavities in the surrounding ICM as no
effective upper limits for the cavity power could be estimated (see
section \ref{sec:cavpow}).  The inclusion of the upper limits produces
a significant shift in the correlation but only a small change in the
observed slope, which is consistent within the error.  The BCES
orthogonal best fit was found to be $\mathrm{log}(P_{\mathrm{cav}}) =
(0.69\pm0.08)\mathrm{log}(L_{\mathrm{X}}) + 15\pm3$ compared to the
\textsc{asurv} Buckley-James best fit of $\mathrm{log}(P_{\mathrm{cav}}) =
(0.7\pm0.1)\mathrm{log}(L_{\mathrm{X}}) + 14$.  The majority of the
sample analysed are in a radiatively inefficient accretion mode where
the mechanical cavity power dominates over the radiative output from
the AGN.  The observed slope shows that the radiative efficiency of
the X-ray nucleus increases with increasing cavity power.

For comparison, we have included three quasars from the literature
that are located in luminous cool core clusters to illustrate this
increase in radiative efficiency.  H1821+643 (\citealt{Russell10}),
IRAS\,09104+4109 (Cavagnolo et al. submitted) and 3C\,186
(\citealt{Siemiginowska05,Siemiginowska10}) each have a total
radiative power of $L\sim10^{47}\ergps$ that exceeds their cavity
power by at least an order of magnitude.  Fig. \ref{fig:cavpow} (left)
shows that these sources appear to form an extension of the trend to
much higher cavity powers and presumably much higher accretion rates.
This implies that as the accretion rate rises black holes become more
radiatively efficient.

Following \citet{Merloni07}, we also estimate the bolometric point
source luminosity for all sources in the sample and scale the
luminosity and cavity power by the Eddington luminosity
(Fig. \ref{fig:cavpow} right).  The SED for low luminosity AGN lacks
the `big blue bump' of emission dominating higher accretion rate
sources and is likely to be dominated by the emission at hard X-ray
energies.  \citet{Vasudevan07} find a typical bolometric correction
for low luminosity sources of order $\sim10$, however for ease of
comparison we adopt the same value as \citet{Merloni07}.
Fig. \ref{fig:cavpow} (right) shows the correlation between the
bolometric point source luminosity and the cavity power where both
quantities are scaled by the Eddington luminosity (section
\ref{sec:eddlumin}).  The non-parametric Buckley-James linear
regression method from \textsc{asurv} software was used to determine
the best-fit relation for both the points and the upper limits
(\citealt{Buckley79}).  The best-fit slope of
$\mathrm{log}(L_{\mathrm{cav}}/L_{\mathrm{Edd}})\propto0.9\mathrm{log}(L_{\mathrm{nuc}}/L_{\mathrm{Edd}})$
is significantly steeper than that found by \citet{Merloni07}
(Fig. \ref{fig:cavpow} right).  Using the BCES estimators linear
regression fit to only the data points we determine that the best fit
is consistent with the \citet{Merloni07} result within the errors.
The inclusion of a large number of upper limits results in a
significant difference in the slope of the correlation.  We show in
section \ref{sec:variab} that 3 systems vary by up to an order of
magnitude over a ten year timespan.  This is apparently a
significantly contributing factor to the scatter to which the
\citet{Merloni07} sample was not sensitive.  Another factor that may
be contributing is beaming of the central X-ray source, which is
discussed in detail in \citet{Merloni07}.

\subsection{Selection effects}
\label{sec:seleff}

It is also clear that the scatter in this correlation between the
nuclear X-ray luminosity and the cavity power is underestimated by our
sample selection.  Sources were selected primarily on the detection of
cavities in X-ray observations and therefore sources with point
sources but no cavities are generally missed from the lower right of
Fig. \ref{fig:cavpow}.  Three such sources, Zw\,2089, A611 and A2667,
were added to the sample to illustrate this selection bias.  Note that
the upper limits on the cavity powers are only illustrative, not
quantitative (section \ref{sec:cavpow}).  Whilst the observation of
A2667 is shallow at only $8\ks$, both A611 and Zw\,2089 have
sufficiently deep X-ray observations that would detect signs of any
cavities or feedback-related substructure in the cluster cores.  In
particular, Zw\,2089 contains a bright X-ray point source and appears
very relaxed with smooth extended X-ray emission and no likely cavity
structures.  However, if cavities in this system are emerging along
our line of sight they will be particularly difficult to detect.
There could be a significant number of similar systems with bright
point sources but no cavity structures, which will tend to increase
the scatter in the observed correlation further.

\begin{figure}
\centering
\includegraphics[width=0.98\columnwidth]{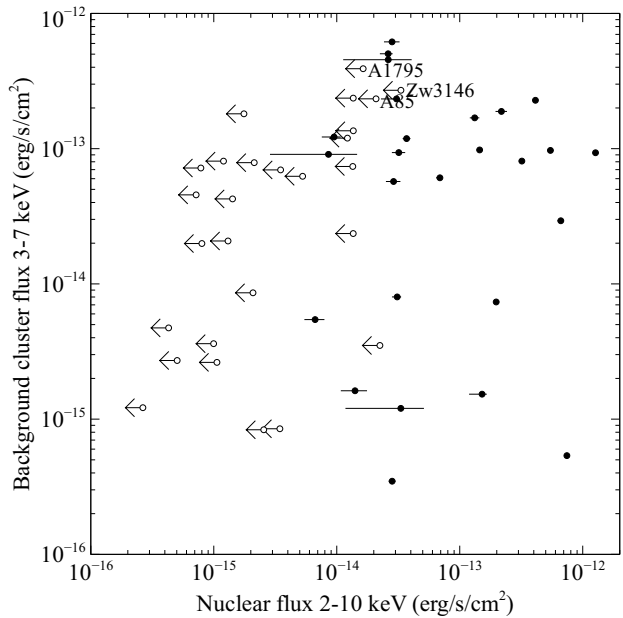}
\caption{Nuclear X-ray flux in the $2-10\keV$ energy band versus the cluster flux from a surrounding annulus in the $3-7\keV$ energy band.}
\label{fig:bias3to7keV}
\end{figure}

The brightness and temperature of the surrounding cluster emission
could also introduce another selection effect to this analysis.  Point
sources are identified by a significant detection of emission in a
hard X-ray band, $3-7\keV$, above the background cluster (section
\ref{sec:psflux}).  Bright, high temperature clusters will have more
emission in this energy band than fainter, cooler systems therefore
potentially making it more difficult to detect a point source above this background.
Fig. \ref{fig:bias3to7keV} shows that for systems with strong
background cluster emission in the $3-7\keV$ band there are several
sources (labelled) with higher upper limits.  This suggests that an
AGN in these BCGs would have to be brighter to be detected than in
BCGs with fainter emission in this energy band.  However, only a
handful of sources in our sample appear to be affected so this will
only slightly reduce the scatter in Fig. \ref{fig:cavpow}.

The selection bias in our sample will have a significant impact on the
best-fit linear relation determined for the correlation in
Fig. \ref{fig:cavpow} (right).  This is illustrated by the difference
in slope between the Buckley-James linear regression in this analysis
and the result found by \citet{Merloni07}.  We have therefore not
drawn any further conclusions from the slope of this correlation but
considered the possible sources of the scatter, which covers at least
three orders of magnitude.

\subsection{Uncertainty in cavity power and black hole masses}

The measured cavity powers available in the literature for a
particular system are often found to differ by factors of a few up to
an order of magnitude.  This can be due to new observations of an
object, which reveal more cavities or better constrain the shapes of
previously known cavities.  It also reflects the inherent systematic
uncertainty and judgement of the extent of the cavity volume (see
eg. \citealt{McNamara12}).  Cavities with bright rims, such as those
in the Hydra A (eg. \citealt{McNamara00}) and A2052
(eg. \citealt{Blanton11}), have a well-defined shape, although some
uncertainty still exists over the line of sight extent and whether the
inner, outer or middle of the rims should be used.  Most cavities do
not have complete rims and their extent is difficult to constrain
given the rapid decline in X-ray surface brightness with radius
(Birzan et al. 2012).  The cavity power for A2390 has a
particularly large error,
$P_{\mathrm{cav}}=1.0^{+1.0}_{-0.9}\times10^{46}\ergps$, because the
extent of cavities is difficult to determine from the X-ray image.  As
discussed in section \ref{sec:cavpow}, the cavity power for 3C295 is
likely to have been significantly underestimated and this is therefore
an outlier.  Therefore, for the majority of the sources, uncertainty
in the cavity power is unlikely to introduce scatter greater than an
order of magnitude.  Also, although the black hole masses are likely
to be a significant source of additional error for
Fig. \ref{fig:cavpow} (right), the scatter in Fig. \ref{fig:cavpow}
(left) is comparably large and does not depend on black hole mass.

\subsection{Absorption}

\begin{figure}
\centering
\includegraphics[width=0.98\columnwidth]{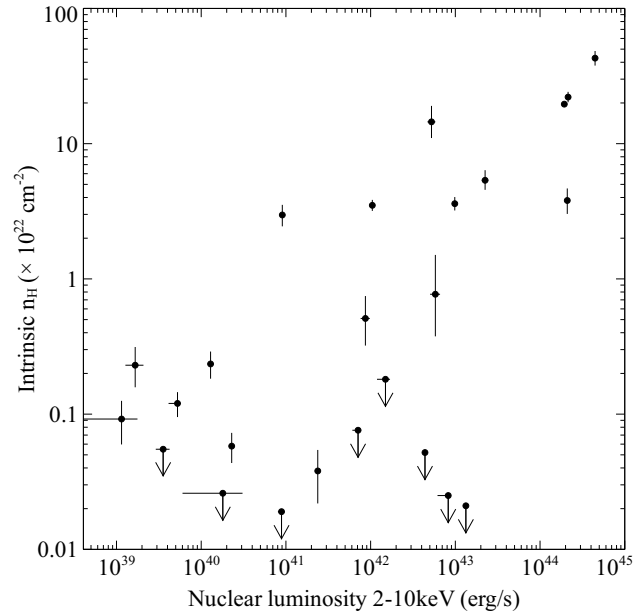}
\caption{Nuclear X-ray luminosity versus the intrinsic absorption of the point source for a subset of objects with sufficient counts for a reasonable spectral fit.}
\label{fig:nHlumin}
\end{figure}

Another significant source of uncertainty in the X-ray point source
fluxes may be attributable to photoelectric absorption by intervening
gas within the galaxy or in a circumnuclear torus.  This would cause a
systematic underestimate of the point source luminosity.  The amount
of intrinsic absorption can be determined from spectral fitting but
only for the brighter sources in our sample.  Fig. \ref{fig:nHlumin}
shows the intrinsic absorbing column density as a function of nuclear
X-ray luminosity for 25 of the 27 detected low luminosity AGN.  There were
insufficient counts for the detections of 3C388 and A2667 to constrain
the intrinsic absorption.  The intrinsic $n_{\mathrm{H}}$ values cover
a range from relatively unobscured sources with
$n_{\mathrm{H}}<10^{21}\pcmsq$, such as M87, up to heavily obscured
narrow line radio galaxies with $n_{\mathrm{H}}>10^{23}\pcmsq$, such
as 3C\,295.  It is therefore plausible that a number of the
non-detected sources could be moderately or heavily absorbed.  This
would account for some of the scatter in Fig. \ref{fig:cavpow} and potentially
render these sources undetectable in the $0.5-7\keV$ energy band
accessible to \textit{Chandra}.  

Fig. \ref{fig:nHlumin} has interesting astrophysical implications as
well.  The most luminous sources in our sample often contain large
columns of cold intervening gas along the line of sight, in some cases
exceeding $10^{23}\mathrm{atoms}\pcmsq$.  It is unknown where along
the line of sight this gas lies but it is likely to be close to the
nucleus.  This implies the existence of a supply of cold gas to fuel
the nucleus.  The range of intrinsic absorption may be related in part
to the geometry of cold gas relative to the central point source.  For
example the point sources in Hydra A and NGC4261 are peering through
circumnuclear gas disks that are highly inclined to the plane of the
sky (eg. \citealt{Jaffe94}; \citealt{Dwarakanath95}).  Furthermore the
path towards the core in Cygnus A is strongly reddened by intervening
dust and presumably accompanying gas located within 800 parsecs of the
nucleus (\citealt{Vestergaard93}).


Using a subset of example upper limits at different redshifts, we
determined that if these undetected sources had an intrinsic
absorption of $10^{22}-10^{23}\pcmsq$ their true luminosity could be
factors of up to a few greater than the observed upper limit.
Intrinsic absorption of $\sim10^{24}\pcmsq$ and above is required for
the true luminosity of the source to be an order of magnitude or more greater
than the observed upper limit.  For a significant amount of the
scatter in Fig. \ref{fig:cavpow} to be generated by intrinsic
absorption, a large fraction of the upper limits must therefore be
Compton thick.  These sources would then represent a very different
population to the detected point sources in this sample.

Although a large fraction of the upper limits in this sample could be
significantly absorbed (eg. \citealt{Maiolino98};
\citealt{Risaliti99}; \citealt{Fabian99}; \citealt{Brandt05};
\citealt{Guainazzi05}), this still seems unlikely to completely
account for the three to four orders of magnitude scatter in
Fig. \ref{fig:cavpow}.  Reliably identifying Compton-thick AGN and
determining their intrinsic luminosity is difficult
(eg. \citealt{Comastri04}; \citealt{Nandra07}; \citealt{Alexander08}).
\citet{Gandhi09} (see also \citealt{Hardcastle09}) found a linear
relation between the $2-10\keV$ X-ray luminosity of the AGN and the
$12\mum$ mid-infrared luminosity.  Somewhat surprisingly,
\citet{Gandhi09} also found that the 8 Compton-thick sources in their
sample did not deviate significantly from this trend.  We might
therefore expect that a large fraction of the undetected sources in
our sample have high infrared luminosities as the absorbed power from the
nucleus is re-radiated at these energies.  

However, for BCGs at the centre of cool core clusters, a significant
fraction of the observed infrared luminosity is likely due to star formation
(eg. \citealt{Egami06}; \citealt{Quillen08}; \citealt{ODea08}).
\citet{Egami06} found that the BCGs in A1835, A2390 and Zw\,3146,
all with point source upper limits in our sample, are infrared bright
and have SEDs typical of star-forming galaxies.  \citet{Quillen08}
identify several BCGs as having very strong AGN contributions in the
infrared, such as Zw\,2089, but in more modest cases this is
difficult to disentangle from star formation.  Interestingly,
\citet{Quillen08} find that the BCG in S\'ersic\,159-03 is infrared
faint with only an upper limit on the IR star formation rate.
Therefore, despite being one of the strongest upper limits in our
sample (Fig. \ref{fig:cavpow}), S\'ersic\,159-03 does not appear to
host a heavily absorbed AGN.  Without detailed SEDs to disentangle the
AGN contribution from star formation in these objects it is difficult
to systematically determine if a significant fraction of the point
source upper limits are heavily absorbed sources.  However, even if
this is the case, it is unlikely to explain the three orders of
magnitude scatter in Fig. \ref{fig:cavpow} (see also \citealt{Evans06}).

\subsection{X-ray variability}
\label{sec:variab}

\begin{figure*}
\begin{minipage}[h!]{\textwidth}
\centering
\includegraphics[width=0.33\columnwidth]{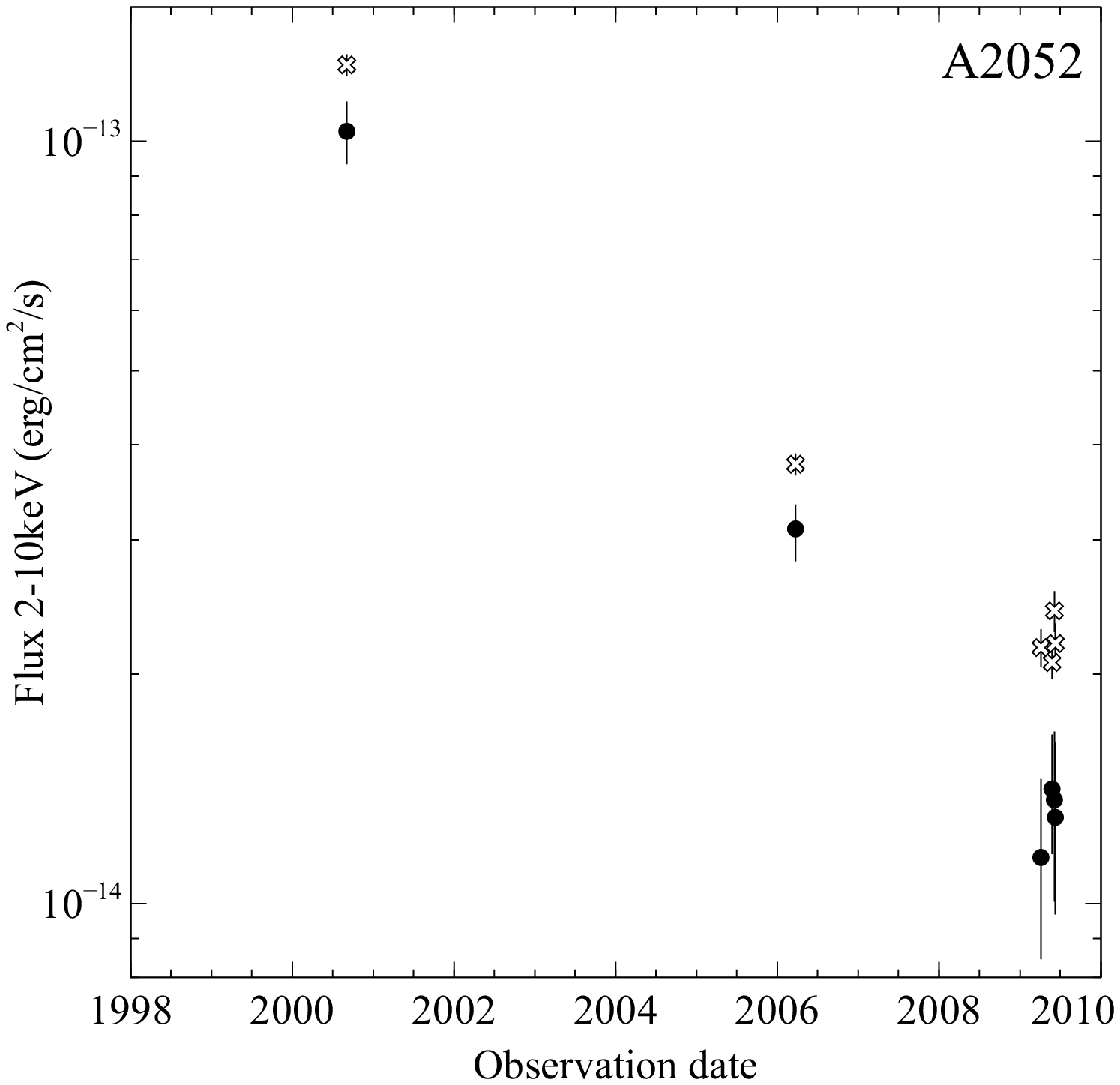}
\includegraphics[width=0.33\columnwidth]{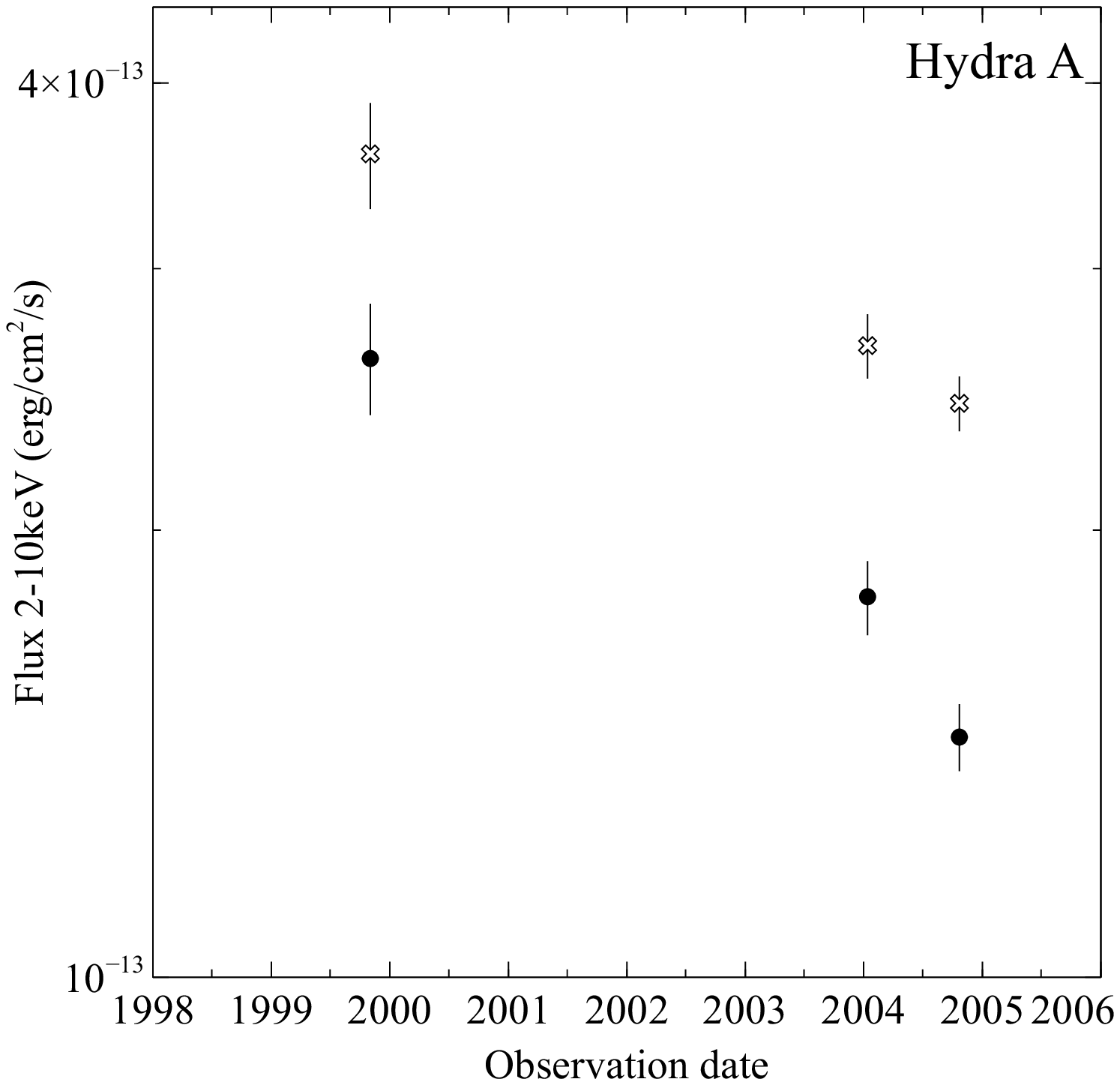}
\includegraphics[width=0.33\columnwidth]{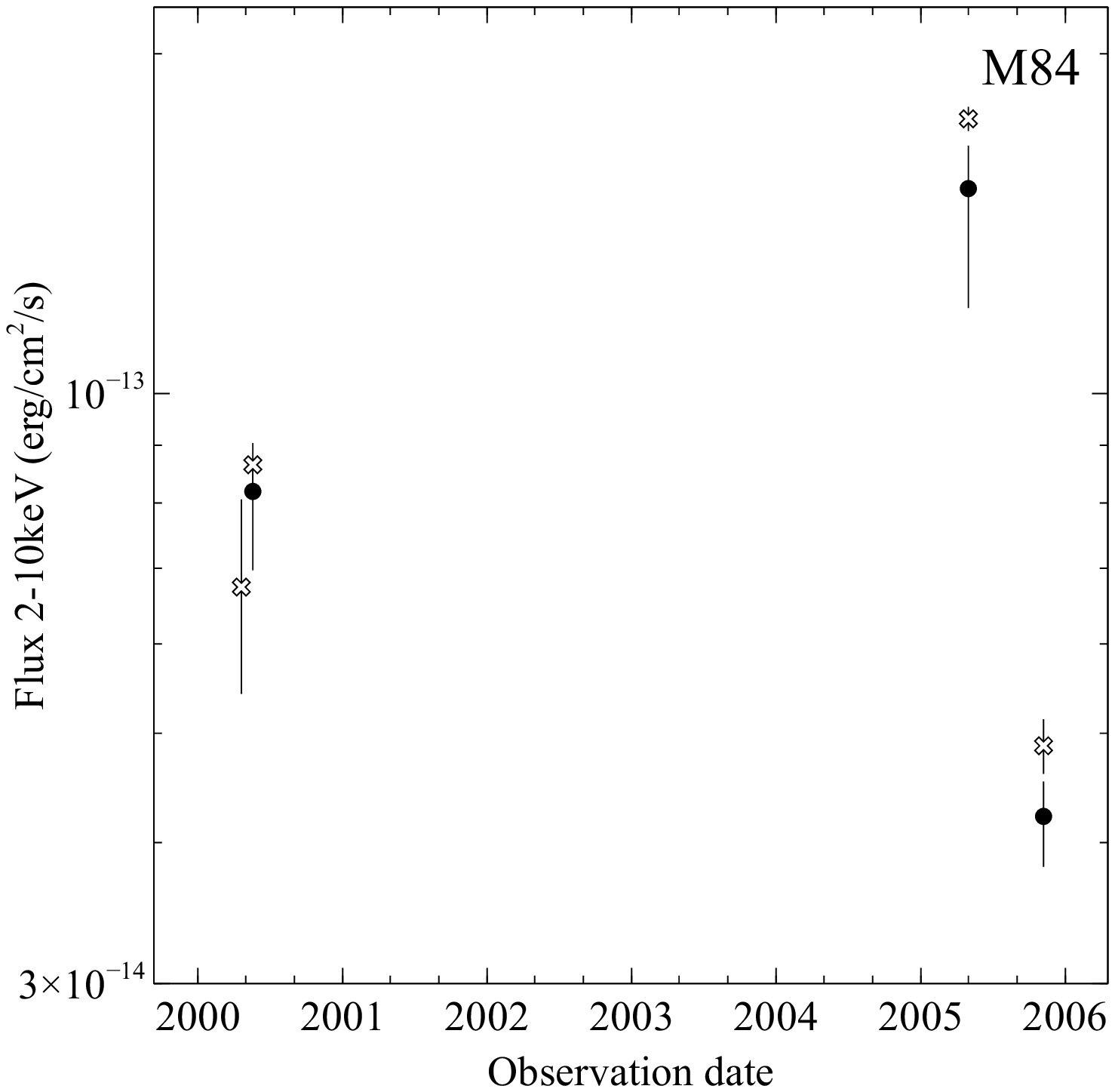}
\caption{AGN flux variability in A2052 (left), Hydra A (centre) and M84 (right).  The point source fluxes are calculated using both the photometric (open crosses) and spectroscopic methods (filled circles).  Note that a spectroscopic flux measurement could not be produced for M84 obs. ID 401 because the $1\ks$ exposure was too short.}
\label{fig:variable}
\end{minipage}
\end{figure*}

\begin{figure*}
\begin{minipage}[h!]{\textwidth}
\centering
\includegraphics[width=0.33\columnwidth]{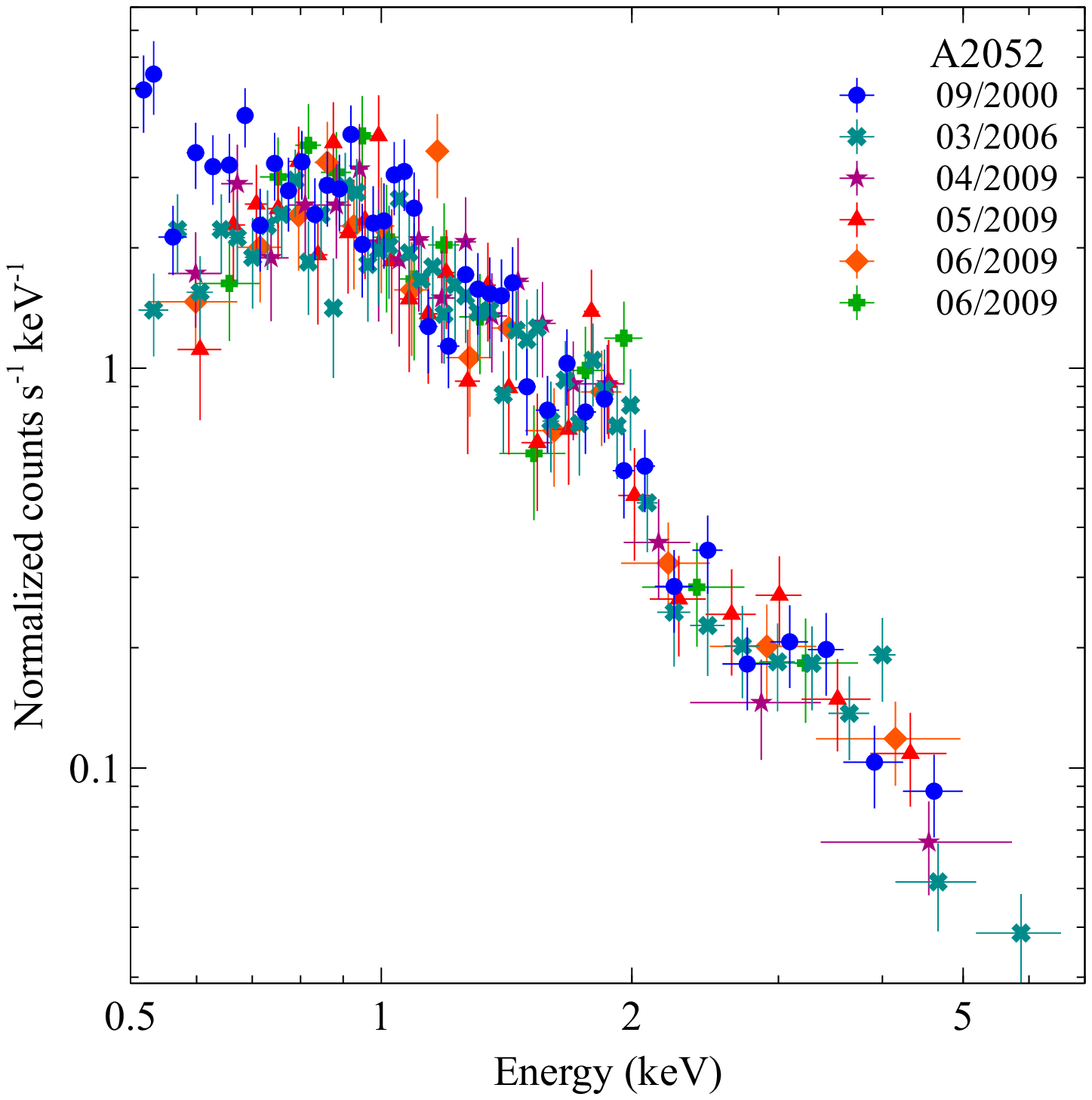}
\includegraphics[width=0.33\columnwidth]{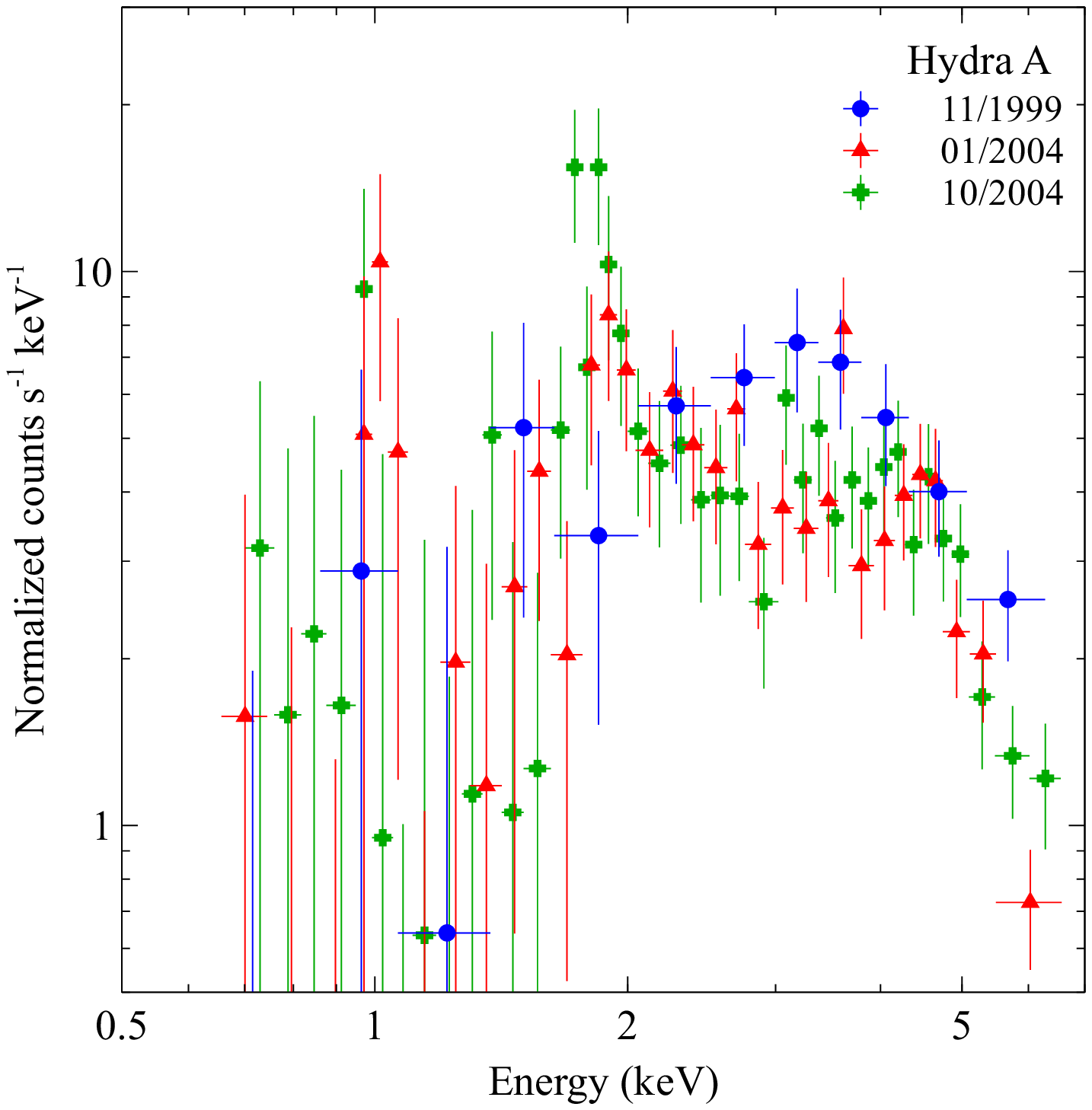}
\includegraphics[width=0.33\columnwidth]{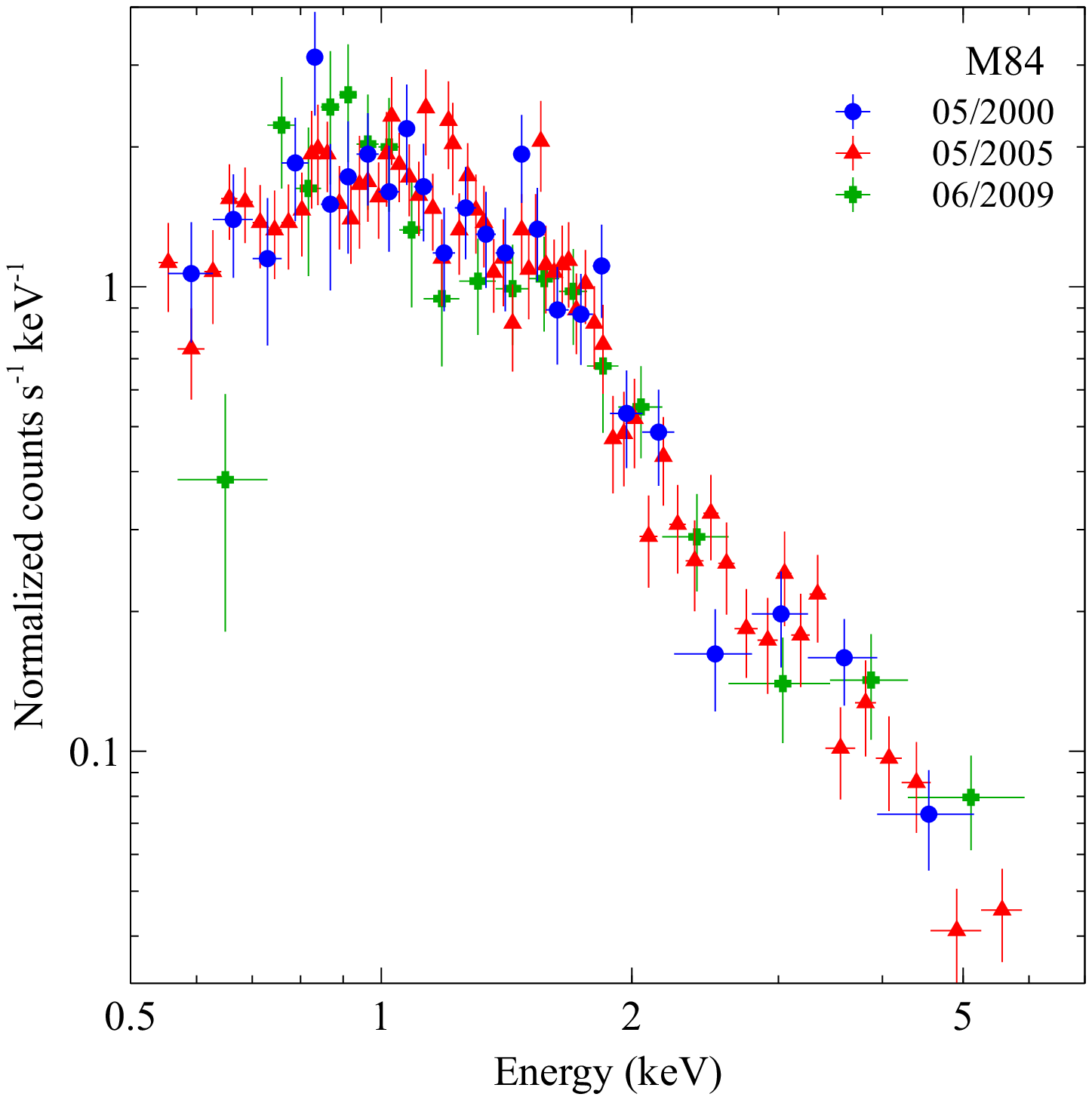}
\caption{Point source spectra normalized to the flux at $2\keV$ for A2052 (left), Hydra A (centre) and M84 (right).  Note that a spectrum could not be produced for M84 obs. ID 401 because the $1\ks$ exposure was too short.}
\label{fig:varspectra}
\end{minipage}
\end{figure*}

Another source of scatter is variability in the nuclear X-ray
luminosity.  The cavity power is averaged over the cavity ages, which
are estimated from the sound crossing time or the buoyant rise time
and are typically $10^7-10^8\yr$ (\citealt{Birzan04}).  However, the
nuclear power is expected to be variable on timescales much shorter
than this.  The effect of variability on the trends in Fig. \ref{fig:cavpow}
is very much like that of relativistic beaming
(\citealt{Merloni07}).  \citet{Merloni07} further pointed out that
variability is likely only to be a problem in the most luminous AGN
such as Seyferts and quasars, which can vary on timescales of weeks.
A subset of the clusters in our sample have multiple observations
spaced by several years in the \textit{Chandra} archive, which we
searched for possible variations in the X-ray point source flux.

The point source flux was calculated for each observation using both
the photometric and spectroscopic methods described in section
\ref{sec:psflux}.  For the spectroscopic method, the differing depth
of the observations could produce variations in the cluster parameters
determined from surrounding annuli, which would add scatter to the
measured point source fluxes.  However, by fixing the cluster
parameters in each observation to those determined from the deepest
exposure available, we verified that this did not significantly alter
the results.  The point source fluxes for this subsample are shown in
Table \ref{tab:variab}.  The best-fit intrinsic absorption and photon
index are shown for each exposure but the fluxes were calculated using
the values of these parameters from the deepest exposure of that
source.  No significant change in the intrinsic absorption or the
photon index was found for the sources analysed.

Fig. \ref{fig:variable} shows that the point source flux was found to
significantly vary in A2052, Hydra A and M84.  The flux in A2052 was
observed to decline by an order of magnitude over the ten years traced
by the \textit{Chandra} archive.  Hydra A shows a more modest decline
by a factor of $\sim2$ over 5 years, whereas for M84 a decline by a
factor of $\sim3.5$ drop in flux occurs in only 6 months.  Note that
for Hydra A the earliest ACIS-I observation in 2000 was not included
(obs. ID 575) as this dataset was taken during the soft proton damage
to the detector.  This analysis is generally limited by the
availability of suitably spaced observations of sufficient depth in
the \textit{Chandra} archive but suggests that a significant fraction
of AGN in BCGs may be varying on timescales of months to a few years.
The bright central point sources in the Perseus cluster and M87 have
long been known to be variable at X-ray wavelengths
(eg. \citealt{Rothschild81}; \citealt{Harris97}).  Another source in
our sample, NGC4261, has been found to be variable on short $3-5\ks$
timescales in a study by \citet{Sambruna03}. This variability is likely to be a
significant source of scatter in the correlation between nuclear X-ray
luminosity and cavity power.  The cavity power is an average of the
AGN activity over $10^7-10^8\yr$ whereas the point source luminosity
is likely to fluctuate significantly on much shorter timescales
potentially by orders of magnitude.

The shape of the nuclear spectrum from Hydra A is dramatically
different from those of A2052 and M84.  Hydra A's spectrum falls
sharply below $2\keV$ while the flux below $2\keV$ in A2052 and M84
continues to rise.  This strong decline in flux shortward of $2\keV$
in Hydra A is due to a large column of intervening gas that may be
associated with the large circumnuclear disk
(eg. \citealt{Dwarakanath95}; Hamer et al. in prep).  We also searched
for a change in the shape of the nuclear spectrum as the sources
varied.  Fig. \ref{fig:varspectra} shows the spectrum for each
observation of each point source found to have significantly varying
X-ray flux.  The cluster background was subtracted from each spectrum
using a surrounding annulus.  The spectra are remarkably consistent
between the observations and suggest that despite the large variations
in flux, particularly in A2052, there has been no significant change
in the shape of the spectrum.


\subsection{Nuclear radio luminosity}
\label{sec:resradio}

\begin{figure*}
\begin{minipage}{\textwidth}
\centering
\includegraphics[width=0.48\columnwidth]{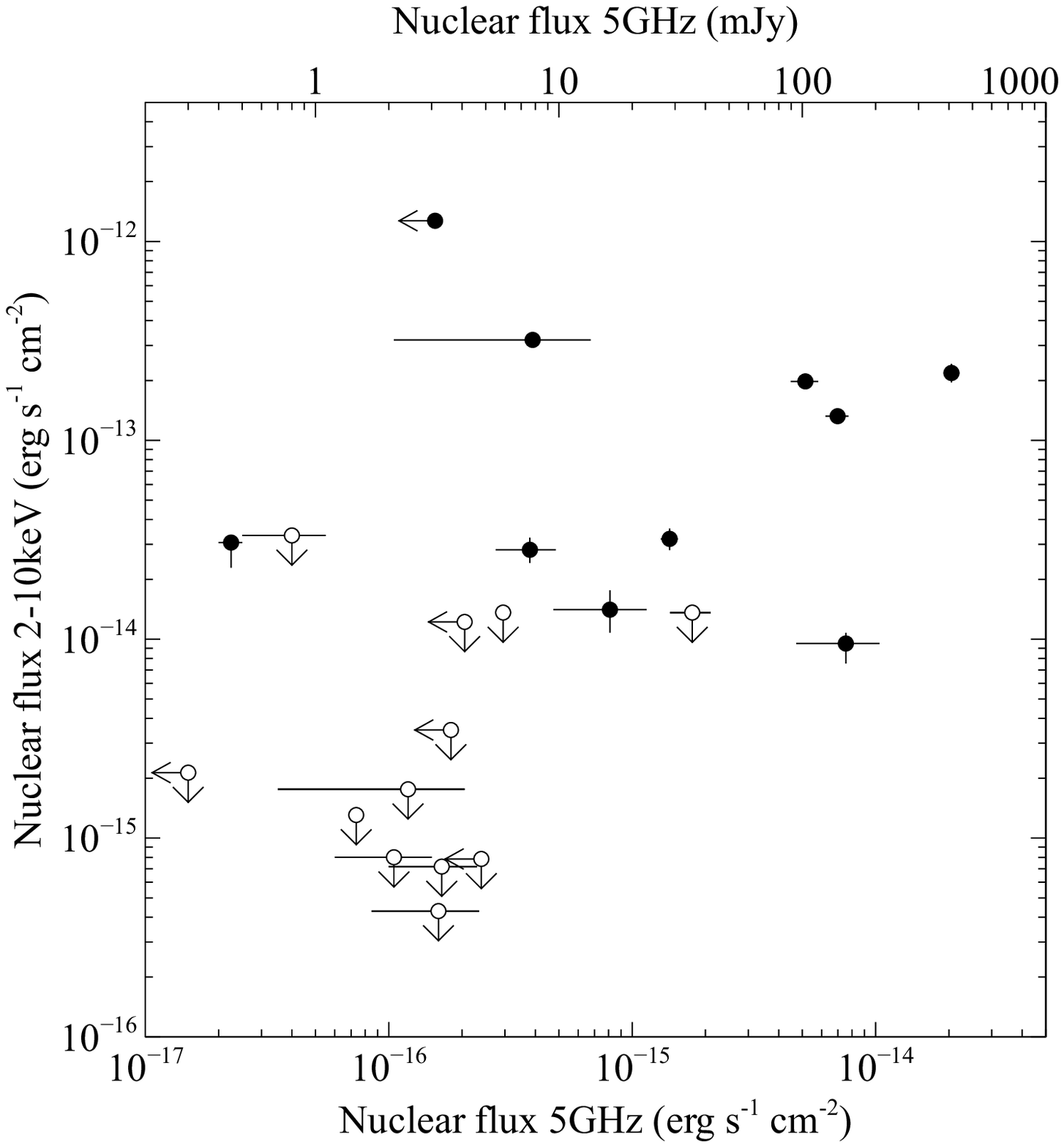}
\includegraphics[width=0.48\columnwidth]{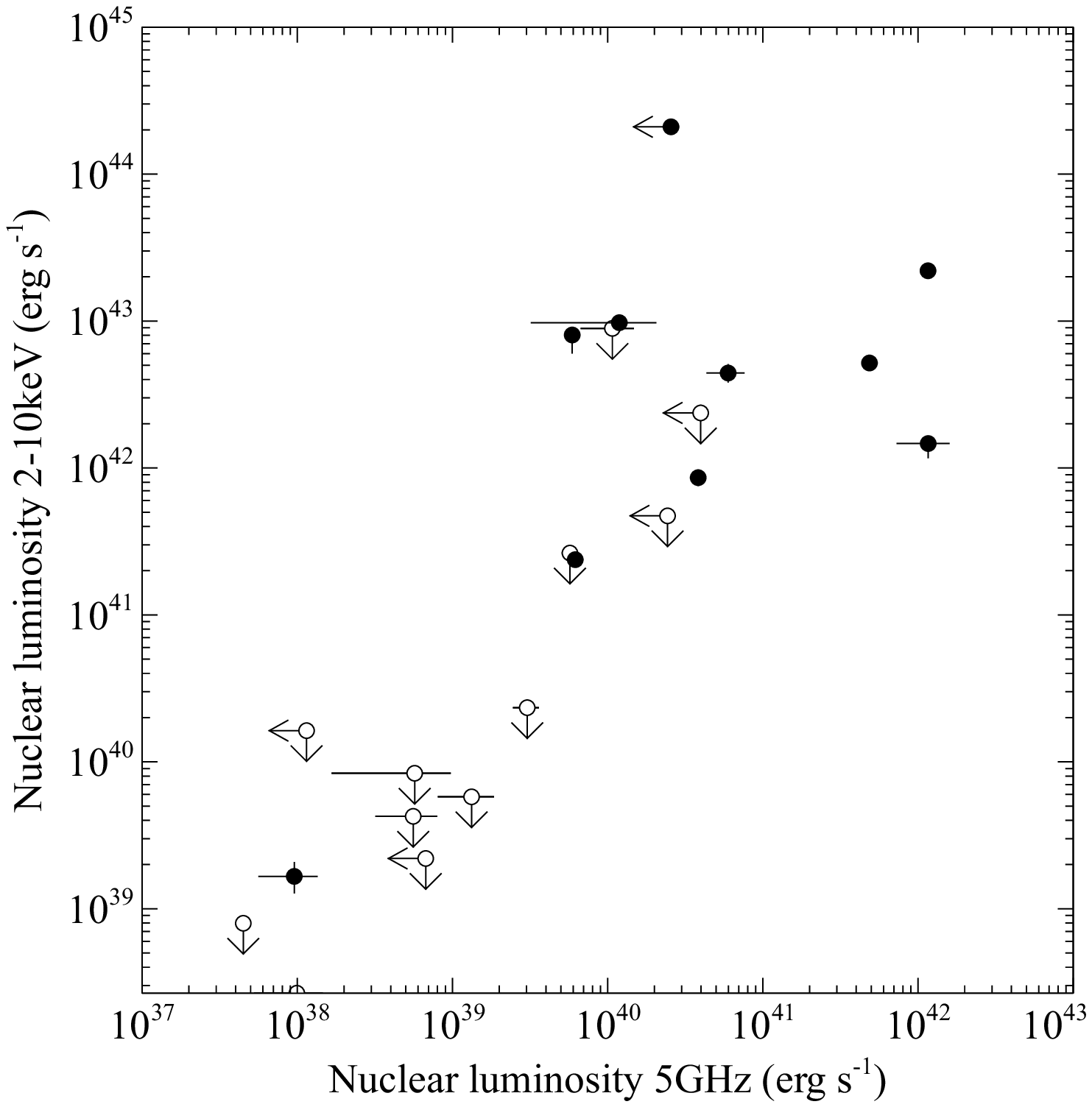}
\caption{X-ray $2-10\keV$ vs. radio $5\GHz$ flux (left) or luminosity (right).  Nuclear X-ray detections are shown by the solid points and X-ray upper limits are shown by the open points.}
\label{fig:xrayradio}
\end{minipage}
\end{figure*}

Fig. \ref{fig:xrayradio} (left) shows no apparent correlation between
the nuclear X-ray flux and the $5\GHz$ radio core flux.  There does
appear to be an approximately linear trend between the nuclear X-ray
luminosity and the radio luminosity (Fig. \ref{fig:xrayradio}, right),
although the X-ray flux is on average an order of magnitude larger.
However, it is highly likely that this trend is due to redshift selection
effects given the lack of a correlation in the flux-flux plot.  

Whilst care was taken to provide reliable core contributions to the
overall radio flux density at C band, there are of course
limitations. A variety of facilities were used to obtain the flux
measurements used in the SEDs.  Whilst this variety was considered in
the decompositions, there will undoubtedly be situations where the
true core contribution is lower than found in this analysis. This is
due to contamination from extended emission in the lower resolution
observations which is not adequately accounted for in the
models. Similarly, for the highly core-dominated sources, large
observed variability may lead to the radio core flux being
underestimated at the epoch of the X-ray observations.  These
shortcomings will be a contributing factor to the scatter seen in
Fig. \ref{fig:xrayradio}.  It should be noted however that the radio core contributions
used here are taken from a larger sample of radio-loud BCGs analysed
by Hogan et al (in prep).  Of this larger sample, 26 are observed with
the VLBA and strong agreement is seen between the direct VLBA core
measurements and the SED-breakdown derived core
contributions. Finally, many of the radio cores are self-absorbed so
the $5\GHz$ flux may significantly underestimate the total radio power
of the core.  There is also likely to be significant scatter due to
variability in both the X-ray and the radio flux.  With no clear trend
between the X-ray and radio nuclear flux it appears less likely that
the X-ray emission originates solely from the base of a jet.

\subsection{Bondi accretion}
\label{sec:resBondi}

The deprojected temperature and electron density profiles for the
Bondi subsample of 13 systems are shown in Fig. \ref{fig:profiles}.
For each cluster we have marked the location of the Bondi radius and
shown that the radial profiles are within roughly an order of
magnitude of this.  The two methods of calculating the deprojected
density profile are consistent as expected.  For the clusters that
overlap with the \citet{Allen06} sample, we generally find good agreement
between the density and temperature profiles.  The Centaurus cluster
profiles were found to differ significantly because \citet{Allen06}
used a $35^{\circ}$ wide sector to the NE of the nucleus compared to our
full annuli, which included the complex structure W of the nucleus.
The temperature profile for NGC5846 is also significantly different in
shape but we note that the central values used for the Bondi analysis
are consistent.  We used a more recent, deep observation of this
source and the results are consistent with the \citet{Machacek11}
analysis.

Inner cavity substructure produced some sharp decreases in the
deprojected density profile in several clusters, including NGC4636 and
NGC5044 (Fig. \ref{fig:profiles}).  The density models were therefore
fitted to all points within the central few kpc to smooth over
substructure that is difficult to correctly deproject.  In general,
the inner radii of the density profiles were well-described by the
three models used to extrapolate to the Bondi radius.  The Bondi
radius, accretion rate and cavity powers calculated from the
temperature and density profiles for each of the selected systems are
shown in Table \ref{tab:bondi}.  Following \citet{Allen06}, we
calculated the cavity power for only the inner two cavities of each
object that are currently being inflated by the central AGN.

\begin{figure*}
\begin{minipage}{\textwidth}
\centering
\includegraphics[width=0.3\columnwidth]{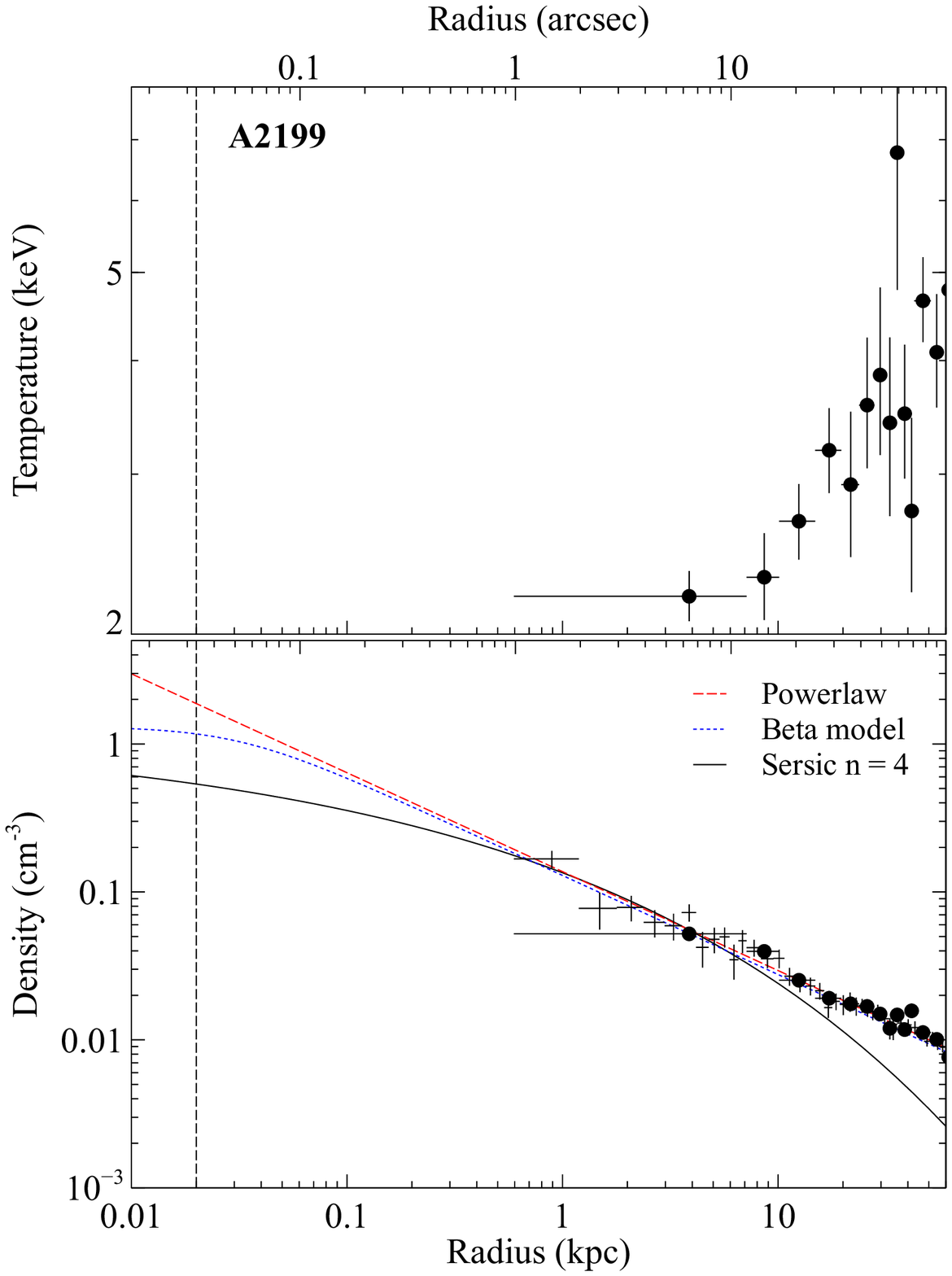}
\includegraphics[width=0.3\columnwidth]{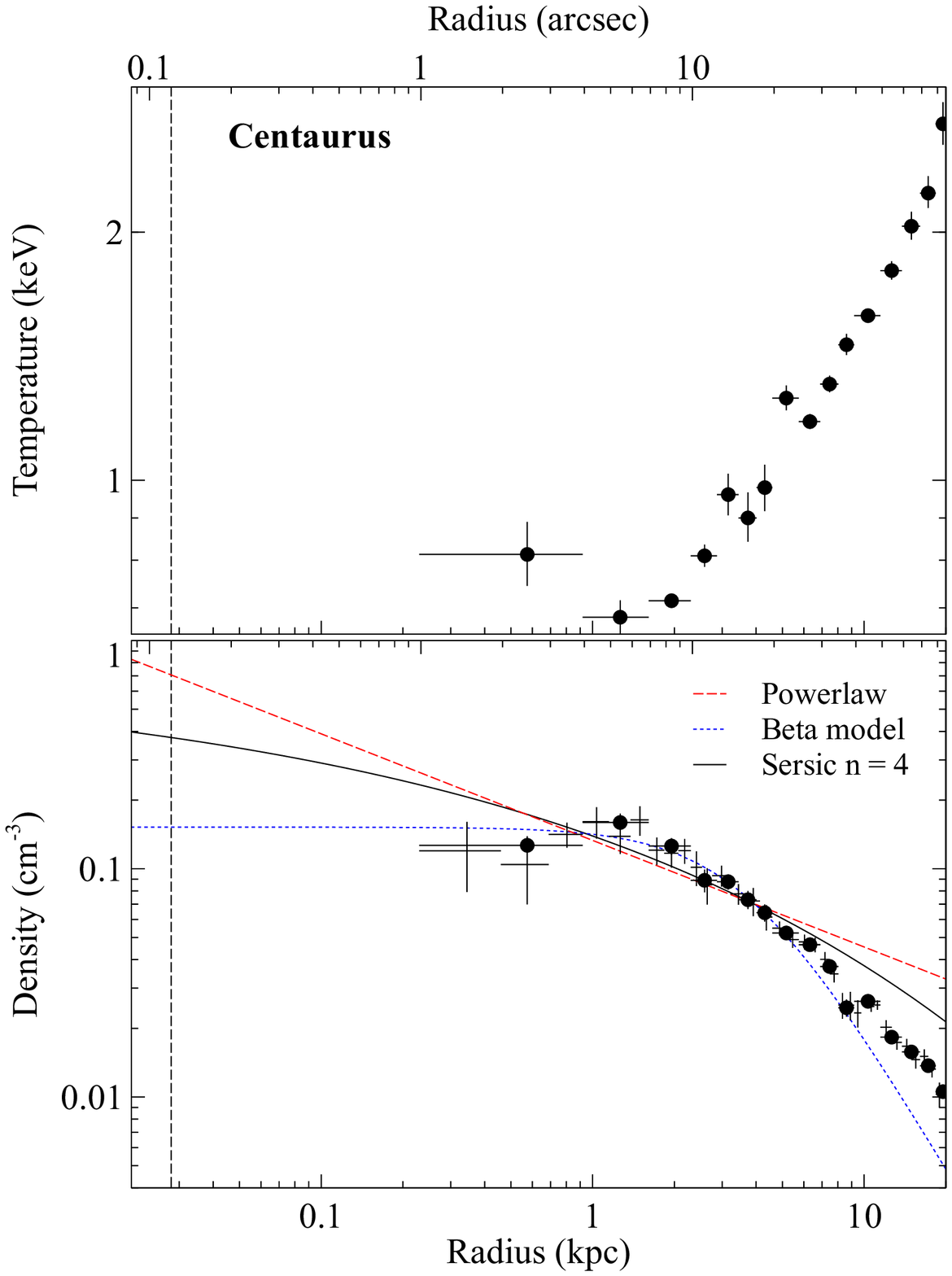}
\includegraphics[width=0.3\columnwidth]{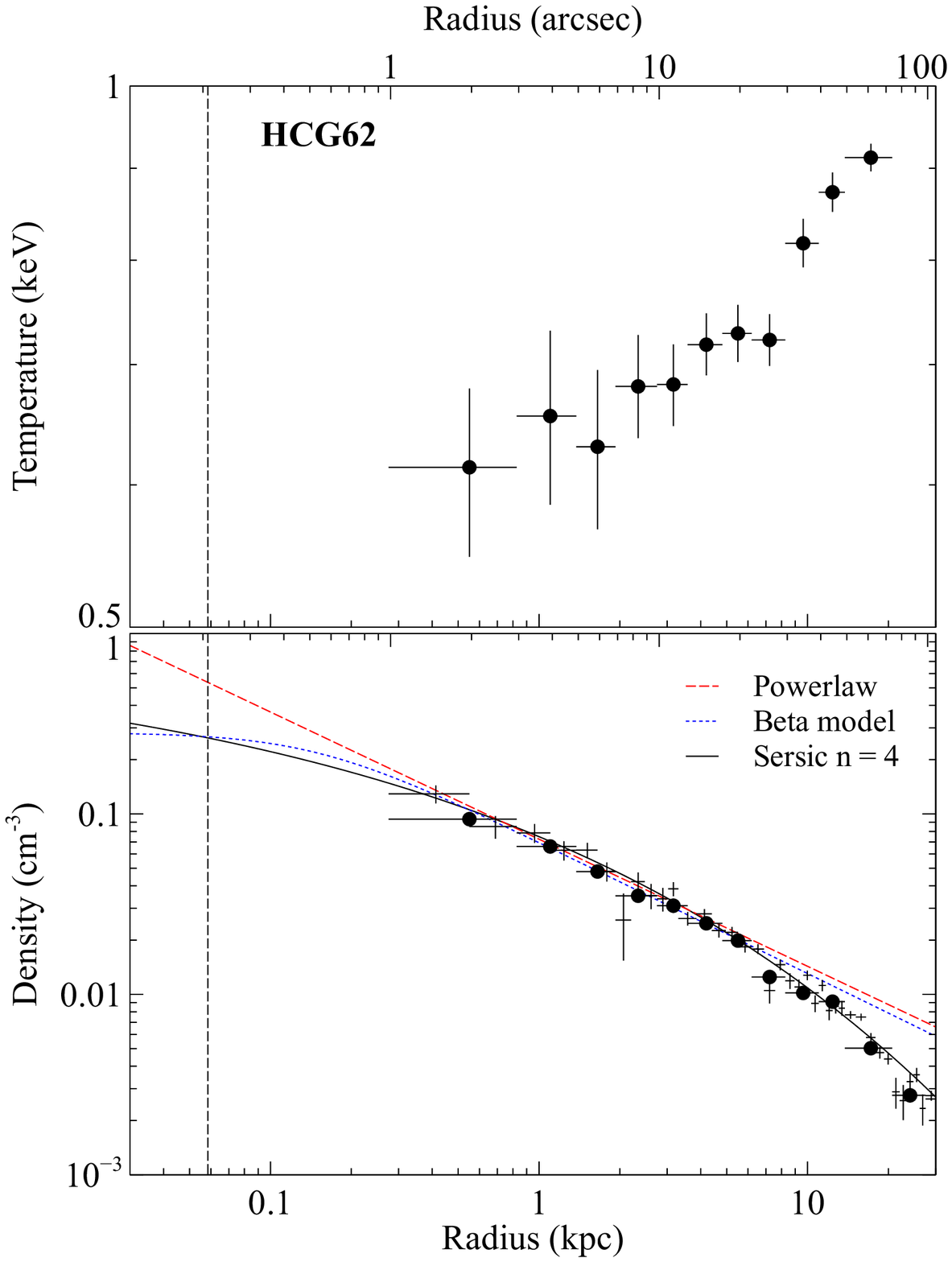}
\includegraphics[width=0.3\columnwidth]{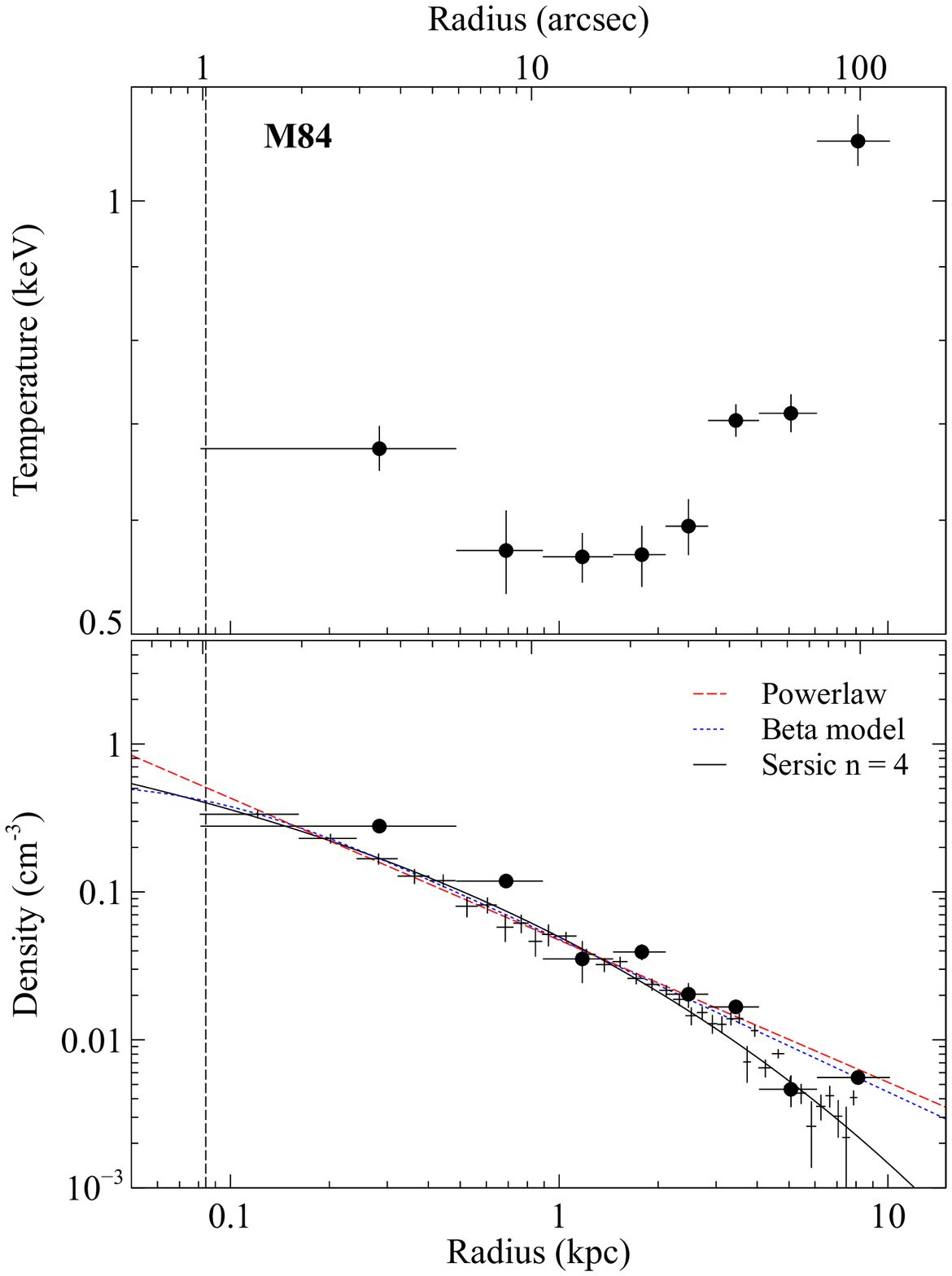}
\includegraphics[width=0.3\columnwidth]{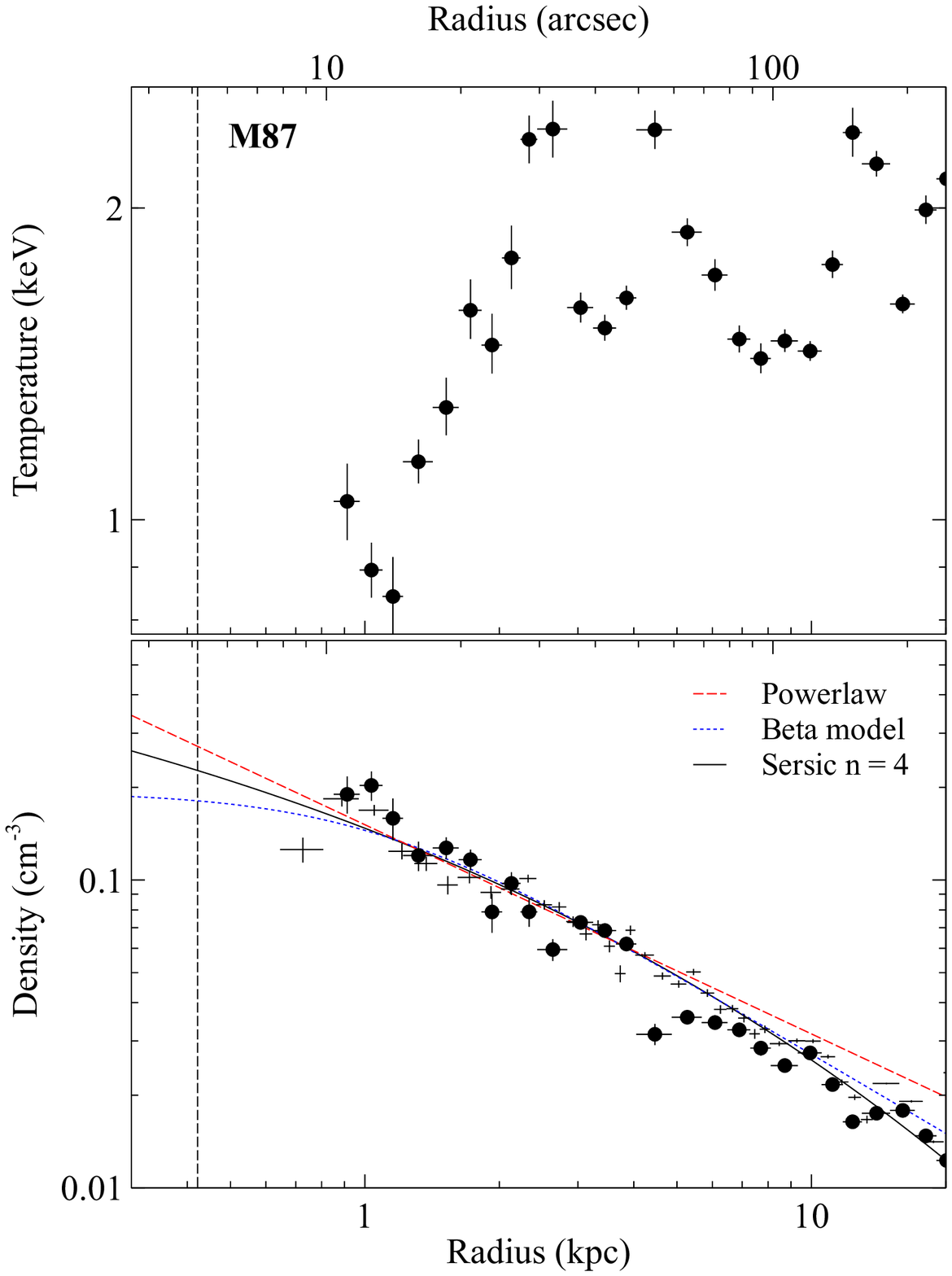}
\includegraphics[width=0.3\columnwidth]{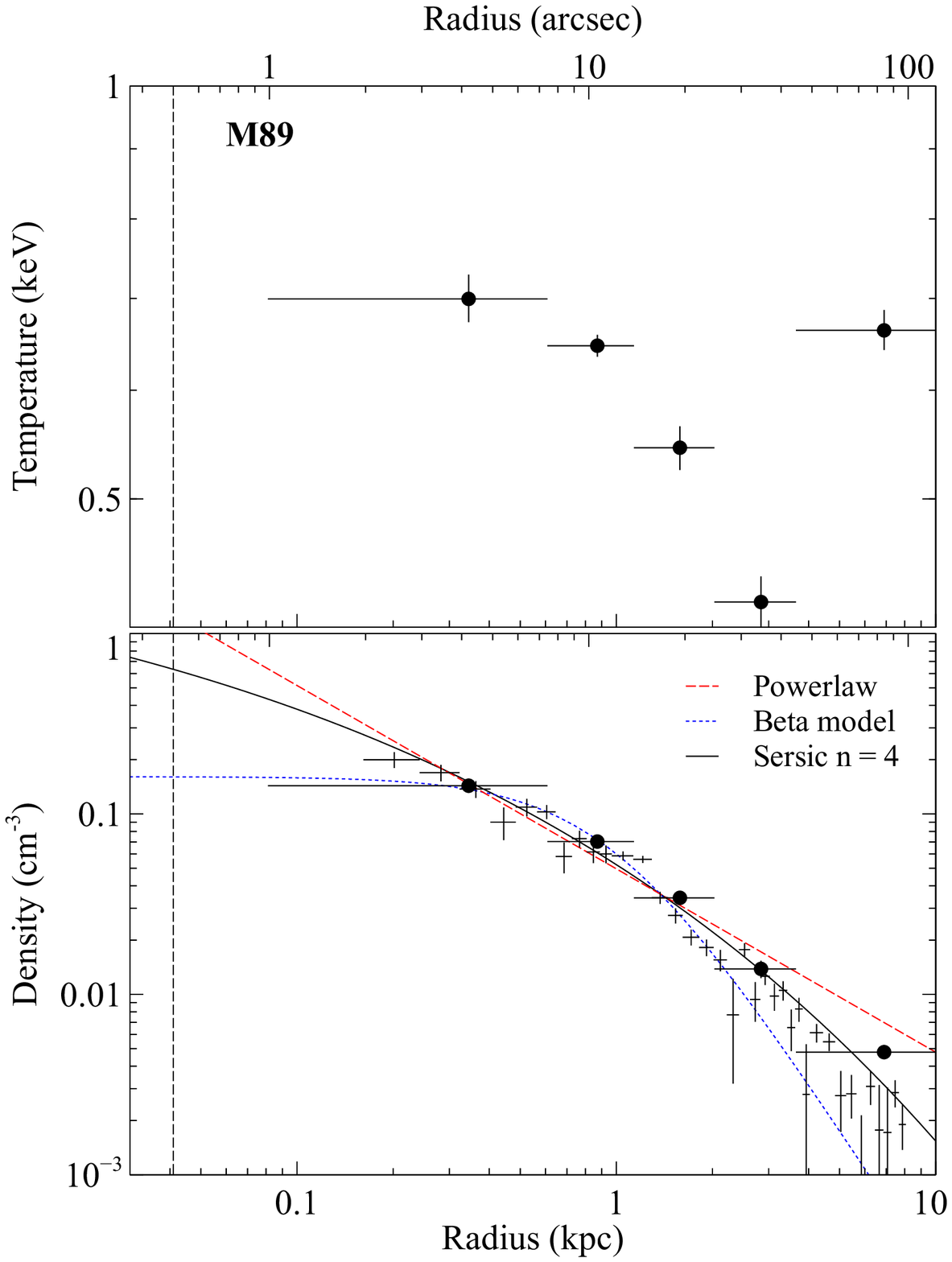}
\includegraphics[width=0.3\columnwidth]{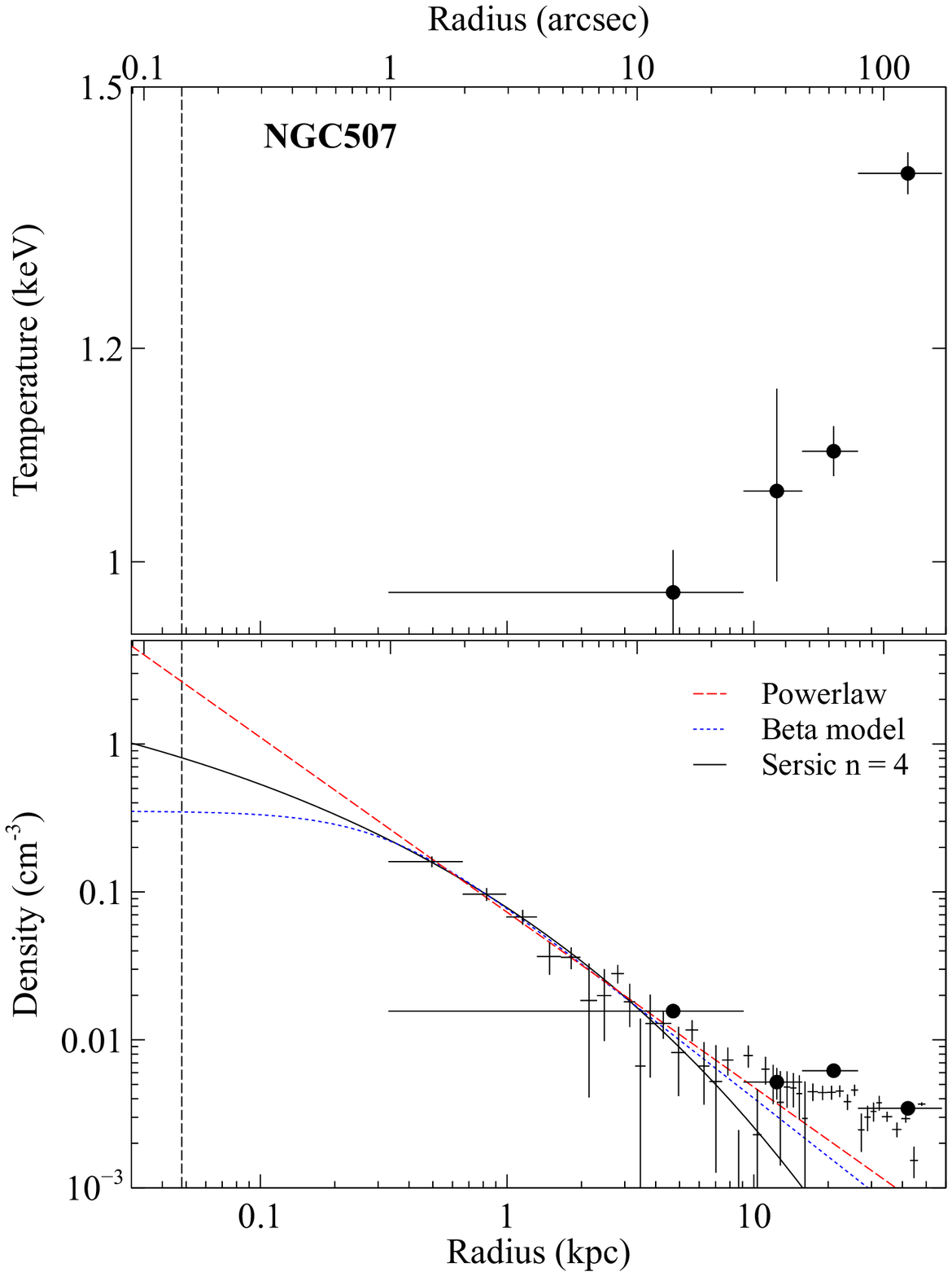}
\includegraphics[width=0.3\columnwidth]{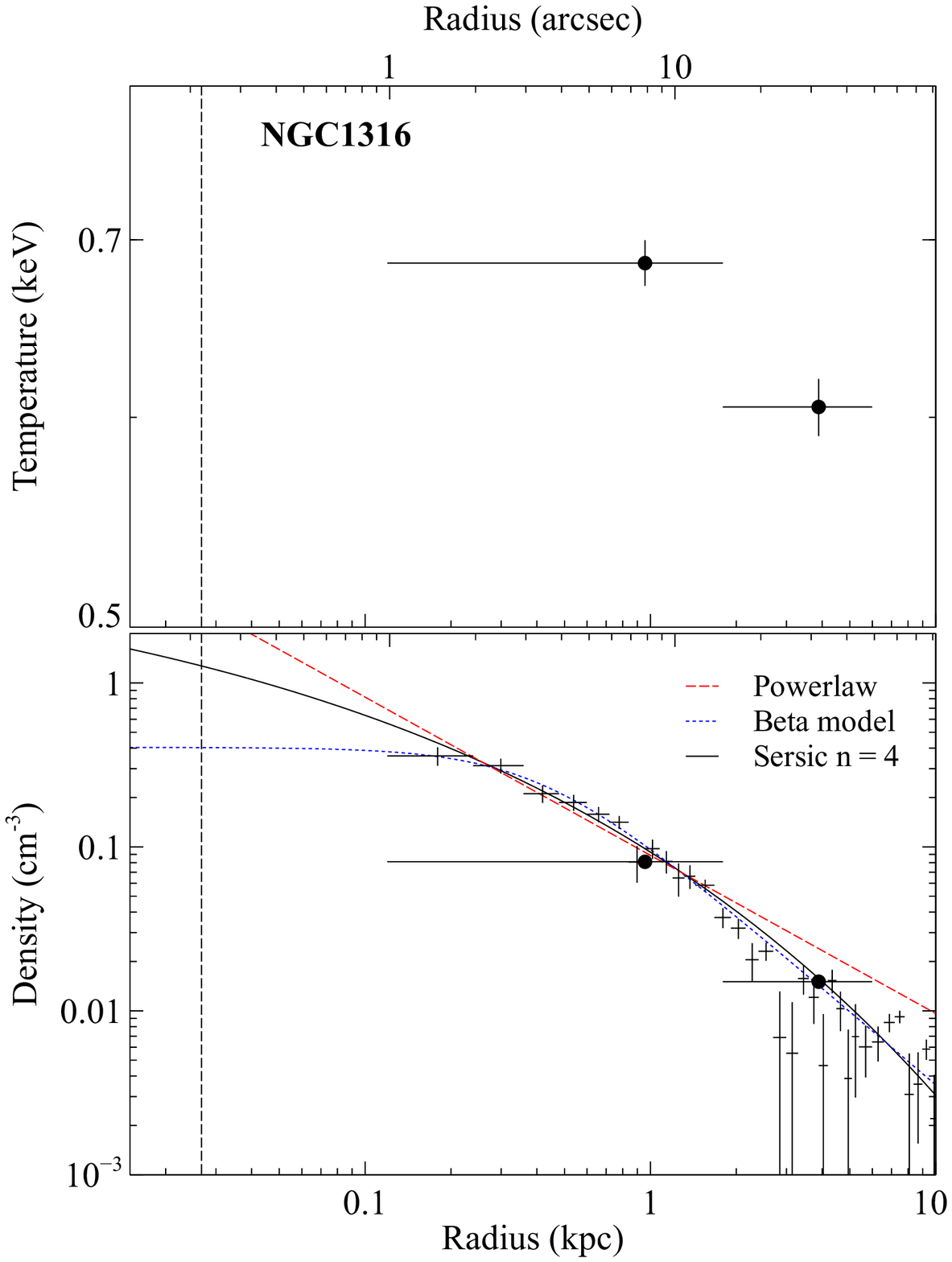}
\includegraphics[width=0.3\columnwidth]{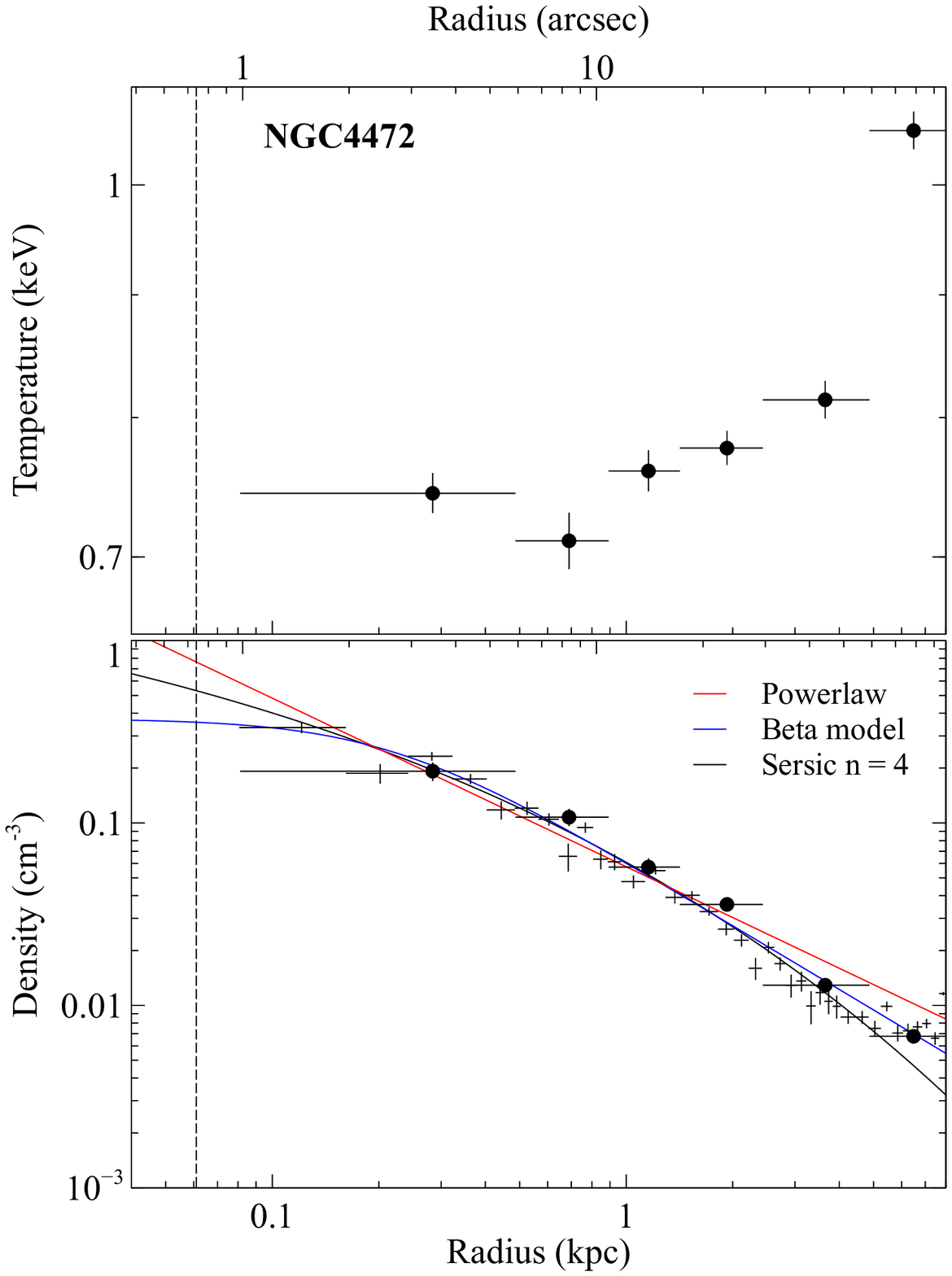}
\captcont{Deprojected temperature and electron density profiles of a
  subset of the cluster sample for which the cluster properties can be
  resolved at radii within an order of magnitude of the Bondi radius (shown by the vertical dashed line).}
\label{fig:profiles}
\end{minipage}
\end{figure*}

\begin{figure*}
\begin{minipage}{\textwidth}
\centering
\includegraphics[width=0.3\columnwidth]{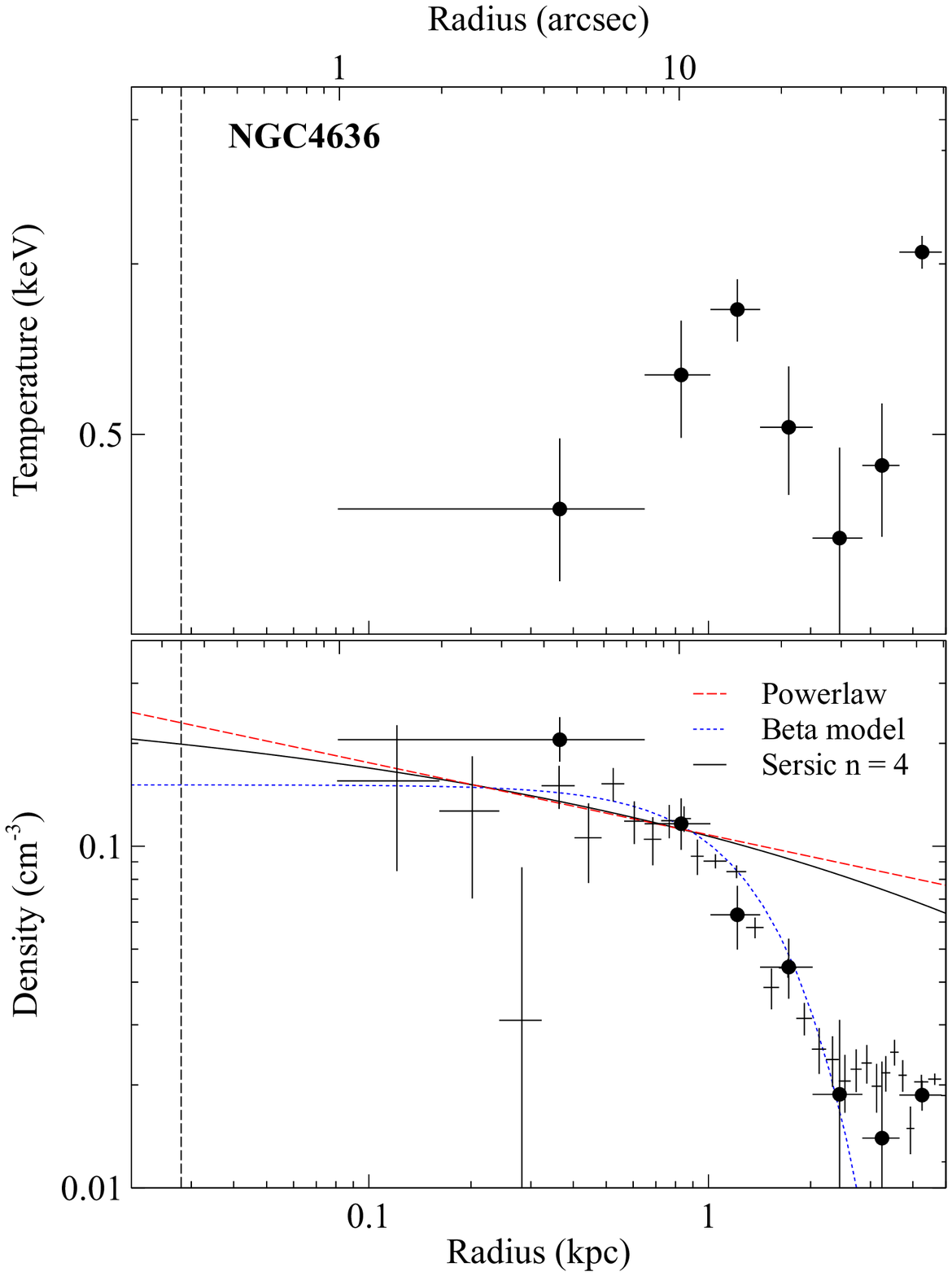}
\includegraphics[width=0.3\columnwidth]{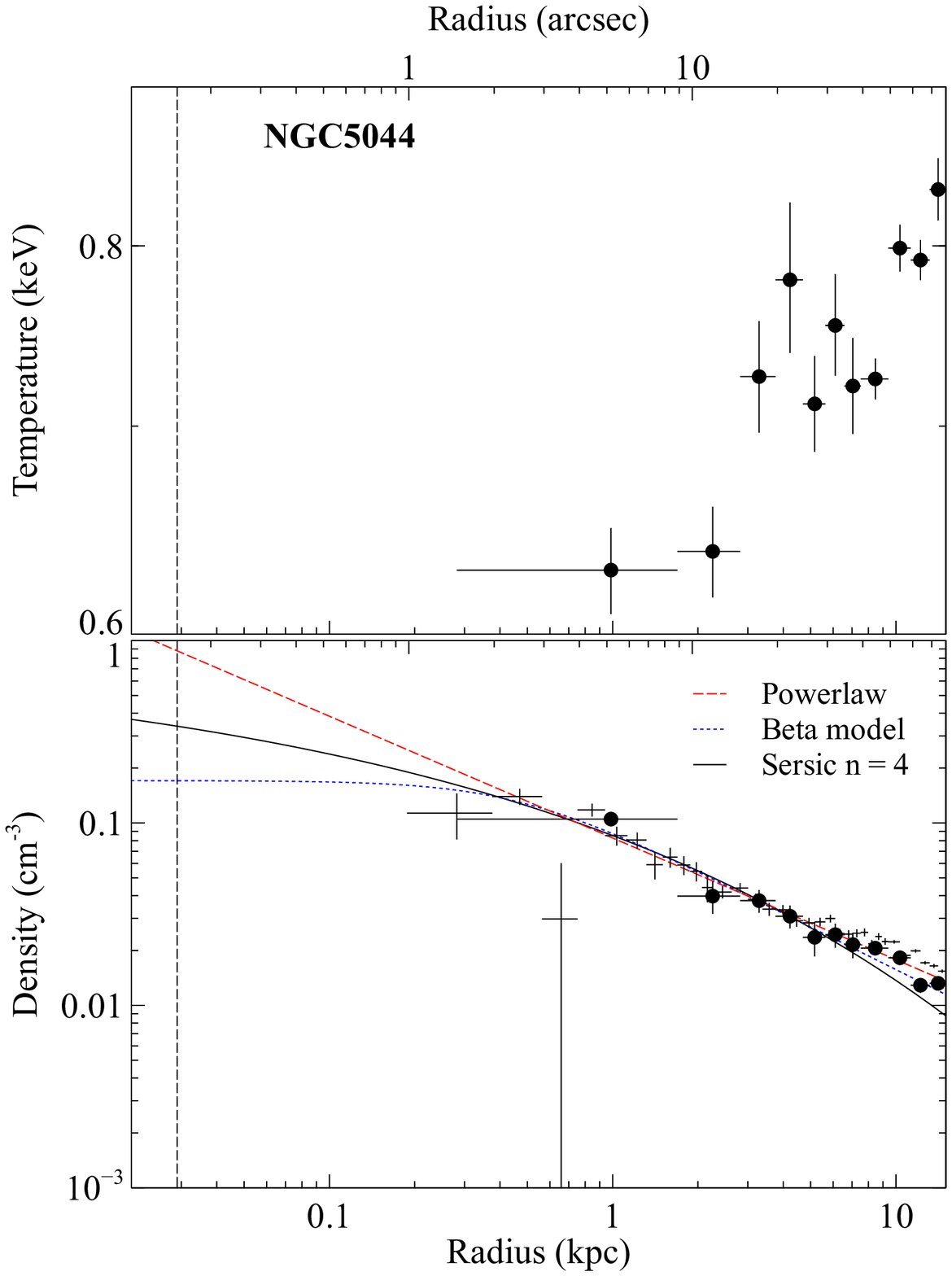} \\
\includegraphics[width=0.3\columnwidth]{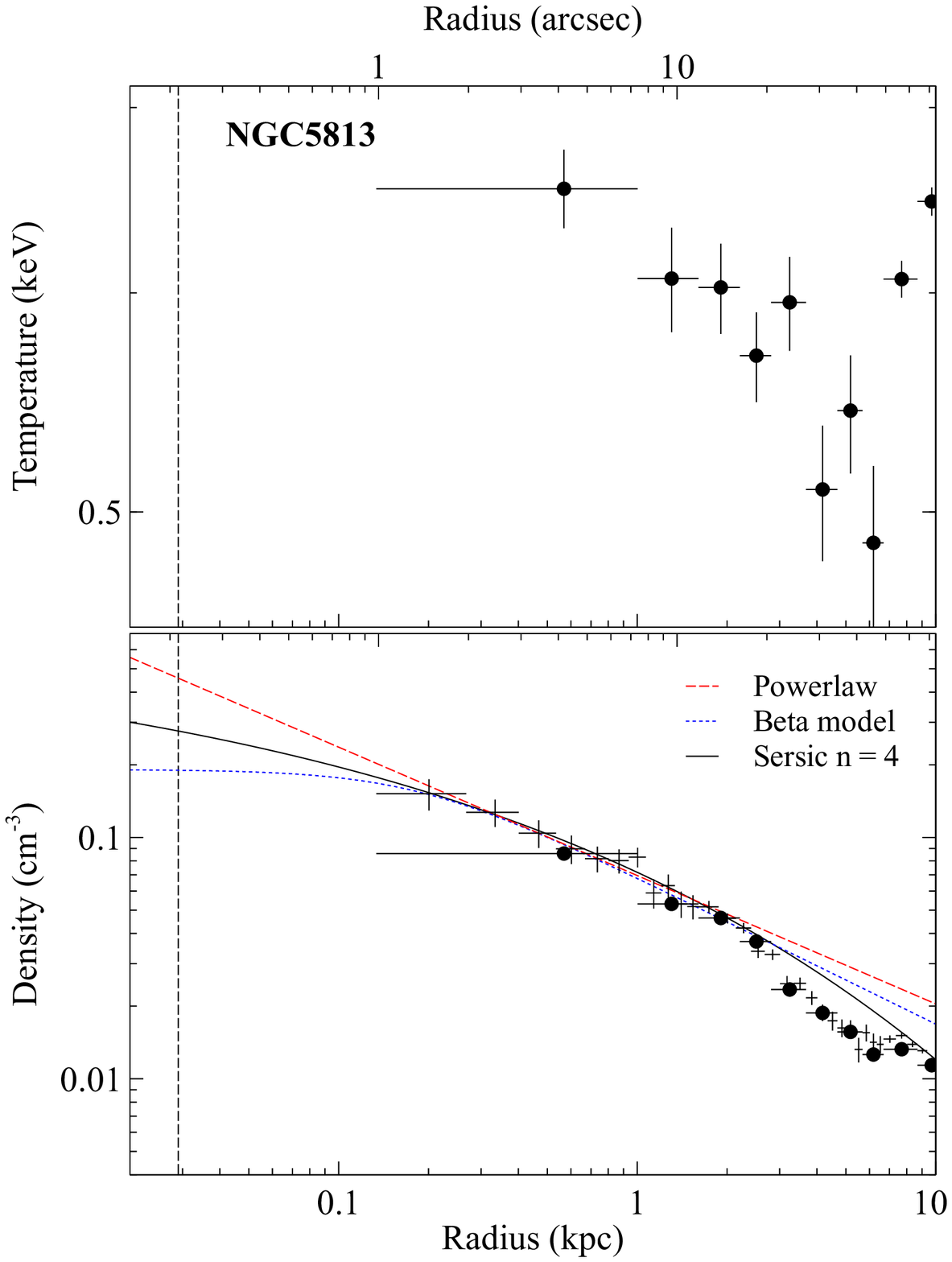}
\includegraphics[width=0.3\columnwidth]{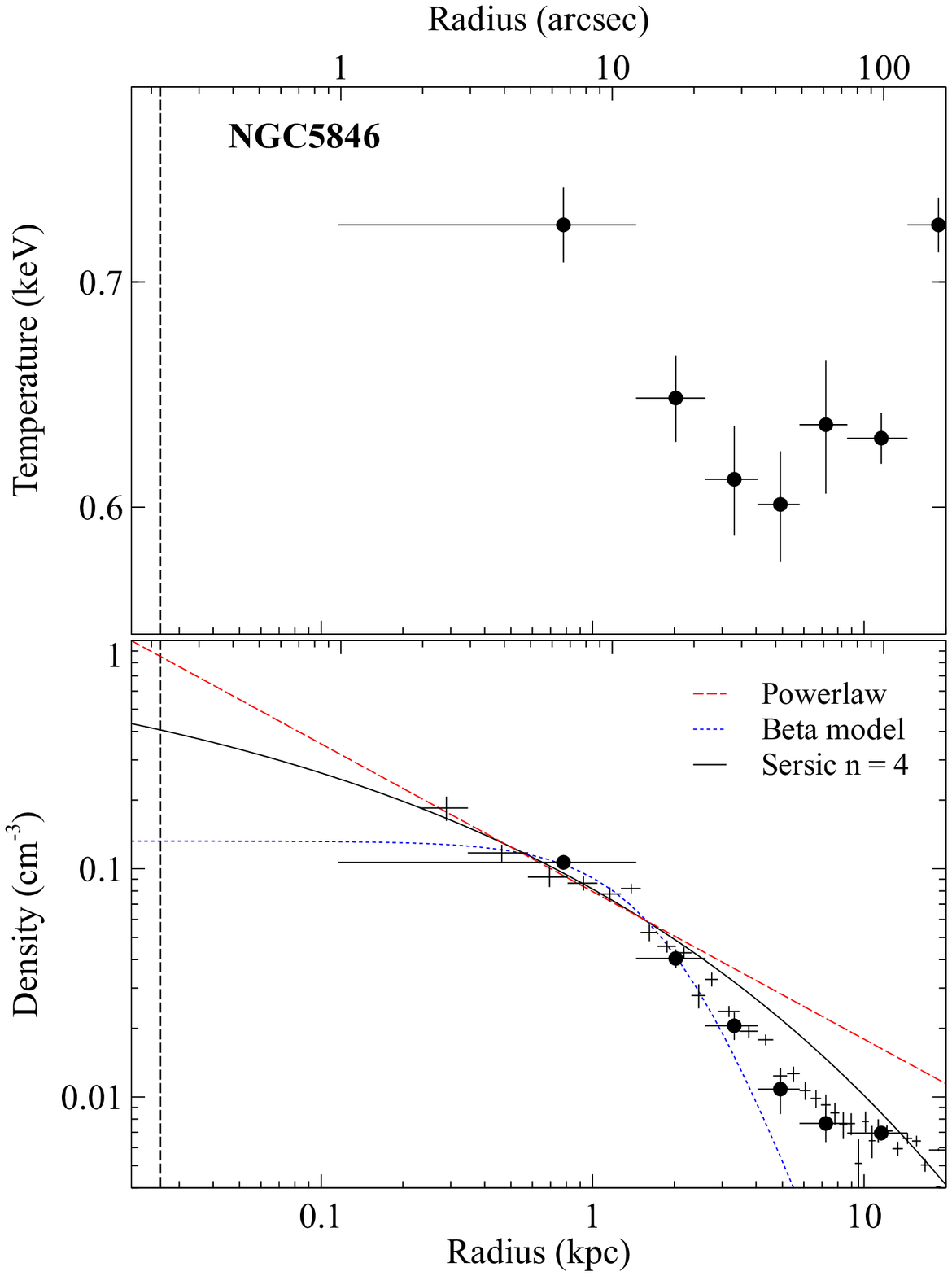}
\caption*{continued.}
\end{minipage}
\end{figure*}

\begin{figure*}
\begin{minipage}{\textwidth}
\centering
\includegraphics[width=0.48\columnwidth]{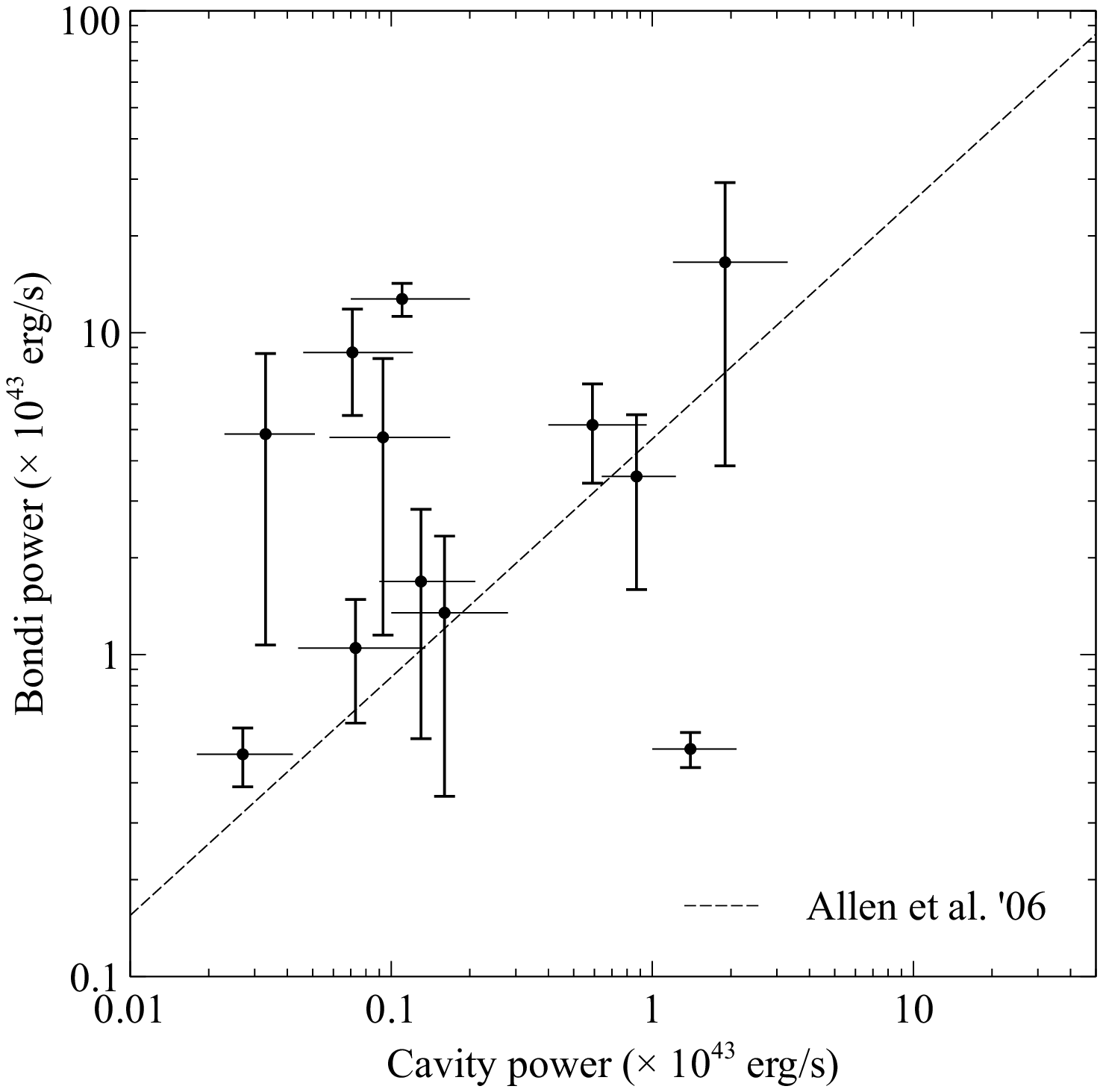}
\includegraphics[width=0.48\columnwidth]{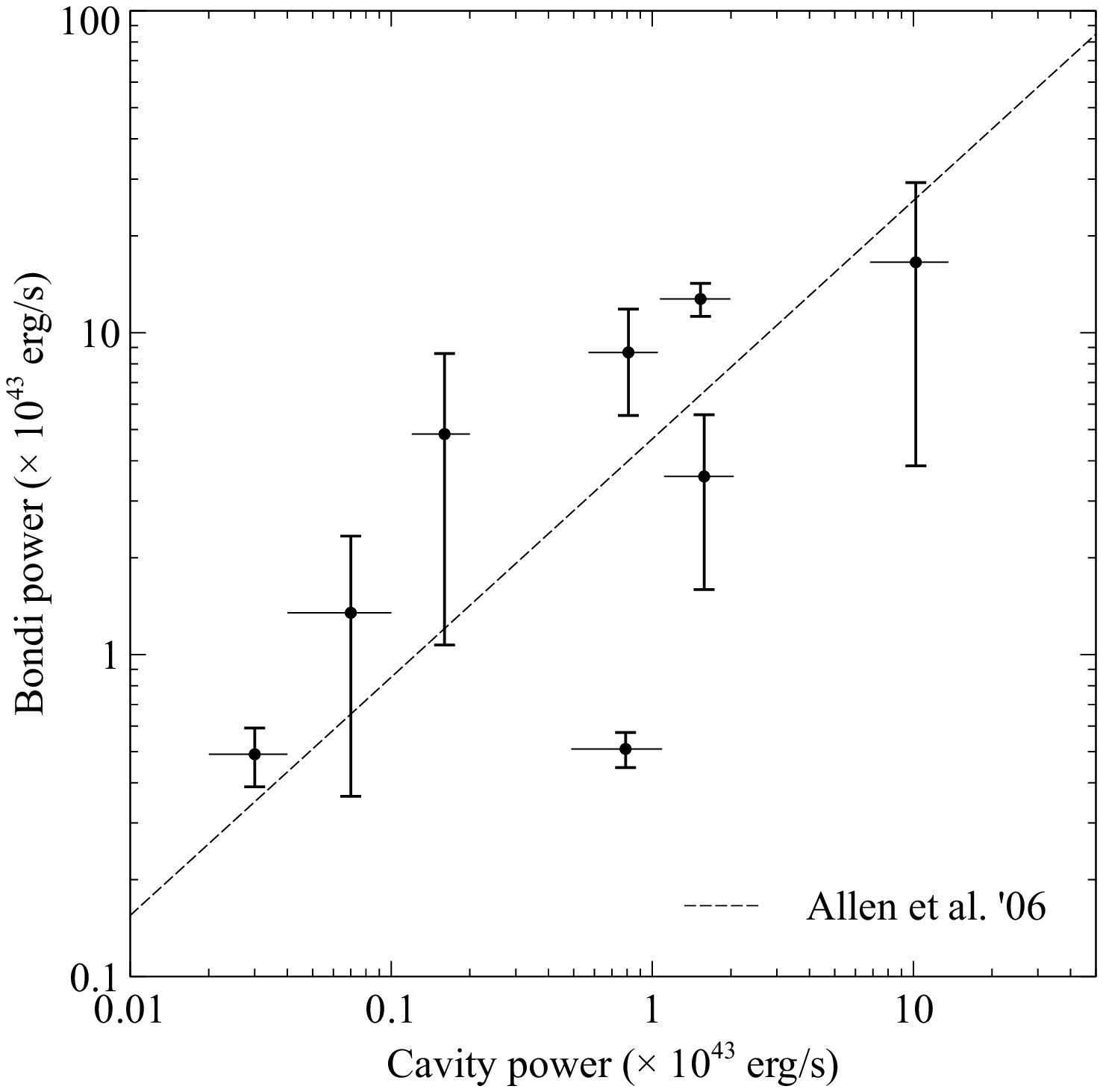}
\caption{Bondi accretion power versus cavity power.  Left: cavity power from this analysis.  Right: cavity power from \citet{Allen06}.  The best-fit relation from \citet{Allen06} is shown as a dashed line.}
\label{fig:Allen}
\end{minipage}
\end{figure*}

Fig. \ref{fig:Allen} (left) shows the Bondi power plotted against the
cavity power generated by the inner two cavities in each system.  This
shows a significant weakening of the trend found by \citet{Allen06}
driven mainly by a difference in the estimates of cavity volume.  For
most of these sources, our cavity powers are consistent with
\citet{Allen06} within the large errors of a factor of $2-3$ on these
values.  However, for M84, M89, NGC4472 and NGC507 our cavity powers
are lower than those of \citet{Allen06} by factors of up to an order
of magnitude (see also \citealt{Merloni07}).  This was partly due to the availability of new, deeper
observations of M84, NGC507 and NGC4472, which more clearly revealed
the cavity extent.  \citet{Allen06} also used $1.4\GHz$ radio images to
determine the edges of the cavities and this may have caused
significant differences from our primarily X-ray method.  Our new
estimates of X-ray cavity power agree with estimates from
\citet{Cavagnolo10}, \citet{Rafferty06} and \citet{OSullivan11}.  It
is not clear whether cavity powers will be more accurate when
calculated using the X-ray or the radio observations, therefore we
have also included Fig. \ref{fig:Allen} (right) showing our analysis
of the Bondi accretion power versus the cavity powers from
\citet{Allen06}.  This plot shows a larger scatter than
\citet{Allen06} found and this scatter is due solely to differences in
how we and \citet{Allen06} calculated the central density.  The
vertical `error bars' should not be interpreted as such.  Instead they
represent the range of Bondi powers from the three best-fit models and
the midpoint is marked as no model provides a significantly better
fit.  The exception is the Centaurus cluster where the $\beta$-model
is significantly preferred over the powerlaw and S\'ersic profiles.
We therefore used the Bondi power from the best-fit $\beta$-model and
its associated errors.

The Kendall's $\tau$ rank correlation was used to determine if these
two measures of cavity power are significantly correlated with the
Bondi accretion rate.  For our estimates of cavity power, we find no
significant correlation with $\tau=0.2$.  For the estimates of cavity
power from \citet{Allen06}, we calculate $\tau=0.7$ and reject the
null hypothesis of no correlation at 95\% confidence but not at 99\%
confidence.  This analysis therefore suggests weaker evidence for a
trend between the cavity power and Bondi accretion power, primarily
due to the uncertainty in estimates of the cavity volumes.

\section{Discussion}


X-ray central point sources appear to be common in BCGs hosting X-ray
cavities.  We find a detection fraction of $\sim50$ per cent for the BCGs in our sample.  The
majority of these sources are radiatively inefficient with required
average accretion rates of only $10^{-5}-10^{-2}\dot{M}_{\mathrm{Edd}}$.  The
nuclear X-ray luminosity for these sources was observed to correlate
with the AGN cavity power, which is surprising given the vastly
different timescales for these quantities.  Cavity power is averaged
over the bubble ages, typically $10^{7}-10^{8}\yr$, while the nuclear
X-ray luminosity is an instantaneous measurement and we have shown
that this can vary significantly on shorter timescales of months to
years.  The scatter in this correlation covers over three orders of
magnitude.  A significant fraction of this scatter is likely due to
X-ray variability but absorption and uncertainty in the cavity power
estimates will also contribute.

The interpretation of these results is complicated by the uncertainty
in the origin of the nuclear X-ray emission.  The X-ray emission may
originate from the accretion disk corona, from the base of a
parsec-scale jet or a combination of the two although another origin
is also possible.  However, the nuclear X-ray emission is generally
considered a probe of accretion power, whether it is from the
accretion flow or from the jet (eg. \citealt{Falcke95};
\citealt{Heinz03}).  Therefore, the observed
$L_{\mathrm{X}}-P_{\mathrm{cav}}$ correlation suggests the accretion
power roughly scales with the cavity power over long timescales with
the large scatter reflecting variability on shorter timescales.

\subsection{Duty cycle of activity}

It is also not clear if systems with only upper limits on the point
source luminosity are simply faint or currently `off'.  As shown in
Fig. \ref{fig:cavpow}, whether the upper limits form the faint end of
the detected source population or are instead a different population
of `off' systems can have a significant impact on the correlation's
slope and scatter.  Approximately $\sim50$ per cent of the sample do
not have detected nuclear X-ray emission.  It is likely that at least
some sources are simply a little too faint to be detected,
particularly if they are embedded in bright cluster emission (see
section \ref{sec:seleff}).  These objects may therefore still be
consistent with the observed $P_{\mathrm{cav}}-L_{\mathrm{X}}$
correlation.  However, objects such as MS0735 and Sersic\,159 have
upper limits on their radiative luminosities a factor of thousand
below that expected from this trend and are effectively `off'.


\citet{HlavacekLarrondo11} considered a sample of highly radiatively
inefficient nuclei in clusters with powerful AGN outbursts, including
MS0735, and suggested several explanations including absorption and
variability.  We have found significant variability for several
sources in a subset of our sample, which could indicate a cycle of
activity, but don't find absorption to be as important.  Although we
caution that this sample is by no means complete, the fraction of
detections to non-detections indicates a duty cycle of at least
$\sim50$ per cent in systems with recent AGN outbursts.  If we
consider only the 31 sources that overlap with the \citet{Rafferty06}
sample, we find a similar detection fraction of at least 40 per cent.
This suggests that roughly half of all systems undergoing an AGN
outburst in the last $\sim10^8\yr$ have evidence of ongoing accretion.
\citet{Mendygral12} found that simulations with a jet duty cycle of
$50$ per cent, cycling on and off with a $26\Myr$ period, produced
multiple cavity pairs with a similar morphology to observations (see
also \citealt{ONeill10}; \citealt{Mendygral11}).  For complete samples
of clusters, the fraction with detected X-ray cavities implies a duty
cycle of at least $\sim60-70$ per cent (\citealt{DunnFabian06};
\citealt{Birzan09}; \citealt{Birzan12}).

\subsection{Radiative efficiency and evidence for a transition luminosity?}

Although it is not clear if the upper limits form the faint end of the
detected source population or are a separate population of `off'
systems, the best fit models for these two possibilities have a
consistent slope in Fig. \ref{fig:cavpow} (left).  This slope shows an
increase in radiative efficiency with the mean accretion rate.  The
quasars included for comparison form an extension of this trend from
cavity power-dominated to radiation-dominated sources.  Studies have
also shown that the radio loudness of low luminosity AGN to luminous
quasars is inversely correlated with the mass accretion rate
(eg. \citealt{Ho02}; \citealt{Terashima03}; \citealt{Panessa07}).
Supermassive black holes appear to become more efficient at releasing
energy through jets as their accretion rate drops.  Hlavacek-Larrondo
et al. (submitted) also find strong evolution in the nuclear X-ray
luminosities of SMBHS hosted by BCGs such that the fraction of BCGs with
radiatively-efficient nuclei is decreasing over time.

Observational evidence suggests that the accretion process is largely
similar for both stellar mass and supermassive black holes and
therefore we could potentially use studies of X-ray binaries to understand accretion
in AGN (eg. \citealt{Maccarone03}; \citealt{Merloni03};
\citealt{Falcke04}).  X-ray binaries are broadly classified into
low-hard and high-soft states, which relate to the accretion disk
properties and variation in the accretion rate can trigger state
transitions (eg. \citealt{Remillard06}).  In the low-hard state, the
accretion rate is low, the accretion disk is optically thin and
radiatively inefficient.  The mechanical power of the radio jet
dominates over the radiative power and the X-ray and radio fluxes are
correlated (eg. \citealt{Gallo03}; \citealt{FenderBelloni04}).   
Observations of X-ray binaries have shown that as the accretion rate
rises above $\sim0.01-0.1\dot{M}_{\mathrm{Edd}}$ the source makes a
spectral transition from the low-hard to the high-soft state
(eg. \citealt{Nowak95}; \citealt{Done07}).  In this state the X-ray
emission is dominated by an optically thick, geometrically thin
accretion disk and the radio emission drops dramatically suggesting
the outflow is suppressed (eg. \citealt{Fender99}).

\begin{figure}
\centering
\includegraphics[width=0.98\columnwidth]{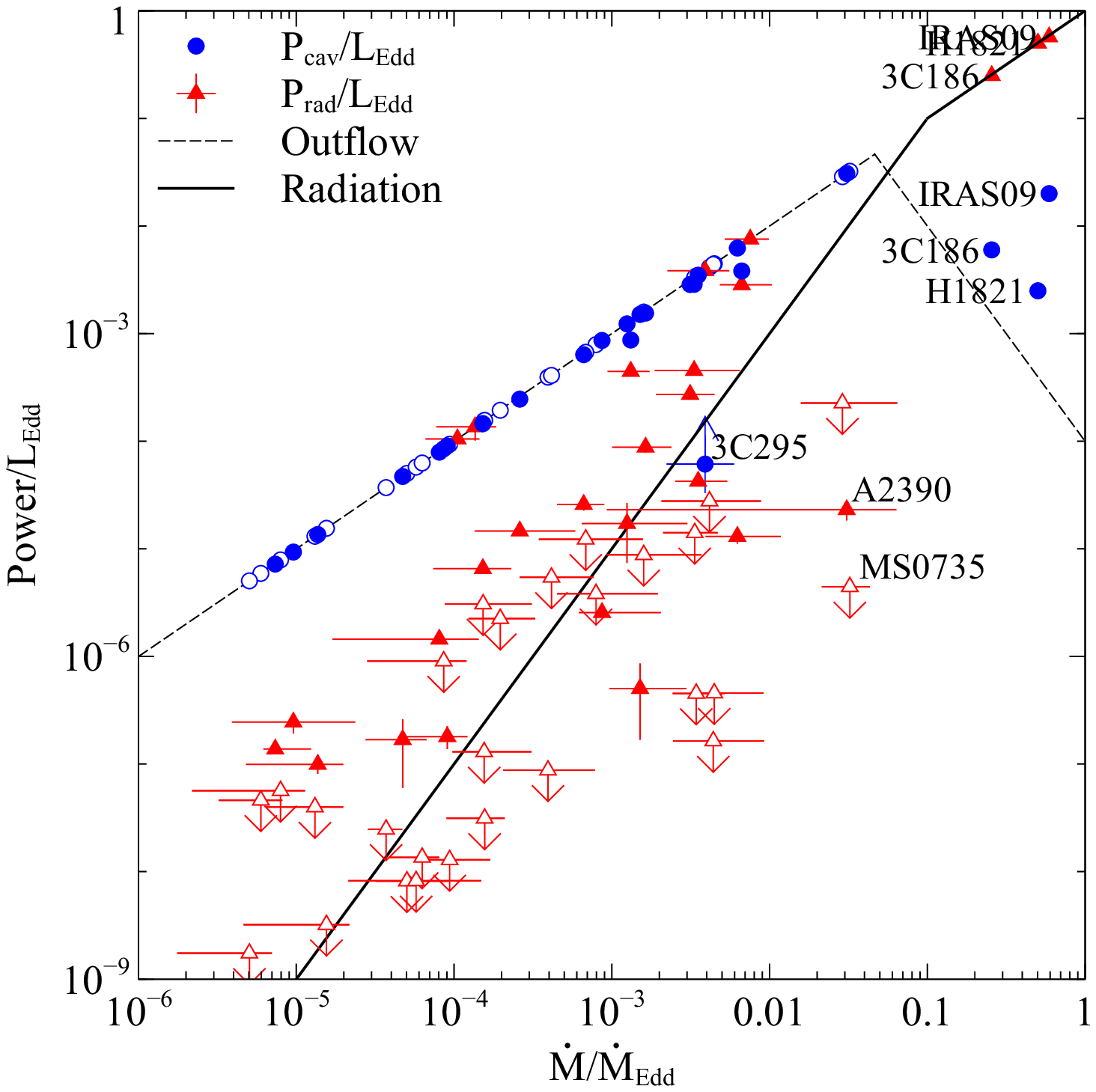}
\caption{The required mean accretion rate scaled by the Eddington rate,
  $\dot{M}/\dot{M}_{\mathrm{Edd}}$, plotted against the cavity power
  (blue circles) and the radiative power (red triangles) scaled by the
  Eddington luminosity.  Therefore, there are two points for each
  source.  Detected nuclear X-ray point sources are shown by the solid
  symbols and upper limits are shown by the open symbols.  Quasar
  sources are labelled.  The radiation and outflow model lines are
  illustrative only and show a transition from outflow domination at
  low accretion rates to radiative domination at high accretion rates
  (from \citealt{Churazov05}, Fig. 1).}
\label{fig:Chrzvplot}
\end{figure}

Fig. \ref{fig:Chrzvplot} shows the radiative and cavity power output
as a function of the required mean accretion rate for our AGN sample, where all quantities are
scaled by the Eddington rate.  The mean accretion rate was calculated from the
cavity power plus the bolometric luminosity of the point source and
scaled by the Eddington accretion rate,

\begin{equation}
\frac{\dot{M}}{\dot{M}_{\mathrm{Edd}}}=\frac{\left(P_{\mathrm{cav}} + L_{\mathrm{bol}}\right)}{L_{\mathrm{Edd}}}.
\label{eq:accrate}
\end{equation}


\noindent $L_{\mathrm{bol}}$ was calculated as shown in section
\ref{sec:radcavpowout} for the low luminosity AGN and was taken from
the literature for the quasars.  Note that for most of the sources
considered $L_{\mathrm{bol}}$ is insignificant compared to
$P_{\mathrm{cav}}$ and the required mean accretion rate is dictated by the
cavity power.  The quasars are the obvious exceptions.
There are two points for each source on the plot showing both
the cavity power and the radiative power.  For sources
where the radiative power or the cavity power dominates the output,
the corresponding points will, by definition, lie on a line of
equality between $\mathrm{Power}/L_{\mathrm{Edd}}$ and
$\dot{M}/\dot{M}_{\mathrm{Edd}}$.  This produces a clear line of
points along $y=x$ in Fig. \ref{fig:Chrzvplot}.

The illustrative model from \citet{Churazov05} of a change from a
radiatively inefficient, outflow dominated mode to a radiation
dominated mode has been shown for comparison in
Fig. \ref{fig:Chrzvplot}.  Fig. \ref{fig:Chrzvplot} shows a trend of
increasing radiative efficiency with mean accretion rate (see also
Fig. \ref{fig:cavpow}).  The radiative and mechanical power outputs
converge and become comparable at an Eddington rate of a few per cent.
For accretion rates below $\sim0.1\dot{M}_{\mathrm{Edd}}$ the cavity
power dominates over the radiative output, which is a factor of
$10-1000$ times lower.  Above $\sim0.1\dot{M}_{\mathrm{Edd}}$, a
transition apparently occurs where mechanical power drops suddenly and
the radiative power strongly dominates.  This strong transition is
seen in three objects: H1821+643, IRAS09104+4109 and 3C\,186.  These
are quasars in the centres of galaxy clusters, few are known but they
show this intriguing and potentially very important effect where they
transition from mechanically dominated to radiation dominated AGN.
This is precisely the behaviour expected when the accretion rate
increases and an object transitions from an ADAF to a geometrically thin
and optically thick disk.  The AGN in our sample therefore appear to
show the same qualitative behaviour with variation in accretion rate
found for stellar mass black holes.

However, there is significant scatter in the trend for radiatively
inefficient sources, some of which may be due to variability in the
X-ray flux.  There are also outliers on the plot, notably MS0735 and
A2390.  A2390 may have overestimated cavity power and thus could move
to the left.  MS0735 was noted as anomalous to Churazov's scenario in
\citet{Churazov05}.  In this case the issue could be related to
powering by the spin of the black hole (\citealt{McNamara09}) where
the spin energy is tapped more efficiently than $mc^2$
(eg. \citealt{McNamara11}; \citealt{Tchekovskoy11}; \citealt{Cao11};
\citealt{McKinney12}).  This would imply greater jet power per
accreted mass than objects powered directly by accretion, moving it to
the left in Fig. \ref{fig:Chrzvplot}. It is also possible that the
unknown value of $\epsilon$, the conversion efficiency between mass
and energy $P={\epsilon}\dot{M}c^2$, is a large source of scatter,
particularly as it is applied to mechanical power.  We have assumed in
eq. \ref{eq:accrate} that $\epsilon$ is tied between the radiation and
cavity power and divides out but this is of course not necessarily
true.  Nevertheless, the increasing nuclear brightness relative to
mechanical power is solid.  And the transition to quasars does depend
on power output and by inference, $\dot{M}$.  This picture is also a
simplification of stellar mass black hole state transitions.  Observed
transitions from the low-hard to the high-soft state in X-ray binaries
are accompanied by an intense and rapid radio outburst, which has no
obvious analogy in our AGN model (eg. \citealt{Fender04}).

\subsection{Accretion power}


The source of fuel for the observed AGN activity has been the subject
of considerable debate.  There is, in general, sufficient cold gas in
BCGs to fuel the range of observed jet powers (\citealt{Edge01};
\citealt{Salome03}; \citealt{Soker08}; \citealt{Donahue11}).  However,
Bondi accretion directly from the cluster's hot atmosphere is
appealing because it can provide both a steady fuel supply and a
simple feedback mechanism.  Although whilst the gas density is high
enough to supply sufficient fuel to low power jet systems through
Bondi accretion (eg. \citealt{Allen06}), this is difficult to achieve
for high power jets ($>10^{45}\ergps$; \citealt{Rafferty06};
\citealt{Hardcastle07}; \citealt{McNamara11}).  Using a sample of
nearby systems, \citet{Allen06} found a correlation between the cavity
power and Bondi accretion rate suggesting that a few per cent of the
rest mass energy of material crossing the Bondi radius emerges in the jets.

In our analysis of 13 systems, including the \citet{Allen06} sample,
we found a significantly larger scatter in the correlation between the
cavity power and Bondi accretion rate.  This was primarily due to
differences in our calculation of the density at the Bondi radius and
estimates of the cavity power.  Calculation of the density at the
Bondi radius required an extrapolation of this profile over an order
of magnitude in radius for each object in our sample.  We used three
different model density profiles for the extrapolation and found each
provided a similarly good fit, with the exception of the Centaurus
cluster where a $\beta$-model was significantly better.  This fit was
extended to cover a few kpc rather than just the central points as
cavity substructure close to the centre affected the deprojection of
the density profile.  The use of three equally plausible
extrapolations of the density profile produced different estimates of
the density at the Bondi radius and increased the scatter in the Bondi
accretion rate compared to \citet{Allen06}.  Deeper observations of
several clusters, showing the cavity extent more clearly,
significantly altered the cavity power measured for those objects.  We
have therefore found weaker evidence for a trend between the cavity
power and Bondi accretion rate.

Bondi accretion is energetically a plausible mechanism for fuelling
the lower-powered radio sources in our sample.  However, it is
insufficient to fuel the most powerful systems
(\citealt{Hardcastle07}; \citealt{McNamara11}).  There are also
theoretical issues with Bondi accretion that include the ability to
shed angular momentum and the zero central pressure requirement
(\citealt{Proga03}; \citealt{PizzolatoSoker05,Pizzolato10};
\citealt{Soker08}; \citealt{Narayan11}).  For a more complete discussion see
\citet{McNamara12}.

In view of the high column densities found for many objects in this
sample, which are consistent with significant levels of cold
circumnuclear gas, and the prevalence of cold molecular gas in cD
galaxies (eg. \citealt{Edge01}; \citealt{Salome03}) we suggest that
cold gas fuelling is a likely source of accretion power in these
objects.  Nevertheless we cannot rule out or exclude Bondi accretion,
which could play a significant role, particularly in low power jets
(\citealt{Allen06}).


\subsection{Nuclear X-ray emission mechanism}
\label{sec:xrayorigin}

The origin of the observed nuclear X-ray emission is not currently
understood.  Observed correlations between the X-ray and radio core
luminosities provide the strongest support for a non-thermal
jet-related origin (eg. \citealt{Fabbiano89}; \citealt{Canosa99};
\citealt{Hardcastle99}).  Radio and optical luminosity correlations
for FR I nuclei also support this conclusion (\citealt{Chiaberge99})
and multi-wavelength spectral energy distributions for these sources
can be modelled by synchrotron and synchrotron self-Compton emission
from a jet (eg. \citealt{Capetti02}; \citealt{Yuan02};
\citealt{Chiaberge03}).  However, \citet{Donato04} show that a
significant fraction of sources with strong optical jet emission do
not have an X-ray component potentially indicating different physical
origins.  The detection of broad Fe\,K$\alpha$ lines and rapid
variability on ks timescales favours an accretion flow origin
(eg. \citealt{Gliozzi03}).  Radio and X-ray correlations do not
necessarily imply a common origin for the emission as accretion
processes and jets are likely to be correlated phenomena
(eg. \citealt{Begelman84}).  \citet{Merloni03} and \citet{Falcke04}
argued that these correlations are part of a `fundamental plane' linking
radio and X-ray emission to black hole mass
but differ on whether this reveals trends in accretion or jet physics
(eg. \citealt{Kording06}; \citealt{Hardcastle09};
\citealt{Plotkin12}).


\subsubsection{Jet origin?}

\citet{Wu07} analysed the spectral energy distributions of eight FR I
sources including two of the variable sources in our sample, A2052 and
M84 (3C\,272.1).  They found that the emission in M84 is dominated by
a jet and the ADAF model predicts too hard a spectrum at X-ray
energies.  A2052 appears to have a comparable contribution from the
jet and the ADAF.  These two sources have very different accretion
rates: for A2052
$\dot{M}/\dot{M}_{\mathrm{Edd}}=9^{+10}_{-2}\times10^{-4}$ and for M84
$\dot{M}/\dot{M}_{\mathrm{Edd}}=5^{+8}_{-4}\times10^{-6}$.  Yet they
both experience significant variations in nuclear flux on timescales
of months to years.

Strong flux variability on timescales of $1-2$ months is seen from
both the core and the jet knot HST-1 in M87 (eg. \citealt{Harris97};
\citealt{Harris06}).  From 2000 to 2009, HST-1 is the site of a
massive X-ray, UV and radio flare.  During this period, the X-ray
emission from HST-1 dominates over the nucleus and rises and falls by
an order of magnitude (\citealt{Harris09}).  The nuclear variability
is characterised by \citealt{Harris09} as `flickering' with changes in
flux of order a factor of a few over timescales of months to years.
It is not clear if the nuclear X-ray emission is due to the inner
unresolved jet or the accretion flow.  The magnitude and timescales of
the X-ray flux variability found in A2052, Hydra A and M84 are
therefore consistent with that observed in M87.  Interestingly, HST UV
observations of the nucleus in A2052 find the luminosity increased by
a factor of ten from 1994 to 1999 (\citealt{Chiaberge02}).  This
period was then followed by a decrease in the X-ray luminosity by a
similar factor from 2000 to 2010, which could indicate a flaring event
similar to that experienced by HST-1.

\citet{Sambruna03} found variability with \textit{XMM-Newton} on
$3-5\ks$ timescales in the FR I radio galaxy NGC4261, which is also
part of our sample.  For an ADAF, the X-ray emission is radiated from
a relatively large volume and variation is expected on timescales
longer than around a day (\citealt{Ptak98}; \citealt{Terashima02}).
The observed variability timescale in NGC4261 is around two orders of
magnitude shorter than the ADAF light crossing time suggesting that
the variable component is more likely to be associated with the inner
jet.  Unfortunately, the count rate in the \textit{Chandra}
observations of our sample is generally not large enough to search the
light curve of each individual observation for flux variation.  We
also do not find any significant variation in the spectral properties
of the three variable sources identified in our subsample.  







\subsubsection{ADAF origin?}

ADAF models predict trends between the nuclear radio and X-ray
luminosities that scale as
$L_{\mathrm{R}}{\propto}L_{\mathrm{X}}^{0.6}$ and
$\dot{M}{\propto}L_{\mathrm{X}}^{0.5}$ (\citealt{YiBoughn98}).
Assuming $\dot{M}{\propto}P_{\mathrm{cav}}$, this scaling is
consistent with the slope shown in Fig. \ref{fig:cavpow}.  Although
the slope between $L_{\mathrm{R}}$ and $L_{\mathrm{X}}$ shown in
Fig. \ref{fig:xrayradio} appears to be steeper than
$L_{\mathrm{R}}{\propto}L_{\mathrm{X}}^{0.6}$, the lack of a flux-flux
correlation and the large uncertainties suggest this trend is unreliable
(see section \ref{sec:resradio}).  It is therefore worthwhile to
consider the consequences of emission from an ADAF.

\begin{figure}
\centering
\includegraphics[width=0.98\columnwidth]{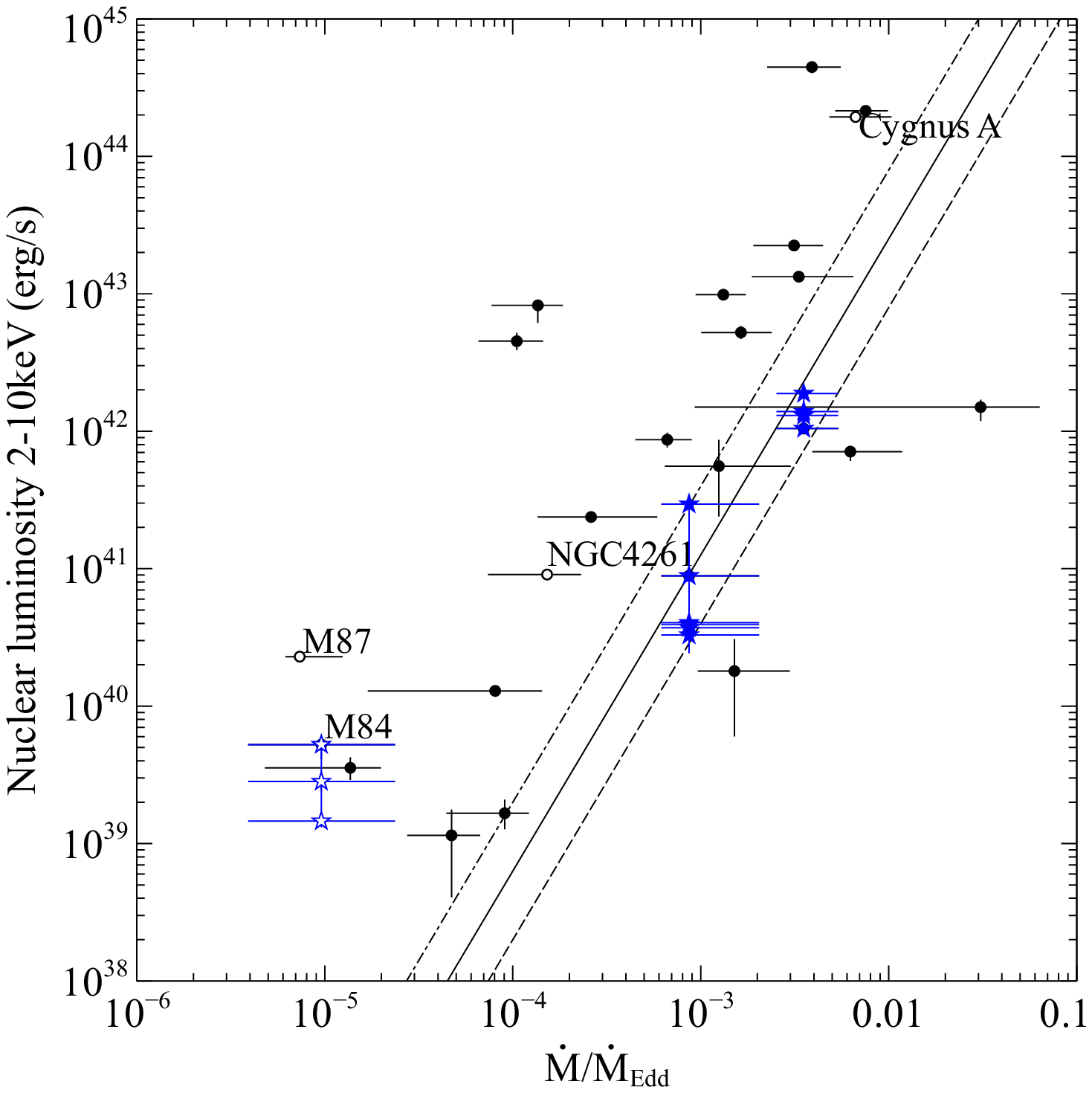}
\caption{Point source X-ray luminosity versus the inferred accretion rate scaled by the Eddington rate.  The lines are ADAF model predictions from \citet{Merloni03} for black hole masses of $5\times10^8\Msun$ (dashed), $1\times10^9\Msun$ (solid) and $5\times10^9\Msun$ (dash-dotted).  The variable point sources are shown by the blue stars.  Sources with dynamical black hole masses are shown by the open symbols (Cygnus A, M84, M87 and NGC4261).}
\label{fig:ADAFmods}
\end{figure}

Fig. \ref{fig:ADAFmods} compares the X-ray point source luminosities
with predictions from ADAF models at different accretion rates
(\citealt{Merloni03}).  The nuclear X-ray luminosity scales very close
to linearly with black hole mass allowing us to scale up these models
to the required $\sim10^{9}\Msun$.  For low accretion rates, the
$2-10\keV$ emission includes inverse Compton scattering of soft
synchrotron or disk photons and a bremsstrahlung component at higher
energies.  At higher accretion rates, the Compton-scattered component
dominates as the optical depth rises and cooling processes become more
efficient.  The exact scaling of $L_{2-10\keV}$ with $\dot{M}/\dot{M}_{\mathrm{Edd}}$ will
depend on the parameters chosen for the model, such as the viscosity,
magnetic pressure and electron heating fraction.  Therefore,
\citet{Merloni03} obtain a single powerlaw fit
$L_{2-10\keV}\propto(\dot{M}/\dot{M}_{\mathrm{Edd}})^{2.3}$ from their ADAF models for comparison
with observational data.

The required mean accretion rate was calculated from the cavity power
plus the radiative power as shown in eq. \ref{eq:accrate}.  For most
of the sources considered $L_{\mathrm{bol}}$ is insignificant compared
to $P_{\mathrm{cav}}$ and the required mean accretion rate is dictated
by the cavity power.  The three quasars are the obvious exceptions to this.
The majority of the observed sources are consistent with the emission
expected from an ADAF for black hole masses from $5\times10^{8}\Msun$
to $5\times10^{9}\Msun$.  Four sources in our sample have more accurate dynamical black hole
masses and therefore provide a more reliable constraint when compared
with ADAF model predictions.  Cygnus A and NGC4261 have mass accretion
rates a factor of $3-5$ smaller than the ADAF model predictions for
their respective black hole masses.  The difference is even greater
for M84 and M87 with over an order of magnitude and close to two
orders of magnitude discrepancy, respectively.  There are several
possible reasons for this.  The cavity power in M84 is particularly
difficult to estimate as the outburst appears to have blown out most
of the X-ray atmosphere.  The cavity volume and surrounding pressure
may therefore have been underestimated.  The total mechanical power in M87 has
been significantly underestimated as the shock produces an additional
$2.4\times10^{43}\ergps$, which is four times greater than the cavity
power.  This would cause M84 and M87 to move to the right in
Fig. \ref{fig:ADAFmods} and closer to the ADAF models.  So this
translates to a similar increase in the mean accretion rate.  The nuclear
X-ray emission from M84 and M87 is therefore likely to be consistent
with an ADAF given the large uncertainties in the ADAF models.  As
previously discussed in section \ref{sec:cavpow}, the cavity power is
likely to have been underestimated for the majority of the systems in
this sample and this will tend to move points to the right in
Fig. \ref{fig:ADAFmods}.

We therefore conclude that it is plausible that the X-ray point source
emission is due to an ADAF but we cannot distinguish between this and
a jet origin with the available data.  Given the lack of a clear trend
between the nuclear radio and X-ray flux it is likely that further
progress on this problem will require modelling of the AGN spectral
energy distribution (eg. \citealt{Wu07}).



\section{Conclusions}


Using archival \textit{Chandra} observations of 57 BCGs, we have
investigated the relationship between nuclear X-ray emission and AGN
radio jet (cavity) power.  Although this is not a complete sample of
objects, we find that nuclear X-ray emission is common with roughly
half of the sample hosting a detectable X-ray point source.  Assuming
nuclear X-ray emission indicates active accretion, our study implies
that the AGN in systems with recent outbursts are `on' at least 50 per
cent of the time.  Furthermore, we examine the correlation between the
nuclear X-ray luminosity and the average accretion rate determined
from the energy required to inflate the X-ray cavities.  This
correlation is consistent with the hypothesis that the nuclear X-ray
emission traces active accretion in these systems.  The majority of
the sources in this sample are radiatively inefficient with required
mean accretion rates of only $10^{-5}-10^{-2}\dot{M}_{\mathrm{Edd}}$.  The
nuclear X-ray sources become more luminous compared to the cavity power as
the average accretion rate increases.  The nuclear X-ray emission
exceeds the cavity power when the average accretion rate rises above a
few percent of the Eddington rate, where the AGN power output appears
to transition from cavity power-dominated to radiation-dominated in
the three BCGs hosting quasars.


A small subset of the clusters in our sample had multiple archival
\textit{Chandra} observations of sufficient depth to search for
variability in the nuclear X-ray flux.  We found that A2052, M84 and
Hydra A were significantly varying by factors of $2-10$ on timescales
of 6 months to ten years.  Despite the large variations in flux, we
did not find significant change in the shape of the nuclear spectra.
This analysis is generally limited by the availability of suitably
spaced observations of sufficient depth in the Chandra archive but
suggests that a significant fraction of AGN in BCGs may be varying on
timescales of months to a few years.  This variability is likely to be
a significant source of the large scatter in the observed correlation
between the nuclear luminosity and cavity power.  Our results suggest
that the accretion power roughly scales with the cavity power over
long timescales with the large scatter reflecting the variability on
shorter timescales.  


The interpretation of these results is complicated by the uncertainty
of the nuclear X-ray emission origin.  This emission may originate
from the accretion disk corona, from the base of a jet or a
combination of both, although other mechanisms are also possible.  We
discuss the similarity in magnitude and timescale of the X-ray
variability found in A2052, M84 and Hydra A to that observed from both
the core and jet knot HST-1 in M87.  We also show that the X-ray
nuclear luminosity and required mean accretion rate of the systems analysed
are consistent with the predictions from ADAF models.  We conclude
that an ADAF is a plausible origin of the X-ray point source emission
but we cannot distinguish between this and a jet origin with the
available data.

We have also considered the longstanding problem of whether jets are
powered by the accretion of cold circumnuclear gas or accretion from
the hot keV atmosphere.  For a subsample of 13 nearby systems, the
Bondi accretion rate was calculated using three equally plausible
model extrapolations of the cluster density profile to the Bondi
radius.  The results suggest weaker evidence for a trend between the
cavity power and the Bondi accretion rate, primarily due to the
uncertainty in the cavity volumes.  Cold gas fuelling may therefore be
a more likely source of accretion power given the high column
densities found for many objects in our sample, which are consistent
with significant quantities of cold circumnuclear gas, and the
prevalance of cold molecular gas in BCGs.  However, we cannot rule out
Bondi accretion, which may play a significant role, particularly in
low power jets.

\section*{Acknowledgements}
HRR and BRM acknowledge generous financial support from the Canadian
Space Agency Space Science Enhancement Program.  RAM and ANV
acknowledge support from the Natural Sciences and Engineering Research
Council of Canada.  We thank Paul Nulsen, Avery Broderick, Roderick
Johnstone and James Taylor for helpful discussions.  This publication
makes use of data products from the Two Micron All Sky Survey, which
is a joint project of the University of Massachusetts and the Infrared
Processing and Analysis Center/California Institute of Technology,
funded by the National Aeronautics and Space Administration and the
National Science Foundation.  This research has made use of the
NASA/IPAC Extragalactic Database (NED) which is operated by the Jet
Propulsion Laboratory, California Institute of Technology, under
contract with the National Aeronautics and Space Administration.

\bibliographystyle{mn2e} 
\bibliography{refs.bib}

\clearpage

\begin{deluxetable}{lcccccccccc}
\rotate
\tablecolumns{11}
\tablewidth{0pc}
\tablecaption{Sample properties and point source fluxes.\label{tab:fluxes}}
\tablehead{
\colhead{Target} & \colhead{Redshift} & \colhead{Obs. ID} & \colhead{Aimpoint} & \colhead{Exposure$^a$} & \colhead{$n_{\mathrm{H}}$$^b$} & \colhead{$n_{\mathrm{H,z}}$} & \colhead{$\Gamma$} & \colhead{$F_{\mathrm{P},2-10\keV}$$^c$} & \colhead{$F_{\mathrm{S},2-10\keV}$$^c$} & \colhead{References} \\
\colhead{} & \colhead{} & \colhead{} & \colhead{} & \colhead{(ks)} & \colhead{($10^{22}\pcmsq$)} & \colhead{($10^{22}\pcmsq$)} & \colhead{} & \colhead{($10^{-14}\ergpcmsqps$)} & \colhead{($10^{-14}\ergpcmsqps$)} & \colhead{}}
\startdata
2A0335+096 & 0.0349 & 7939 & S3 & 49.5 & 0.2218$^{*}$ & 0 & 1.9 & $<2.3$ & $<0.1$ & [1]\\
3C295 & 0.4641 & 2254 & I3 & 75.6 & 0.0134 & $43^{+6}_{-5}$ & 1.9 & $76\pm4$ & $55\pm3$ & [2],[3]\\
3C388 & 0.0917 & 5295 & I3 & 26.3 & 0.0555 & 0 & 1.9 & $4.0\pm0.3$ & $2\pm1$ & [4],[1]\\
3C401 & 0.2011 & 4370 & S3 & 22.7 & 0.0582 & 0 & 1.9 & $3.7\pm0.3$ & $3.7\pm0.3$ & [2],[1]\\
4C55.16 & 0.2412 & 4940 & S3 & 64.5 & 0.0429 & 0 & $1.55^{+0.09}_{-0.08}$ & $7.3\pm0.3$ & $6.9^{+0.3}_{-0.5}$ & [5],[1]\\
A85 & 0.0551 & 904 & I0 & 38.2 & 0.0278 & 0 & 1.9 & $<2.2$ & $<2.1$ & [1]\\
A133 & 0.0566 & 9897 & I3 & 67.9 & 0.0159 & 0 & 1.9 & $<0.8$ & $<0.2$ & [1]\\
A262 & 0.0166 & 7921 & S3 & 108.6 & 0.0567 & 0 & 1.9 & $<0.49$ & $<0.04$ & [1]\\
A478 & 0.0881 & 1669 & S3 & 39.3 & 0.2778$^{*}$ & 0 & 1.9 & $<3.2$ & $<1.4$ & [6],[1]\\
A611 & 0.2880 & 3194 & S3 & 32.0 & 0.0447 & 0 & 1.9 & $3.0\pm0.2$ & $3.1^{+0.2}_{-0.8}$ & - \\
A1795 & 0.0625 & 10900 & S3 & 15.8 & 0.0119 & 0 & 1.9 & $<1.8$ & $<1.6$ & [1]\\
A1835 & 0.2523 & 6880 & I3 & 109.2 & 0.0204 & 0 & 1.9 & $<3.1$ & $<1.2$ & [6],[1]\\
A2029 & 0.0773 & 4977 & S3 & 74.7 & 0.0325 & 0 & 1.9 & $<4.5$ & $<1.4$ & [1]\\
A2052 & 0.0351 & 5807 & S3 & 123.8 & 0.0272 & 0 & 1.9 & $3.8\pm0.1$ & $3.1^{+0.2}_{-0.3}$ & [7],[1]\\
A2199 & 0.0302 & 10748 & I3 & 40.6 & 0.0089 & 0 & 1.9 & $3.1\pm0.3$ & $0.9\pm0.6$ & [8],[1]\\
A2390 & 0.2280 & 4193 & S3 & 70.6 & 0.0768$^{*}$ & 0 & 1.9 & $1.4\pm0.1$ & $1.0^{+0.1}_{-0.2}$ & - \\
A2597 & 0.0852 & 7329 & S3 & 56.3 & 0.0248 & 0 & 1.9 & $<2.3$ & $<1.3$ & [1]\\
A2667 & 0.2300 & 2214 & S3 & 8.4 & 0.0173 & 0 & 1.9 & $2.4\pm0.4$ & $2.8\pm0.4$ & - \\
A4059 & 0.0475 & 5785 & S3 & 85.9 & 0.0121 & 0 & 1.9 & $<0.72$ & $<0.08$ & [1]\\
Centaurus & 0.0114 & 4954 & S3 & 87.2 & 0.0948$^{*}$ & 0 & 1.9 & $<1.6$ & $<0.10$ & [6],[1]\\
Cygnus A & 0.0561 & 1707 & S3 & 9.2 & 0.272 & $19.6\pm0.7$ & 1.9 & $3160\pm70$ & $2560\pm60$ & [6],[1]\\
HCG62 & 0.0137 & 10462 & S3 & 65.1 & 0.0332 & 0 & 1.9 & $<0.67$ & $<0.03$ & [1]\\
Hercules A & 0.1540 & 6257 & S3 & 48.5 & 0.0618 & 0 & 1.9 & $<0.94$ & $<0.53$ & [1]\\
Hydra A & 0.0549 & 4970 & S3 & 94.6 & 0.0468 & $3.5\pm0.3$ & 1.9 & $24\pm1$ & $14.5^{+0.8}_{-0.7}$ & [9],[1]\\
M84 & 0.0035 & 5908 & S3 & 45.1 & 0.0299 & $0.12^{+0.03}_{-0.02}$ & $2.13\pm0.09$ & $17.5\pm0.4$ & $15^{+1}_{-3}$ & [10],[1]\\
M87 & 0.0044 & 1808 & S3 & 12.8 & 0.0194 & $0.06\pm0.01$ & $2.37\pm0.07$ & $67\pm1$ & $66\pm4$ & [11],[1]\\
M89 & 0.0011 & 2072 & S3 & 53.4 & 0.0262 & $0.092\pm0.03$ & $2.3\pm0.1$ & $5.5\pm0.2$ & $3\pm2$ & [12],[13]\\
MKW3S & 0.0450 & 900 & I3 & 51.9 & 0.0268 & 0 & 1.9 & $<1.0$ & $<0.2$ & [1]\\
MS0735.6+7421 & 0.2160 & 10470 & I3 & 133.8 & 0.0328 & 0 & 1.9 & $<0.95$ & $<0.12$ & [14]\\
NGC507 & 0.0165 & 2882 & I3 & 40.0 & 0.0525 & 0 & 1.9 & $<1.1$ & $<0.1$ & [15]\\
NGC1316 & 0.0059 & 2022 & S3 & 20.2 & 0.0240 & 0 & 1.9 & $<3.0$ & $<2.2$ & [15]\\
NGC1600 & 0.0156 & 4283 & S3 & 20.5 & 0.0349 & 0 & 1.9 & $<0.6$ & $<0.2$ & [15]\\
NGC4261 & 0.0075 & 9569 & S3 & 99.9 & 0.0175 & $3.0^{+0.6}_{-0.5}$ & $0.9\pm0.2$ & $126\pm2$ & $74\pm2$ & [16],[17]\\
NGC4472 & 0.0033 & 11274 & S3 & 39.7 & 0.0153 & 0 & 1.9 & $<2.4$ & $<0.1$ & [15]\\
NGC4636 & 0.0031 & 3926 & I3 & 67.8 & 0.0190 & 0 & 1.9 & $<0.9$ & $<0.3$ & [15]\\
NGC4782 & 0.0154 & 3220 & S3 & 49.3 & 0.0337 & 0 & 1.9 & $0.84\pm0.09$ & $0.7\pm0.1$ & [18],[15]\\
NGC5044 & 0.0093 & 9399 & S3 & 82.5 & 0.0487 & $0.23^{+0.08}_{-0.07}$ & 1.9 & $2.6\pm0.2$ & $1.4^{+0.4}_{-0.3}$ & [19],[15]\\
NGC5813 & 0.0066 & 9517 & S3 & 98.8 & 0.0918$^{*}$ & 0 & 1.9 & $<1.02$ & $<0.3$ & [20]\\
NGC5846 & 0.0057 & 7923 & I3 & 84.7 & 0.0429 & 0 & 1.9 & $<0.72$ & $<0.05$ & [15]\\
NGC6269 & 0.0348 & 4972 & I3 & 35.3 & 0.0510 & 0 & 1.9 & $<2.4$ & $<0.08$ & [15]\\
NGC6338 & 0.0274 & 4194 & I3 & 44.5 & 0.0223 & 0 & 1.9 & $<2.2$ & $<1.4$ & [15]\\
PKS\,0745-191 & 0.1028 & 12881 & S3 & 116.0 & 0.38 & 0 & 1.9 & $4.8\pm0.3$ & $2.6^{+0.2}_{-0.4}$ & [1]\\
PKS\,1404-267 & 0.0230 & 12884 & S3 & 83.5 & 0.0578$^{*}$ & $0.04\pm0.02$ & $2.01\pm0.06$ & $21.4\pm0.4$ & $19.8\pm0.7$ & [21],[1]\\
RBS797 & 0.3540 & 7902 & S3 & 38.3 & 0.0228 & $3.8^{+0.9}_{-0.8}$ & $1.4\pm0.2$ & $61\pm2$ & $41\pm2$ & [22]\\
RXCJ0352.9+1941 & 0.1090 & 10466 & S3 & 27.2 & 0.1440$^{*}$ & $3.6\pm0.4$ & 1.9 & $36\pm2$ & $32\pm2$ & -\\
RXCJ1459.4-1811 & 0.2357 & 9428 & S3 & 39.5 & 0.0736 & $5.4^{+1.0}_{-0.8}$ & 1.9 & $21\pm1$ & $13\pm1$ & -\\
RXCJ1524.2-3154 & 0.1028 & 9401 & S3 & 40.9 & 0.1107$^{*}$ & $0.5\pm0.2$ & 1.9 & $5.7\pm0.5$ & $3.2\pm0.4$ & -\\
RXCJ1558.3-1410 & 0.0970 & 9402 & S3 & 36.0 & 0.1060 & $15^{+5}_{-3}$ & 1.9 & $41\pm3$ & $22\pm2$ & -\\
S\'ersic\,159-03 & 0.0580 & 11758 & I3 & 93.3 & 0.0114 & 0 & 1.9 & $<1.0$ & $<0.07$ & [1]\\
UGC408 & 0.0147 & 11389 & S3 & 93.8 & 0.0256 & $0.23^{+0.06}_{-0.05}$ & $2.2\pm0.2$ & $3.3\pm0.1$ & $2.8\pm0.2$ & [20]\\
Zw\,348 & 0.2535 & 10465 & S3 & 48.9 & 0.0250 & $0.8^{+0.7}_{-0.4}$ & 1.9 & $4.4\pm0.4$ & $2.9\pm0.4$ & -\\
Zw\,2089 & 0.2350 & 10463 & S3 & 38.6 & 0.0286 & $22\pm2$ & 1.9 & $203\pm7$ & $127^{+6}_{-5}$ & -\\
Zw\,2701 & 0.2150 & 12903 & S3 & 95.8 & 0.0075 & 0 & 1.9 & $<0.53$ & $<0.3$ & [6],[23]\\
Zw\,3146 & 0.2906 & 9371 & I3 & 34.5 & 0.0246 & 0 & 1.9 & $<5.0$ & $<3.3$ & [6],[1]\\
\enddata
\tablecomments{$^a$Final cleaned exposure times.  $^b$Column densities marked with $^*$ were found to be higher than the Galactic value measured by \citet{Kalberla05} (see section \ref{sec:psflux}).  $^c$Errors and upper limits on the photometric $F_{\mathrm{P},2-10\keV}$ and spectroscopic $F_{\mathrm{S},2-10\keV}$ flux values are $1\sigma$.  References: [1] \citet{Rafferty06}, [2] \citet{Hardcastle06}, [3] \citet{HlavacekLarrondo12}, [4] \citet{Evans06}, [5] \citet{HlavacekLarrondo4C11}, [6] \citet{HlavacekLarrondo11}, [7] \citet{Blanton03}, [8] \citet{DiMatteo01}, [9] \citet{McNamara00}, [10] \citet{Balmaverde06}, [11] \citet{DiMatteo03}, [12] \citet{Machacek06}, [13] \citet{Allen06}, [14] McNamara et al. in prep, [15] \citet{Cavagnolo10}, [16] \citet{Chiabergengc426103}, [17] \citet{OSullivanngc426111}, [18] \citet{Machacek07}, [19] \citet{David09}, [20] \citet{OSullivan11}, [21] \citet{Johnstone05}, [22] \citet{Cavagnolo11}, [23] McNamara et al. in prep}
\end{deluxetable}

\begin{table*}
\begin{minipage}{\textwidth}
\caption{Bondi parameters for the selected subsample of targets.}
\begin{center}
\begin{tabular}{l c c c c c c c}
\hline
Target & $D_{\mathrm{L}}$ & $T(r_{\mathrm{B}})$ & $r_{\mathrm{B}}$ & $n_e(r_{\mathrm{B}})$ & $\dot{M}_{\mathrm{B}}$ & $P_{\mathrm{B}}$ & $P_{\mathrm{cav}}$ \\
 & (Mpc) & (keV) & (kpc) & ($\pcmcu$) & ($\Msunpyr$) & ($10^{43}\ergps$) & ($10^{42}\ergps$) \\
\hline
A2199 & 132.3 & $1.10\pm0.07$ & $0.020\pm0.001$ & $1.1^{+0.8}_{-0.6}$ & $0.006^{+0.004}_{-0.003}$ & $4\pm2$ &$9^{+4}_{-2}$\\
Centaurus & 49.3 & $0.41\pm0.04$ & $0.028\pm0.002$ & $0.4^{+0.3}_{-0.2}$ & $0.002^{+0.002}_{-0.001}$ & $0.51\pm0.06$ & $14^{+7}_{-4}$\\
HCG62 & 59.3 & $0.31\pm0.03$ & $0.058\pm0.006$ & $0.3^{+0.3}_{-0.004}$ & $0.006^{+0.006}_{-0.000}$ & $5\pm2$ & $6^{+4}_{-2}$\\
M84 & 17.0 & $0.34\pm0.01$ & $0.084\pm0.003$ & $0.41^{+0.10}_{-0.01}$ & $0.020^{+0.005}_{-0.001}$ & $13\pm1$ & $1.1^{+0.9}_{-0.4}$\\
M87 & 17.0 & $0.52\pm0.04$ & $0.42\pm0.04$ & $0.23\pm0.05$ & $0.35\pm0.07$ & $200\pm40$ & $8^{+7}_{-3}$\\
M89 & 17.0 & $0.35\pm0.01$ & $0.041\pm0.002$ & $0.6^{+0.7}_{-0.5}$ & $0.007^{+0.008}_{-0.006}$ & $5\pm4$ & $0.3^{+0.2}_{-0.1}$\\
NGC507 & 71.6 & $0.49\pm0.02$ & $0.048\pm0.002$ & $0.8^{+1.8}_{-0.5}$ & $0.02^{+0.04}_{-0.01}$ & $20\pm10$ & $19^{+14}_{-7}$\\
NGC1316 & 25.4 & $0.343\pm0.007$ & $0.0267\pm0.0005$ & $1.3^{+1.6}_{-0.9}$ & $0.006^{+0.008}_{-0.004}$ & $5\pm4$ & $0.9^{+0.8}_{-0.4}$\\
NGC4472 & 17.0 & $0.372\pm0.007$ & $0.061\pm0.001$ & $0.5\pm0.2$ & $0.014^{+0.006}_{-0.005}$ & $9\pm3$ & $0.7^{+0.5}_{-0.3}$\\
NGC4636 & 17.0 & $0.23\pm0.02$ & $0.028\pm0.002$ & $0.20^{+0.03}_{-0.05}$ & $0.0009^{+0.0001}_{-0.0002}$ & $0.5\pm0.1$ & $0.27^{+0.15}_{-0.09}$\\
NGC5044 & 40.1 & $0.31\pm0.01$ & $0.0290\pm0.0009$ & $0.3^{+0.5}_{-0.1}$ & $0.002^{+0.003}_{-0.001}$ & $2\pm1$ & $1.3^{+0.8}_{-0.4}$\\
NGC5813 & 28.4 & $0.33\pm0.01$ & $0.0288\pm0.0009$ & $0.28^{+0.18}_{-0.09}$ & $0.0016^{+0.0010}_{-0.0005}$ & $1.0\pm0.4$ & $0.7^{+0.6}_{-0.3}$\\
NGC5846 & 24.5 & $0.378\pm0.009$ & $0.0256\pm0.0006$ & $0.4^{+0.4}_{-0.3}$ & $0.002^{+0.002}_{-0.001}$ & $1.3\pm1.0$ & $1.6^{+1.2}_{-0.6}$\\
\hline
\end{tabular}
\end{center}
\label{tab:bondi}
\end{minipage}
\end{table*}

\begin{table*}
\begin{minipage}{\textwidth}
\caption{Point source fluxes and key parameters for each source tested
  for variability.  The photometric point source flux,
  $F_{\mathrm{P}}$, and the spectroscopic point source flux,
  $F_{\mathrm{S}}$, are both given in the $2-10\keV$ energy band.  A
  spectroscopic flux measurement could not be produced for M84 obs. ID
  401 because the $1\ks$ exposure was too short. }
\begin{center}
\begin{tabular}{l c c c c c c c c}
\hline
Target & Obs. ID & Date & Aimpoint & Exposure & $n_{\mathrm{H,z}}$ & $\Gamma$ & $F_{\mathrm{P}}$ & $F_{\mathrm{S}}$ \\
 & & & & (ks) & ($10^{22}\pcmsq$) & & ($10^{-14}\ergpcmsqps$) & ($10^{-14}\ergpcmsqps$) \\
\hline
A2052 & 890 & 03/09/2000 & S3 & 30.5 & $<0.014$ & $2.06^{+0.08}_{-0.06}$ & $12.6\pm0.4$ & $10\pm1$\\
 & 5807 & 24/03/2006 & S3 & 123.8 & $<0.019$ & $2.00^{+0.09}_{-0.07}$ & $3.8\pm0.1$ & $3.1^{+0.2}_{-0.3}$\\
 & 10879 & 05/04/2009 & S3 & 80.0 & $0.16^{+0.08}_{-0.07}$ & $3.2\pm0.4$ & $2.2\pm0.1$ & $1.1\pm0.3$\\
 & 10478 & 25/05/2009 & S3 & 119.0 & $0.03^{+0.05}_{-0.03}$ & $2.2\pm0.2$ & $2.1\pm0.1$ & $1.4\pm0.3$ \\
 & 10477 & 05/06/2009 & S3 & 59.0 & $0.11^{+0.09}_{-0.08}$ & $2.8^{+0.5}_{-0.4}$ & $2.4\pm0.1$ & $1.4^{+0.3}_{-0.4}$ \\
 & 10479 & 09/06/2009 & S3 & 63.9 & $0.08^{+0.08}_{-0.07}$ & $2.8\pm0.4$ & $2.2\pm0.1$ & $1.3^{+0.3}_{-0.3}$ \\
A2390 & 500 & 08/10/2000 & S3 & 8.8 & 0 & 1.9 & $0.9\pm0.4$ & $<1.1$ \\
 & 4193 & 11/09/2003 & S3 & 70.6 & $0.1^{+0.2}_{-0.1}$ & 1.9 & $1.4\pm0.1$ & $0.9^{+0.1}_{-0.2}$ \\
Hydra A & 576 & 02/11/1999 & S3 & 17.4 & $4.3^{+0.8}_{-0.6}$ & 1.9 & $36\pm3$ & $26\pm2$ \\
 & 4969 & 13/01/2004 & S3 & 62.1 & $3.0^{+0.8}_{-0.7}$ & $1.7^{+0.4}_{-0.3}$ & $27\pm1$ & $18\pm1$ \\
 & 4970 & 22/10/2004 & S3 & 94.6 & $2.2^{+0.6}_{-0.5}$ & $1.2\pm0.3$ & $24\pm1$ & $14.5^{+0.8}_{-0.7}$ \\
M84 & 401 & 20/04/2000 & S3 & 1.7 & 0 & 1.9 & $7\pm1$ & - \\
 & 803 & 19/05/2000 & S3 & 25.5 & $0.17\pm0.04$ & $2.1\pm0.1$ & $8.6\pm0.4$ & $8.2^{+0.4}_{-1.2}$ \\
 & 5908 & 01/05/2005 & S3 & 45.1 & $0.12^{+0.03}_{-0.02}$ & $2.13\pm0.09$ & $17.5\pm0.4$ & $15^{+1}_{-3}$ \\
 & 6131 & 07/11/2005 & S3 & 35.8 & $0.17^{+0.08}_{-0.07}$ & $2.2\pm0.3$ & $4.9\pm0.3$ & $4.2^{+0.3}_{-0.4}$ \\
NGC5044 & 798 & 19/03/2000 & S3 & 14.3 & 0 & 1.9 & $2.4\pm0.4$ & $<0.2$ \\
 & 9399 & 07/03/2008 & S3 & 82.5 & $0.23^{+0.08}_{-0.07}$ & 1.9 & $2.6\pm0.2$ & $1.4^{+0.4}_{-0.3}$ \\
PKS0745 & 2427 & 16/06/2001 & S3 & 17.9 & 0 & 1.9 & $3.0\pm0.6$ & $2.0^{+0.6}_{-1.8}$ \\
 & 12881 & 27/01/2011 & S3 & 116.0 & $<0.08$ & $1.8\pm0.2$ & $4.8\pm0.3$ & $2.6^{+0.2}_{-0.4}$ \\
PKS1404 & 1650 & 07/06/2001 & S3 & 7.1 & $0.07\pm0.05$ & $2.4\pm0.2$ & $24\pm1$ & $22\pm2$ \\
 & 12884 & 03/01/2011 & S3 & 83.5 & $0.04\pm0.02$ & $2.01\pm0.06$ & $21.4\pm0.4$ & $19.8\pm0.7$ \\
\hline
\end{tabular}
\end{center}
\label{tab:variab}
\end{minipage}
\end{table*}

\clearpage

\end{document}

%% file: defn.tex
\newcommand{\Mpc}{\rm\; Mpc}
\newcommand{\kpc}{\rm\; kpc}

\newcommand{\km}{\rm\; km}

\newcommand{\cm}{\rm\; cm}
\newcommand{\mum}{\hbox{$\rm\; \mu m\,$}}
\newcommand{\pix}{\rm\; pixel}

\newcommand{\ppix}{\hbox{$\pix^{-1}\,$}}

\newcommand{\yr}{\rm\; yr}

\newcommand{\Myr}{\rm\; Myr}
\newcommand{\s}{\rm\; s}

\newcommand{\ks}{\rm\; ks}

\newcommand{\GHz}{\rm\; GHz}

\newcommand{\Msun}{\hbox{$\rm\thinspace M_{\odot}$}}

\newcommand{\Msunpyr}{\hbox{$\Msun\yr^{-1}\,$}}

\newcommand{\keV}{\rm\; keV}

\newcommand{\erg}{\rm\; erg}

\newcommand{\ergpcmsqps}{\hbox{$\erg\cm^{-2}\s^{-1}\,$}}

\newcommand{\ergps}{\hbox{$\erg\s^{-1}\,$}}

\newcommand{\expmapcorr}{\hbox{$\rm\thinspace photons\pcmsq\ps\ppix$}}

\newcommand{\kmps}{\hbox{$\km\s^{-1}\,$}}

\newcommand{\kmpspMpc}{\hbox{$\kmps\Mpc^{-1}\,$}}

\newcommand{\dex}{\rm\; dex}

\newcommand{\asec}{\rm\; arcsec}

\newcommand{\pcmsq}{\hbox{$\cm^{-2}\,$}}
\newcommand{\pcmcu}{\hbox{$\cm^{-3}\,$}}

\newcommand{\ps}{\hbox{$\s^{-1}\,$}}




%% file: AGNPS.bbl
\begin{thebibliography}{181}
\expandafter\ifx\csname natexlab\endcsname\relax\def\natexlab#1{#1}\fi

\bibitem[{{Abramowicz} {et~al}\mbox{.}(1995){Abramowicz}, {Chen}, {Kato},
  {Lasota}, \& {Regev}}]{Abramowicz95}
{Abramowicz} M.~A., {Chen} X., {Kato} S., {Lasota} J.-P., {Regev} O., 1995,
  \apjl, 438, L37

\bibitem[{{Akritas} \& {Bershady}(1996)}]{Akritas96}
{Akritas} M.~G., {Bershady} M.~A., 1996, \apj, 470, 706

\bibitem[{{Alexander} {et~al}\mbox{.}(2008){Alexander}, {Chary}, {Pope},
  {Bauer}, {Brandt}, {Daddi}, {Dickinson}, {Elbaz}, \& {Reddy}}]{Alexander08}
{Alexander} D.~M. {et~al.}, 2008, \apj, 687, 835

\bibitem[{{Allen} {et~al}\mbox{.}(2006){Allen}, {Dunn}, {Fabian}, {Taylor}, \&
  {Reynolds}}]{Allen06}
{Allen} S.~W., {Dunn} R.~J.~H., {Fabian} A.~C., {Taylor} G.~B., {Reynolds}
  C.~S., 2006, \mnras, 372, 21

\bibitem[{{Allen}, {Ettori} \& {Fabian}(2001){Allen}, {Ettori}, \&
  {Fabian}}]{Allen01}
{Allen} S.~W., {Ettori} S., {Fabian} A.~C., 2001, \mnras, 324, 877

\bibitem[{{Anders} \& {Grevesse}(1989)}]{AndersGrevesse89}
{Anders} E., {Grevesse} N., 1989, \gca, 53, 197

\bibitem[{{Arnaud}(1996)}]{Arnaud96}
{Arnaud} K.~A., 1996, in Astronomical Society of the Pacific Conference Series,
  Vol. 101, Astronomical Data Analysis Software and Systems V, {Jacoby} G.~H.,
  {Barnes} J., eds., pp. 17--+

\bibitem[{{Balmaverde}, {Capetti} \& {Grandi}(2006){Balmaverde}, {Capetti}, \&
  {Grandi}}]{Balmaverde06}
{Balmaverde} B., {Capetti} A., {Grandi} P., 2006, \aap, 451, 35

\bibitem[{{Balucinska-Church} \& {McCammon}(1992)}]{Balucinska92}
{Balucinska-Church} M., {McCammon} D., 1992, \apj, 400, 699

\bibitem[{{Batcheldor} {et~al}\mbox{.}(2007){Batcheldor}, {Marconi}, {Merritt},
  \& {Axon}}]{Batcheldor07}
{Batcheldor} D., {Marconi} A., {Merritt} D., {Axon} D.~J., 2007, \apjl, 663,
  L85

\bibitem[{{Begelman}, {Blandford} \& {Rees}(1984){Begelman}, {Blandford}, \&
  {Rees}}]{Begelman84}
{Begelman} M.~C., {Blandford} R.~D., {Rees} M.~J., 1984, Reviews of Modern
  Physics, 56, 255

\bibitem[{{Belsole} {et~al}\mbox{.}(2007){Belsole}, {Worrall}, {Hardcastle}, \&
  {Croston}}]{Belsole07}
{Belsole} E., {Worrall} D.~M., {Hardcastle} M.~J., {Croston} J.~H., 2007,
  \mnras, 381, 1109

\bibitem[{{B{\^i}rzan} {et~al}\mbox{.}(2009){B{\^i}rzan}, {Rafferty},
  {McNamara}, {Nulsen}, \& {Wise}}]{Birzan09}
{B{\^i}rzan} L., {Rafferty} D.~A., {McNamara} B.~R., {Nulsen} P.~E.~J., {Wise}
  M.~W., 2009, in American Institute of Physics Conference Series, Vol. 1201,
  American Institute of Physics Conference Series, {Heinz} S., {Wilcots} E.,
  eds., pp. 301--304

\bibitem[{{B{\^i}rzan} {et~al}\mbox{.}(2004){B{\^i}rzan}, {Rafferty},
  {McNamara}, {Wise}, \& {Nulsen}}]{Birzan04}
{B{\^i}rzan} L., {Rafferty} D.~A., {McNamara} B.~R., {Wise} M.~W., {Nulsen}
  P.~E.~J., 2004, \apj, 607, 800

\bibitem[{{B{\^i}rzan} {et~al}\mbox{.}(2012){B{\^i}rzan}, {Rafferty}, {Nulsen},
  {McNamara}, {R{\"o}ttgering}, {Wise}, \& {Mittal}}]{Birzan12}
{B{\^i}rzan} L., {Rafferty} D.~A., {Nulsen} P.~E.~J., {McNamara} B.~R.,
  {R{\"o}ttgering} H.~J.~A., {Wise} M.~W., {Mittal} R., 2012, ArXiv:1210.7100

\bibitem[{{Blandford} \& {Begelman}(1999)}]{Blandford99}
{Blandford} R.~D., {Begelman} M.~C., 1999, \mnras, 303, L1

\bibitem[{{Blandford} \& {Begelman}(2004)}]{Blandford04}
{Blandford} R.~D., {Begelman} M.~C., 2004, \mnras, 349, 68

\bibitem[{{Blanton} {et~al}\mbox{.}(2011){Blanton}, {Randall}, {Clarke},
  {Sarazin}, {McNamara}, {Douglass}, \& {McDonald}}]{Blanton11}
{Blanton} E.~L., {Randall} S.~W., {Clarke} T.~E., {Sarazin} C.~L., {McNamara}
  B.~R., {Douglass} E.~M., {McDonald} M., 2011, \apj, 737, 99

\bibitem[{{Blanton} {et~al}\mbox{.}(2009){Blanton}, {Randall}, {Douglass},
  {Sarazin}, {Clarke}, \& {McNamara}}]{Blanton09}
{Blanton} E.~L., {Randall} S.~W., {Douglass} E.~M., {Sarazin} C.~L., {Clarke}
  T.~E., {McNamara} B.~R., 2009, \apjl, 697, L95

\bibitem[{{Blanton}, {Sarazin} \& {McNamara}(2003){Blanton}, {Sarazin}, \&
  {McNamara}}]{Blanton03}
{Blanton} E.~L., {Sarazin} C.~L., {McNamara} B.~R., 2003, \apj, 585, 227

\bibitem[{{Bondi}(1952)}]{Bondi52}
{Bondi} H., 1952, \mnras, 112, 195

\bibitem[{{Bower} {et~al}\mbox{.}(2006){Bower}, {Benson}, {Malbon}, {Helly},
  {Frenk}, {Baugh}, {Cole}, \& {Lacey}}]{Bower06}
{Bower} R.~G., {Benson} A.~J., {Malbon} R., {Helly} J.~C., {Frenk} C.~S.,
  {Baugh} C.~M., {Cole} S., {Lacey} C.~G., 2006, \mnras, 370, 645

\bibitem[{{Brandt} \& {Hasinger}(2005)}]{Brandt05}
{Brandt} W.~N., {Hasinger} G., 2005, \araa, 43, 827

\bibitem[{{Brown}, {Hollander} \& {Korwar}(1974){Brown}, {Hollander}, \&
  {Korwar}}]{Brown74}
{Brown} B.~W.~M., {Hollander} M., {Korwar} R.~M., 1974, in Reliability and
  Biometry, {Proschan} F., {Serfling} R.~J., eds., Society for Industrial and
  Applied Mathematics, Philadelphia, pp. 327--354

\bibitem[{{Buckley} \& {James}(1979)}]{Buckley79}
{Buckley} J., {James} I., 1979, Biometrika, 66, 429

\bibitem[{{Canosa} {et~al}\mbox{.}(1999){Canosa}, {Worrall}, {Hardcastle}, \&
  {Birkinshaw}}]{Canosa99}
{Canosa} C.~M., {Worrall} D.~M., {Hardcastle} M.~J., {Birkinshaw} M., 1999,
  \mnras, 310, 30

\bibitem[{{Cao}(2011)}]{Cao11}
{Cao} X., 2011, \apj, 737, 94

\bibitem[{{Capetti} {et~al}\mbox{.}(2002){Capetti}, {Trussoni}, {Celotti},
  {Feretti}, \& {Chiaberge}}]{Capetti02}
{Capetti} A., {Trussoni} E., {Celotti} A., {Feretti} L., {Chiaberge} M., 2002,
  \nar, 46, 335

\bibitem[{{Cash}(1979)}]{Cash79}
{Cash} W., 1979, \apj, 228, 939

\bibitem[{{Cavagnolo} {et~al}\mbox{.}(2009){Cavagnolo}, {Donahue}, {Voit}, \&
  {Sun}}]{Cavagnolo09}
{Cavagnolo} K.~W., {Donahue} M., {Voit} G.~M., {Sun} M., 2009, \apjs, 182, 12

\bibitem[{{Cavagnolo} {et~al}\mbox{.}(2010){Cavagnolo}, {McNamara}, {Nulsen},
  {Carilli}, {Jones}, \& {B{\^i}rzan}}]{Cavagnolo10}
{Cavagnolo} K.~W., {McNamara} B.~R., {Nulsen} P.~E.~J., {Carilli} C.~L.,
  {Jones} C., {B{\^i}rzan} L., 2010, \apj, 720, 1066

\bibitem[{{Cavagnolo} {et~al}\mbox{.}(2011){Cavagnolo}, {McNamara}, {Wise},
  {Nulsen}, {Br{\"u}ggen}, {Gitti}, \& {Rafferty}}]{Cavagnolo11}
{Cavagnolo} K.~W., {McNamara} B.~R., {Wise} M.~W., {Nulsen} P.~E.~J.,
  {Br{\"u}ggen} M., {Gitti} M., {Rafferty} D.~A., 2011, \apj, 732, 71

\bibitem[{{Chiaberge}, {Capetti} \& {Celotti}(1999){Chiaberge}, {Capetti}, \&
  {Celotti}}]{Chiaberge99}
{Chiaberge} M., {Capetti} A., {Celotti} A., 1999, \aap, 349, 77

\bibitem[{{Chiaberge}, {Capetti} \& {Macchetto}(2005){Chiaberge}, {Capetti}, \&
  {Macchetto}}]{Chiaberge05}
{Chiaberge} M., {Capetti} A., {Macchetto} F.~D., 2005, \apj, 625, 716

\bibitem[{{Chiaberge} {et~al}\mbox{.}(2003{\natexlab{a}}){Chiaberge}, {Gilli},
  {Capetti}, \& {Macchetto}}]{Chiaberge03}
{Chiaberge} M., {Gilli} R., {Capetti} A., {Macchetto} F.~D.,
  2003{\natexlab{a}}, \apj, 597, 166

\bibitem[{{Chiaberge} {et~al}\mbox{.}(2003{\natexlab{b}}){Chiaberge}, {Gilli},
  {Macchetto}, {Sparks}, \& {Capetti}}]{Chiabergengc426103}
{Chiaberge} M., {Gilli} R., {Macchetto} F.~D., {Sparks} W.~B., {Capetti} A.,
  2003{\natexlab{b}}, \apj, 582, 645

\bibitem[{{Chiaberge} {et~al}\mbox{.}(2002){Chiaberge}, {Macchetto}, {Sparks},
  {Capetti}, {Allen}, \& {Martel}}]{Chiaberge02}
{Chiaberge} M., {Macchetto} F.~D., {Sparks} W.~B., {Capetti} A., {Allen} M.~G.,
  {Martel} A.~R., 2002, \apj, 571, 247

\bibitem[{{Churazov} {et~al}\mbox{.}(2000){Churazov}, {Forman}, {Jones}, \&
  {B{\"o}hringer}}]{Churazov00}
{Churazov} E., {Forman} W., {Jones} C., {B{\"o}hringer} H., 2000, \aap, 356,
  788

\bibitem[{{Churazov} {et~al}\mbox{.}(2005){Churazov}, {Sazonov}, {Sunyaev},
  {Forman}, {Jones}, \& {B{\"o}hringer}}]{Churazov05}
{Churazov} E., {Sazonov} S., {Sunyaev} R., {Forman} W., {Jones} C.,
  {B{\"o}hringer} H., 2005, \mnras, 363, L91

\bibitem[{{Churazov} {et~al}\mbox{.}(2002){Churazov}, {Sunyaev}, {Forman}, \&
  {B{\"o}hringer}}]{Churazov02}
{Churazov} E., {Sunyaev} R., {Forman} W., {B{\"o}hringer} H., 2002, \mnras,
  332, 729

\bibitem[{{Comastri}(2004)}]{Comastri04}
{Comastri} A., 2004, in Astrophysics and Space Science Library, Vol. 308,
  Supermassive Black Holes in the Distant Universe, {Barger} A.~J., ed., p. 245

\bibitem[{{Crawford} {et~al}\mbox{.}(1999){Crawford}, {Lehmann}, {Fabian},
  {Bremer}, \& {Hasinger}}]{Crawfordqsos99}
{Crawford} C.~S., {Lehmann} I., {Fabian} A.~C., {Bremer} M.~N., {Hasinger} G.,
  1999, \mnras, 308, 1159

\bibitem[{{Croton} {et~al}\mbox{.}(2006){Croton}, {Springel}, {White}, {De
  Lucia}, {Frenk}, {Gao}, {Jenkins}, {Kauffmann}, {Navarro}, \&
  {Yoshida}}]{Croton06}
{Croton} D.~J. {et~al.}, 2006, \mnras, 365, 11

\bibitem[{{Dalla Bont{\`a}} {et~al}\mbox{.}(2009){Dalla Bont{\`a}},
  {Ferrarese}, {Corsini}, {Miralda-Escud{\'e}}, {Coccato}, {Sarzi}, {Pizzella},
  \& {Beifiori}}]{DallaBonta09}
{Dalla Bont{\`a}} E., {Ferrarese} L., {Corsini} E.~M., {Miralda-Escud{\'e}} J.,
  {Coccato} L., {Sarzi} M., {Pizzella} A., {Beifiori} A., 2009, \apj, 690, 537

\bibitem[{{David} {et~al}\mbox{.}(2009){David}, {Jones}, {Forman}, {Nulsen},
  {Vrtilek}, {O'Sullivan}, {Giacintucci}, \& {Raychaudhury}}]{David09}
{David} L.~P., {Jones} C., {Forman} W., {Nulsen} P., {Vrtilek} J., {O'Sullivan}
  E., {Giacintucci} S., {Raychaudhury} S., 2009, \apj, 705, 624

\bibitem[{{Davis}(2001)}]{Davis01}
{Davis} J.~E., 2001, \apj, 562, 575

\bibitem[{{Di Matteo} {et~al}\mbox{.}(2003){Di Matteo}, {Allen}, {Fabian},
  {Wilson}, \& {Young}}]{DiMatteo03}
{Di Matteo} T., {Allen} S.~W., {Fabian} A.~C., {Wilson} A.~S., {Young} A.~J.,
  2003, \apj, 582, 133

\bibitem[{{Di Matteo} {et~al}\mbox{.}(2001){Di Matteo}, {Johnstone}, {Allen},
  \& {Fabian}}]{DiMatteo01}
{Di Matteo} T., {Johnstone} R.~M., {Allen} S.~W., {Fabian} A.~C., 2001, \apjl,
  550, L19

\bibitem[{{Di Matteo} {et~al}\mbox{.}(2000){Di Matteo}, {Quataert}, {Allen},
  {Narayan}, \& {Fabian}}]{DiMatteo00}
{Di Matteo} T., {Quataert} E., {Allen} S.~W., {Narayan} R., {Fabian} A.~C.,
  2000, \mnras, 311, 507

\bibitem[{{Di Matteo}, {Springel} \& {Hernquist}(2005){Di Matteo}, {Springel},
  \& {Hernquist}}]{DiMatteo05}
{Di Matteo} T., {Springel} V., {Hernquist} L., 2005, \nat, 433, 604

\bibitem[{{Donahue} {et~al}\mbox{.}(2011){Donahue}, {de Messi{\`e}res},
  {O'Connell}, {Voit}, {Hoffer}, {McNamara}, \& {Nulsen}}]{Donahue11}
{Donahue} M., {de Messi{\`e}res} G.~E., {O'Connell} R.~W., {Voit} G.~M.,
  {Hoffer} A., {McNamara} B.~R., {Nulsen} P.~E.~J., 2011, \apj, 732, 40

\bibitem[{{Donato}, {Sambruna} \& {Gliozzi}(2004){Donato}, {Sambruna}, \&
  {Gliozzi}}]{Donato04}
{Donato} D., {Sambruna} R.~M., {Gliozzi} M., 2004, \apj, 617, 915

\bibitem[{{Done}, {Gierli{\'n}ski} \& {Kubota}(2007){Done}, {Gierli{\'n}ski},
  \& {Kubota}}]{Done07}
{Done} C., {Gierli{\'n}ski} M., {Kubota} A., 2007, \aapr, 15, 1

\bibitem[{{Dunn} \& {Fabian}(2004)}]{Dunn04}
{Dunn} R.~J.~H., {Fabian} A.~C., 2004, \mnras, 355, 862

\bibitem[{{Dunn} \& {Fabian}(2006)}]{DunnFabian06}
{Dunn} R.~J.~H., {Fabian} A.~C., 2006, \mnras, 373, 959

\bibitem[{{Dwarakanath}, {Owen} \& {van Gorkom}(1995){Dwarakanath}, {Owen}, \&
  {van Gorkom}}]{Dwarakanath95}
{Dwarakanath} K.~S., {Owen} F.~N., {van Gorkom} J.~H., 1995, \apjl, 442, L1

\bibitem[{{Edge}(2001)}]{Edge01}
{Edge} A.~C., 2001, \mnras, 328, 762

\bibitem[{{Egami} {et~al}\mbox{.}(2006){Egami}, {Misselt}, {Rieke}, {Wise},
  {Neugebauer}, {Kneib}, {Le Floc'h}, {Smith}, {Blaylock}, {Dole}, {Frayer},
  {Huang}, {Krause}, {Papovich}, {P{\'e}rez-Gonz{\'a}lez}, \&
  {Rigby}}]{Egami06}
{Egami} E. {et~al.}, 2006, \apj, 647, 922

\bibitem[{{Evans} {et~al}\mbox{.}(2006){Evans}, {Worrall}, {Hardcastle},
  {Kraft}, \& {Birkinshaw}}]{Evans06}
{Evans} D.~A., {Worrall} D.~M., {Hardcastle} M.~J., {Kraft} R.~P., {Birkinshaw}
  M., 2006, \apj, 642, 96

\bibitem[{{Fabbiano}, {Gioia} \& {Trinchieri}(1989){Fabbiano}, {Gioia}, \&
  {Trinchieri}}]{Fabbiano89}
{Fabbiano} G., {Gioia} I.~M., {Trinchieri} G., 1989, \apj, 347, 127

\bibitem[{{Fabian} {et~al}\mbox{.}(1981){Fabian}, {Hu}, {Cowie}, \&
  {Grindlay}}]{Fabian81}
{Fabian} A.~C., {Hu} E.~M., {Cowie} L.~L., {Grindlay} J., 1981, \apj, 248, 47

\bibitem[{{Fabian} \& {Iwasawa}(1999)}]{Fabian99}
{Fabian} A.~C., {Iwasawa} K., 1999, \mnras, 303, L34

\bibitem[{{Fabian} \& {Rees}(1995)}]{Fabian95}
{Fabian} A.~C., {Rees} M.~J., 1995, \mnras, 277, L55

\bibitem[{{Fabian} {et~al}\mbox{.}(2011){Fabian}, {Sanders}, {Allen},
  {Canning}, {Churazov}, {Crawford}, {Forman}, {Gabany}, {Hlavacek-Larrondo},
  {Johnstone}, {Russell}, {Reynolds}, {Salom{\'e}}, {Taylor}, \&
  {Young}}]{Fabian11}
{Fabian} A.~C. {et~al.}, 2011, \mnras, 418, 2154

\bibitem[{{Fabian} {et~al}\mbox{.}(2003){Fabian}, {Sanders}, {Allen},
  {Crawford}, {Iwasawa}, {Johnstone}, {Schmidt}, \& {Taylor}}]{FabianPer03}
{Fabian} A.~C., {Sanders} J.~S., {Allen} S.~W., {Crawford} C.~S., {Iwasawa} K.,
  {Johnstone} R.~M., {Schmidt} R.~W., {Taylor} G.~B., 2003, \mnras, 344, L43

\bibitem[{{Fabian} {et~al}\mbox{.}(2000){Fabian}, {Sanders}, {Ettori},
  {Taylor}, {Allen}, {Crawford}, {Iwasawa}, {Johnstone}, \&
  {Ogle}}]{FabianPer00}
{Fabian} A.~C. {et~al.}, 2000, \mnras, 318, L65

\bibitem[{{Fabian} {et~al}\mbox{.}(2005){Fabian}, {Sanders}, {Taylor}, \&
  {Allen}}]{Fabian05}
{Fabian} A.~C., {Sanders} J.~S., {Taylor} G.~B., {Allen} S.~W., 2005, \mnras,
  360, L20

\bibitem[{{Fabian} {et~al}\mbox{.}(2006){Fabian}, {Sanders}, {Taylor}, {Allen},
  {Crawford}, {Johnstone}, \& {Iwasawa}}]{FabianPer06}
{Fabian} A.~C., {Sanders} J.~S., {Taylor} G.~B., {Allen} S.~W., {Crawford}
  C.~S., {Johnstone} R.~M., {Iwasawa} K., 2006, \mnras, 366, 417

\bibitem[{{Falcke} \& {Biermann}(1995)}]{Falcke95}
{Falcke} H., {Biermann} P.~L., 1995, \aap, 293, 665

\bibitem[{{Falcke}, {K{\"o}rding} \& {Markoff}(2004){Falcke}, {K{\"o}rding}, \&
  {Markoff}}]{Falcke04}
{Falcke} H., {K{\"o}rding} E., {Markoff} S., 2004, \aap, 414, 895

\bibitem[{{Fender} \& {Belloni}(2004)}]{FenderBelloni04}
{Fender} R., {Belloni} T., 2004, \araa, 42, 317

\bibitem[{{Fender} {et~al}\mbox{.}(1999){Fender}, {Corbel}, {Tzioumis},
  {McIntyre}, {Campbell-Wilson}, {Nowak}, {Sood}, {Hunstead}, {Harmon},
  {Durouchoux}, \& {Heindl}}]{Fender99}
{Fender} R. {et~al.}, 1999, \apjl, 519, L165

\bibitem[{{Fender}, {Belloni} \& {Gallo}(2004){Fender}, {Belloni}, \&
  {Gallo}}]{Fender04}
{Fender} R.~P., {Belloni} T.~M., {Gallo} E., 2004, \mnras, 355, 1105

\bibitem[{{Ferrarese}, {Ford} \& {Jaffe}(1996){Ferrarese}, {Ford}, \&
  {Jaffe}}]{Ferrarese96}
{Ferrarese} L., {Ford} H.~C., {Jaffe} W., 1996, \apj, 470, 444

\bibitem[{{Forman} {et~al}\mbox{.}(2005){Forman}, {Nulsen}, {Heinz}, {Owen},
  {Eilek}, {Vikhlinin}, {Markevitch}, {Kraft}, {Churazov}, \&
  {Jones}}]{FormanM8705}
{Forman} W. {et~al.}, 2005, \apj, 635, 894

\bibitem[{{Frank}, {King} \& {Raine}(2002){Frank}, {King}, \&
  {Raine}}]{Frank02}
{Frank} J., {King} A., {Raine} D.~J., 2002, {Accretion Power in Astrophysics:
  Third Edition}. Cambridge University Press

\bibitem[{{Freeman} {et~al}\mbox{.}(2002){Freeman}, {Kashyap}, {Rosner}, \&
  {Lamb}}]{Freeman02}
{Freeman} P.~E., {Kashyap} V., {Rosner} R., {Lamb} D.~Q., 2002, \apjs, 138, 185

\bibitem[{{Gallo}, {Fender} \& {Pooley}(2003){Gallo}, {Fender}, \&
  {Pooley}}]{Gallo03}
{Gallo} E., {Fender} R.~P., {Pooley} G.~G., 2003, \mnras, 344, 60

\bibitem[{{Gandhi} {et~al}\mbox{.}(2009){Gandhi}, {Horst}, {Smette},
  {H{\"o}nig}, {Comastri}, {Gilli}, {Vignali}, \& {Duschl}}]{Gandhi09}
{Gandhi} P., {Horst} H., {Smette} A., {H{\"o}nig} S., {Comastri} A., {Gilli}
  R., {Vignali} C., {Duschl} W., 2009, \aap, 502, 457

\bibitem[{{Gebhardt} {et~al}\mbox{.}(2011){Gebhardt}, {Adams}, {Richstone},
  {Lauer}, {Faber}, {G{\"u}ltekin}, {Murphy}, \& {Tremaine}}]{Gebhardt11}
{Gebhardt} K., {Adams} J., {Richstone} D., {Lauer} T.~R., {Faber} S.~M.,
  {G{\"u}ltekin} K., {Murphy} J., {Tremaine} S., 2011, \apj, 729, 119

\bibitem[{{Gilli}, {Comastri} \& {Hasinger}(2007){Gilli}, {Comastri}, \&
  {Hasinger}}]{Gilli07}
{Gilli} R., {Comastri} A., {Hasinger} G., 2007, \aap, 463, 79

\bibitem[{{Gliozzi}, {Sambruna} \& {Brandt}(2003){Gliozzi}, {Sambruna}, \&
  {Brandt}}]{Gliozzi03}
{Gliozzi} M., {Sambruna} R.~M., {Brandt} W.~N., 2003, \aap, 408, 949

\bibitem[{{Graham}(2007)}]{Graham07}
{Graham} A.~W., 2007, \mnras, 379, 711

\bibitem[{{Guainazzi}, {Matt} \& {Perola}(2005){Guainazzi}, {Matt}, \&
  {Perola}}]{Guainazzi05}
{Guainazzi} M., {Matt} G., {Perola} G.~C., 2005, \aap, 444, 119

\bibitem[{{Haehnelt}, {Natarajan} \& {Rees}(1998){Haehnelt}, {Natarajan}, \&
  {Rees}}]{Haehnelt98}
{Haehnelt} M.~G., {Natarajan} P., {Rees} M.~J., 1998, \mnras, 300, 817

\bibitem[{{Hardcastle}, {Evans} \& {Croston}(2006){Hardcastle}, {Evans}, \&
  {Croston}}]{Hardcastle06}
{Hardcastle} M.~J., {Evans} D.~A., {Croston} J.~H., 2006, \mnras, 370, 1893

\bibitem[{{Hardcastle}, {Evans} \& {Croston}(2007){Hardcastle}, {Evans}, \&
  {Croston}}]{Hardcastle07}
{Hardcastle} M.~J., {Evans} D.~A., {Croston} J.~H., 2007, \mnras, 376, 1849

\bibitem[{{Hardcastle}, {Evans} \& {Croston}(2009){Hardcastle}, {Evans}, \&
  {Croston}}]{Hardcastle09}
{Hardcastle} M.~J., {Evans} D.~A., {Croston} J.~H., 2009, \mnras, 396, 1929

\bibitem[{{Hardcastle} \& {Worrall}(1999)}]{Hardcastle99}
{Hardcastle} M.~J., {Worrall} D.~M., 1999, \mnras, 309, 969

\bibitem[{{Harris}, {Biretta} \& {Junor}(1997){Harris}, {Biretta}, \&
  {Junor}}]{Harris97}
{Harris} D.~E., {Biretta} J.~A., {Junor} W., 1997, \mnras, 284, L21

\bibitem[{{Harris} {et~al}\mbox{.}(2006){Harris}, {Cheung}, {Biretta},
  {Sparks}, {Junor}, {Perlman}, \& {Wilson}}]{Harris06}
{Harris} D.~E., {Cheung} C.~C., {Biretta} J.~A., {Sparks} W.~B., {Junor} W.,
  {Perlman} E.~S., {Wilson} A.~S., 2006, \apj, 640, 211

\bibitem[{{Harris} {et~al}\mbox{.}(2009){Harris}, {Cheung}, {Stawarz},
  {Biretta}, \& {Perlman}}]{Harris09}
{Harris} D.~E., {Cheung} C.~C., {Stawarz} {\L}., {Biretta} J.~A., {Perlman}
  E.~S., 2009, \apj, 699, 305

\bibitem[{{Harris} {et~al}\mbox{.}(2000){Harris}, {Nulsen}, {Ponman}, {Bautz},
  {Cameron}, {David}, {Donnelly}, {Forman}, {Grego}, {Hardcastle}, {Henry},
  {Jones}, {Leahy}, {Markevitch}, {Martel}, {McNamara}, {Mazzotta}, {Tucker},
  {Virani}, \& {Vrtilek}}]{Harris00}
{Harris} D.~E. {et~al.}, 2000, \apjl, 530, L81

\bibitem[{{Heinz} \& {Sunyaev}(2003)}]{Heinz03}
{Heinz} S., {Sunyaev} R.~A., 2003, \mnras, 343, L59

\bibitem[{{Hlavacek-Larrondo} \& {Fabian}(2011)}]{HlavacekLarrondo11}
{Hlavacek-Larrondo} J., {Fabian} A.~C., 2011, \mnras, 413, 313

\bibitem[{{Hlavacek-Larrondo} {et~al}\mbox{.}(2012){Hlavacek-Larrondo},
  {Fabian}, {Edge}, {Ebeling}, {Sanders}, {Hogan}, \&
  {Taylor}}]{HlavacekLarrondo12}
{Hlavacek-Larrondo} J., {Fabian} A.~C., {Edge} A.~C., {Ebeling} H., {Sanders}
  J.~S., {Hogan} M.~T., {Taylor} G.~B., 2012, \mnras, 421, 1360

\bibitem[{{Hlavacek-Larrondo} {et~al}\mbox{.}(2011){Hlavacek-Larrondo},
  {Fabian}, {Sanders}, \& {Taylor}}]{HlavacekLarrondo4C11}
{Hlavacek-Larrondo} J., {Fabian} A.~C., {Sanders} J.~S., {Taylor} G.~B., 2011,
  \mnras, 415, 3520

\bibitem[{{Ho}(2002)}]{Ho02}
{Ho} L.~C., 2002, \apj, 564, 120

\bibitem[{{Hopkins} {et~al}\mbox{.}(2006){Hopkins}, {Hernquist}, {Cox}, {Di
  Matteo}, {Robertson}, \& {Springel}}]{Hopkins06}
{Hopkins} P.~F., {Hernquist} L., {Cox} T.~J., {Di Matteo} T., {Robertson} B.,
  {Springel} V., 2006, \apjs, 163, 1

\bibitem[{{Isobe} \& {Feigelson}(1990)}]{Isobe90}
{Isobe} T., {Feigelson} E.~D., 1990, in Bulletin of the American Astronomical
  Society, Vol.~22, Bulletin of the American Astronomical Society, pp. 917--918

\bibitem[{{Isobe}, {Feigelson} \& {Nelson}(1986){Isobe}, {Feigelson}, \&
  {Nelson}}]{Isobe86}
{Isobe} T., {Feigelson} E.~D., {Nelson} P.~I., 1986, \apj, 306, 490

\bibitem[{{Jaffe} \& {McNamara}(1994)}]{Jaffe94}
{Jaffe} W., {McNamara} B.~R., 1994, \apj, 434, 110

\bibitem[{{Johnstone} {et~al}\mbox{.}(2005){Johnstone}, {Fabian}, {Morris}, \&
  {Taylor}}]{Johnstone05}
{Johnstone} R.~M., {Fabian} A.~C., {Morris} R.~G., {Taylor} G.~B., 2005,
  \mnras, 356, 237

\bibitem[{{Kaastra}(1992)}]{Kaastra92}
{Kaastra} J.~S., 1992, in Internal SRON-Leiden Report, updated version 2.0

\bibitem[{{Kalberla} {et~al}\mbox{.}(2005){Kalberla}, {Burton}, {Hartmann},
  {Arnal}, {Bajaja}, {Morras}, \& {P{\"o}ppel}}]{Kalberla05}
{Kalberla} P.~M.~W., {Burton} W.~B., {Hartmann} D., {Arnal} E.~M., {Bajaja} E.,
  {Morras} R., {P{\"o}ppel} W.~G.~L., 2005, \aap, 440, 775

\bibitem[{{K{\"o}rding}, {Falcke} \& {Corbel}(2006){K{\"o}rding}, {Falcke}, \&
  {Corbel}}]{Kording06}
{K{\"o}rding} E., {Falcke} H., {Corbel} S., 2006, \aap, 456, 439

\bibitem[{{Kormendy} \& {Gebhardt}(2001)}]{Kormendy01}
{Kormendy} J., {Gebhardt} K., 2001, in American Institute of Physics Conference
  Series, Vol. 586, 20th Texas Symposium on relativistic astrophysics,
  {Wheeler} J.~C., {Martel} H., eds., pp. 363--381

\bibitem[{{Kriss}, {Cioffi} \& {Canizares}(1983){Kriss}, {Cioffi}, \&
  {Canizares}}]{Kriss83}
{Kriss} G.~A., {Cioffi} D.~F., {Canizares} C.~R., 1983, \apj, 272, 439

\bibitem[{{Lauer} {et~al}\mbox{.}(2007){Lauer}, {Faber}, {Richstone},
  {Gebhardt}, {Tremaine}, {Postman}, {Dressler}, {Aller}, {Filippenko},
  {Green}, {Ho}, {Kormendy}, {Magorrian}, \& {Pinkney}}]{Lauer07}
{Lauer} T.~R. {et~al.}, 2007, \apj, 662, 808

\bibitem[{{Lavalley}, {Isobe} \& {Feigelson}(1992){Lavalley}, {Isobe}, \&
  {Feigelson}}]{Lavalley92}
{Lavalley} M.~P., {Isobe} T., {Feigelson} E.~D., 1992, in Bulletin of the
  American Astronomical Society, Vol.~24, Bulletin of the American Astronomical
  Society, pp. 839--840

\bibitem[{{Liedahl}, {Osterheld} \& {Goldstein}(1995){Liedahl}, {Osterheld}, \&
  {Goldstein}}]{Liedahl95}
{Liedahl} D.~A., {Osterheld} A.~L., {Goldstein} W.~H., 1995, \apjl, 438, L115

\bibitem[{{Maccarone}, {Gallo} \& {Fender}(2003){Maccarone}, {Gallo}, \&
  {Fender}}]{Maccarone03}
{Maccarone} T.~J., {Gallo} E., {Fender} R., 2003, \mnras, 345, L19

\bibitem[{{Machacek} {et~al}\mbox{.}(2006){Machacek}, {Jones}, {Forman}, \&
  {Nulsen}}]{Machacek06}
{Machacek} M., {Jones} C., {Forman} W.~R., {Nulsen} P., 2006, \apj, 644, 155

\bibitem[{{Machacek} {et~al}\mbox{.}(2011){Machacek}, {Jerius}, {Kraft},
  {Forman}, {Jones}, {Randall}, {Giacintucci}, \& {Sun}}]{Machacek11}
{Machacek} M.~E., {Jerius} D., {Kraft} R., {Forman} W.~R., {Jones} C.,
  {Randall} S., {Giacintucci} S., {Sun} M., 2011, \apj, 743, 15

\bibitem[{{Machacek} {et~al}\mbox{.}(2007){Machacek}, {Kraft}, {Jones},
  {Forman}, \& {Hardcastle}}]{Machacek07}
{Machacek} M.~E., {Kraft} R.~P., {Jones} C., {Forman} W.~R., {Hardcastle}
  M.~J., 2007, \apj, 664, 804

\bibitem[{{Magorrian} {et~al}\mbox{.}(1998){Magorrian}, {Tremaine},
  {Richstone}, {Bender}, {Bower}, {Dressler}, {Faber}, {Gebhardt}, {Green},
  {Grillmair}, {Kormendy}, \& {Lauer}}]{Magorrian98}
{Magorrian} J. {et~al.}, 1998, \aj, 115, 2285

\bibitem[{{Maiolino} {et~al}\mbox{.}(1998){Maiolino}, {Salvati}, {Bassani},
  {Dadina}, {della Ceca}, {Matt}, {Risaliti}, \& {Zamorani}}]{Maiolino98}
{Maiolino} R., {Salvati} M., {Bassani} L., {Dadina} M., {della Ceca} R., {Matt}
  G., {Risaliti} G., {Zamorani} G., 1998, \aap, 338, 781

\bibitem[{{McKinney}, {Tchekhovskoy} \& {Blandford}(2012){McKinney},
  {Tchekhovskoy}, \& {Blandford}}]{McKinney12}
{McKinney} J.~C., {Tchekhovskoy} A., {Blandford} R.~D., 2012, \mnras, 423, 3083

\bibitem[{{McNamara} {et~al}\mbox{.}(2009){McNamara}, {Kazemzadeh}, {Rafferty},
  {B{\^i}rzan}, {Nulsen}, {Kirkpatrick}, \& {Wise}}]{McNamara09}
{McNamara} B.~R., {Kazemzadeh} F., {Rafferty} D.~A., {B{\^i}rzan} L., {Nulsen}
  P.~E.~J., {Kirkpatrick} C.~C., {Wise} M.~W., 2009, \apj, 698, 594

\bibitem[{{McNamara} \& {Nulsen}(2007)}]{McNamaraNulsen07}
{McNamara} B.~R., {Nulsen} P.~E.~J., 2007, \araa, 45, 117

\bibitem[{{McNamara} \& {Nulsen}(2012)}]{McNamara12}
{McNamara} B.~R., {Nulsen} P.~E.~J., 2012, New Journal of Physics, 14, 055023

\bibitem[{{McNamara}, {Rohanizadegan} \& {Nulsen}(2011){McNamara},
  {Rohanizadegan}, \& {Nulsen}}]{McNamara11}
{McNamara} B.~R., {Rohanizadegan} M., {Nulsen} P.~E.~J., 2011, \apj, 727, 39

\bibitem[{{McNamara} {et~al}\mbox{.}(2000){McNamara}, {Wise}, {Nulsen},
  {David}, {Sarazin}, {Bautz}, {Markevitch}, {Vikhlinin}, {Forman}, {Jones}, \&
  {Harris}}]{McNamara00}
{McNamara} B.~R. {et~al.}, 2000, \apjl, 534, L135

\bibitem[{{Mendygral}, {Jones} \& {Dolag}(2012){Mendygral}, {Jones}, \&
  {Dolag}}]{Mendygral12}
{Mendygral} P.~J., {Jones} T.~W., {Dolag} K., 2012, \apj, 750, 166

\bibitem[{{Mendygral}, {O'Neill} \& {Jones}(2011){Mendygral}, {O'Neill}, \&
  {Jones}}]{Mendygral11}
{Mendygral} P.~J., {O'Neill} S.~M., {Jones} T.~W., 2011, \apj, 730, 100

\bibitem[{{Merloni} \& {Heinz}(2007)}]{Merloni07}
{Merloni} A., {Heinz} S., 2007, \mnras, 381, 589

\bibitem[{{Merloni}, {Heinz} \& {di Matteo}(2003){Merloni}, {Heinz}, \& {di
  Matteo}}]{Merloni03}
{Merloni} A., {Heinz} S., {di Matteo} T., 2003, \mnras, 345, 1057

\bibitem[{{Merritt} \& {Ferrarese}(2001)}]{Merritt01}
{Merritt} D., {Ferrarese} L., 2001, \apj, 547, 140

\bibitem[{{Mewe}, {Gronenschild} \& {van den Oord}(1985){Mewe}, {Gronenschild},
  \& {van den Oord}}]{Mewe85}
{Mewe} R., {Gronenschild} E.~H.~B.~M., {van den Oord} G.~H.~J., 1985, \aaps,
  62, 197

\bibitem[{{Mewe}, {Lemen} \& {van den Oord}(1986){Mewe}, {Lemen}, \& {van den
  Oord}}]{Mewe86}
{Mewe} R., {Lemen} J.~R., {van den Oord} G.~H.~J., 1986, \aaps, 65, 511

\bibitem[{{Nandra} \& {Iwasawa}(2007)}]{Nandra07}
{Nandra} K., {Iwasawa} K., 2007, \mnras, 382, L1

\bibitem[{{Narayan} \& {Fabian}(2011)}]{Narayan11}
{Narayan} R., {Fabian} A.~C., 2011, \mnras, 415, 3721

\bibitem[{{Narayan} \& {McClintock}(2008)}]{Narayan08}
{Narayan} R., {McClintock} J.~E., 2008, \nar, 51, 733

\bibitem[{{Narayan} \& {Yi}(1994)}]{Narayan94}
{Narayan} R., {Yi} I., 1994, \apjl, 428, L13

\bibitem[{{Novikov} \& {Thorne}(1973)}]{Novikov73}
{Novikov} I.~D., {Thorne} K.~S., 1973, in Black Holes (Les Astres Occlus),
  {Dewitt} C., {Dewitt} B.~S., eds., pp. 343--450

\bibitem[{{Nowak}(1995)}]{Nowak95}
{Nowak} M.~A., 1995, \pasp, 107, 1207

\bibitem[{{O'Dea} {et~al}\mbox{.}(2008){O'Dea}, {Baum}, {Privon}, {Noel-Storr},
  {Quillen}, {Zufelt}, {Park}, {Edge}, {Russell}, {Fabian}, {Donahue},
  {Sarazin}, {McNamara}, {Bregman}, \& {Egami}}]{ODea08}
{O'Dea} C.~P. {et~al.}, 2008, \apj, 681, 1035

\bibitem[{{O'Neill} \& {Jones}(2010)}]{ONeill10}
{O'Neill} S.~M., {Jones} T.~W., 2010, \apj, 710, 180

\bibitem[{{O'Sullivan} {et~al}\mbox{.}(2011{\natexlab{a}}){O'Sullivan},
  {Giacintucci}, {David}, {Gitti}, {Vrtilek}, {Raychaudhury}, \&
  {Ponman}}]{OSullivan11}
{O'Sullivan} E., {Giacintucci} S., {David} L.~P., {Gitti} M., {Vrtilek} J.~M.,
  {Raychaudhury} S., {Ponman} T.~J., 2011{\natexlab{a}}, \apj, 735, 11

\bibitem[{{O'Sullivan} {et~al}\mbox{.}(2011{\natexlab{b}}){O'Sullivan},
  {Worrall}, {Birkinshaw}, {Trinchieri}, {Wolter}, {Zezas}, \&
  {Giacintucci}}]{OSullivanngc426111}
{O'Sullivan} E., {Worrall} D.~M., {Birkinshaw} M., {Trinchieri} G., {Wolter}
  A., {Zezas} A., {Giacintucci} S., 2011{\natexlab{b}}, \mnras, 416, 2916

\bibitem[{{Panessa} {et~al}\mbox{.}(2007){Panessa}, {Barcons}, {Bassani},
  {Cappi}, {Carrera}, {Ho}, \& {Pellegrini}}]{Panessa07}
{Panessa} F., {Barcons} X., {Bassani} L., {Cappi} M., {Carrera} F.~J., {Ho}
  L.~C., {Pellegrini} S., 2007, \aap, 467, 519

\bibitem[{{Pellegrini} {et~al}\mbox{.}(2003){Pellegrini}, {Venturi},
  {Comastri}, {Fabbiano}, {Fiore}, {Vignali}, {Morganti}, \&
  {Trinchieri}}]{Pellegrini03}
{Pellegrini} S., {Venturi} T., {Comastri} A., {Fabbiano} G., {Fiore} F.,
  {Vignali} C., {Morganti} R., {Trinchieri} G., 2003, \apj, 585, 677

\bibitem[{{Pizzolato} \& {Soker}(2005)}]{PizzolatoSoker05}
{Pizzolato} F., {Soker} N., 2005, \apj, 632, 821

\bibitem[{{Pizzolato} \& {Soker}(2010)}]{Pizzolato10}
{Pizzolato} F., {Soker} N., 2010, \mnras, 408, 961

\bibitem[{{Plotkin} {et~al}\mbox{.}(2012){Plotkin}, {Markoff}, {Kelly},
  {K{\"o}rding}, \& {Anderson}}]{Plotkin12}
{Plotkin} R.~M., {Markoff} S., {Kelly} B.~C., {K{\"o}rding} E., {Anderson}
  S.~F., 2012, \mnras, 419, 267

\bibitem[{{Poggianti}(1997)}]{Poggianti97}
{Poggianti} B.~M., 1997, \aaps, 122, 399

\bibitem[{{Proga} \& {Begelman}(2003)}]{Proga03}
{Proga} D., {Begelman} M.~C., 2003, \apj, 592, 767

\bibitem[{{Ptak} {et~al}\mbox{.}(1998){Ptak}, {Yaqoob}, {Mushotzky},
  {Serlemitsos}, \& {Griffiths}}]{Ptak98}
{Ptak} A., {Yaqoob} T., {Mushotzky} R., {Serlemitsos} P., {Griffiths} R., 1998,
  \apjl, 501, L37

\bibitem[{{Quillen} {et~al}\mbox{.}(2008){Quillen}, {Zufelt}, {Park}, {O'Dea},
  {Baum}, {Privon}, {Noel-Storr}, {Edge}, {Russell}, {Fabian}, {Donahue},
  {Bregman}, {McNamara}, \& {Sarazin}}]{Quillen08}
{Quillen} A.~C. {et~al.}, 2008, \apjs, 176, 39

\bibitem[{{Rafferty} {et~al}\mbox{.}(2006){Rafferty}, {McNamara}, {Nulsen}, \&
  {Wise}}]{Rafferty06}
{Rafferty} D.~A., {McNamara} B.~R., {Nulsen} P.~E.~J., {Wise} M.~W., 2006,
  \apj, 652, 216

\bibitem[{{Remillard} \& {McClintock}(2006)}]{Remillard06}
{Remillard} R.~A., {McClintock} J.~E., 2006, \araa, 44, 49

\bibitem[{{Risaliti}, {Maiolino} \& {Salvati}(1999){Risaliti}, {Maiolino}, \&
  {Salvati}}]{Risaliti99}
{Risaliti} G., {Maiolino} R., {Salvati} M., 1999, \apj, 522, 157

\bibitem[{{Rothschild} {et~al}\mbox{.}(1981){Rothschild}, {Baity}, {Marscher},
  \& {Wheaton}}]{Rothschild81}
{Rothschild} R.~E., {Baity} W.~A., {Marscher} A.~P., {Wheaton} W.~A., 1981,
  \apjl, 243, L9

\bibitem[{{Russell} {et~al}\mbox{.}(2010){Russell}, {Fabian}, {Sanders},
  {Johnstone}, {Blundell}, {Brandt}, \& {Crawford}}]{Russell10}
{Russell} H.~R., {Fabian} A.~C., {Sanders} J.~S., {Johnstone} R.~M., {Blundell}
  K.~M., {Brandt} W.~N., {Crawford} C.~S., 2010, \mnras, 402, 1561

\bibitem[{{Russell}, {Sanders} \& {Fabian}(2008){Russell}, {Sanders}, \&
  {Fabian}}]{Russell08}
{Russell} H.~R., {Sanders} J.~S., {Fabian} A.~C., 2008, \mnras, 390, 1207

\bibitem[{{Salom{\'e}} \& {Combes}(2003)}]{Salome03}
{Salom{\'e}} P., {Combes} F., 2003, \aap, 412, 657

\bibitem[{{Sambruna} {et~al}\mbox{.}(2000){Sambruna}, {Chartas}, {Eracleous},
  {Mushotzky}, \& {Nousek}}]{Sambruna00}
{Sambruna} R.~M., {Chartas} G., {Eracleous} M., {Mushotzky} R.~F., {Nousek}
  J.~A., 2000, \apjl, 532, L91

\bibitem[{{Sambruna} {et~al}\mbox{.}(2003){Sambruna}, {Gliozzi}, {Eracleous},
  {Brandt}, \& {Mushotzky}}]{Sambruna03}
{Sambruna} R.~M., {Gliozzi} M., {Eracleous} M., {Brandt} W.~N., {Mushotzky} R.,
  2003, \apjl, 586, L37

\bibitem[{{Sanders} \& {Fabian}(2007)}]{SandersFabian07}
{Sanders} J.~S., {Fabian} A.~C., 2007, \mnras, 381, 1381

\bibitem[{{Schlegel}, {Finkbeiner} \& {Davis}(1998){Schlegel}, {Finkbeiner}, \&
  {Davis}}]{Schlegel98}
{Schlegel} D.~J., {Finkbeiner} D.~P., {Davis} M., 1998, \apj, 500, 525

\bibitem[{{Shakura} \& {Sunyaev}(1973)}]{Shakura73}
{Shakura} N.~I., {Sunyaev} R.~A., 1973, \aap, 24, 337

\bibitem[{{Siemiginowska} {et~al}\mbox{.}(2010){Siemiginowska}, {Burke},
  {Aldcroft}, {Worrall}, {Allen}, {Bechtold}, {Clarke}, \&
  {Cheung}}]{Siemiginowska10}
{Siemiginowska} A., {Burke} D.~J., {Aldcroft} T.~L., {Worrall} D.~M., {Allen}
  S., {Bechtold} J., {Clarke} T., {Cheung} C.~C., 2010, \apj, 722, 102

\bibitem[{{Siemiginowska} {et~al}\mbox{.}(2005){Siemiginowska}, {Cheung},
  {LaMassa}, {Burke}, {Aldcroft}, {Bechtold}, {Elvis}, \&
  {Worrall}}]{Siemiginowska05}
{Siemiginowska} A., {Cheung} C.~C., {LaMassa} S., {Burke} D.~J., {Aldcroft}
  T.~L., {Bechtold} J., {Elvis} M., {Worrall} D.~M., 2005, \apj, 632, 110

\bibitem[{{Sijacki} \& {Springel}(2006)}]{Sijacki06}
{Sijacki} D., {Springel} V., 2006, \mnras, 366, 397

\bibitem[{{Silk} \& {Rees}(1998)}]{Silk98}
{Silk} J., {Rees} M.~J., 1998, \aap, 331, L1

\bibitem[{{Skrutskie} {et~al}\mbox{.}(2006){Skrutskie}, {Cutri}, {Stiening},
  {Weinberg}, {Schneider}, {Carpenter}, {Beichman}, {Capps}, {Chester},
  {Elias}, {Huchra}, {Liebert}, {Lonsdale}, {Monet}, {Price}, {Seitzer},
  {Jarrett}, {Kirkpatrick}, {Gizis}, {Howard}, {Evans}, {Fowler}, {Fullmer},
  {Hurt}, {Light}, {Kopan}, {Marsh}, {McCallon}, {Tam}, {Van Dyk}, \&
  {Wheelock}}]{Skrutskie06}
{Skrutskie} M.~F. {et~al.}, 2006, \aj, 131, 1163

\bibitem[{{Soker}(2008)}]{Soker08}
{Soker} N., 2008, \apjl, 684, L5

\bibitem[{{Springel}, {Di Matteo} \& {Hernquist}(2005){Springel}, {Di Matteo},
  \& {Hernquist}}]{Springel05}
{Springel} V., {Di Matteo} T., {Hernquist} L., 2005, \mnras, 361, 776

\bibitem[{{Tadhunter} {et~al}\mbox{.}(2003){Tadhunter}, {Marconi}, {Axon},
  {Wills}, {Robinson}, \& {Jackson}}]{Tadhunter03}
{Tadhunter} C., {Marconi} A., {Axon} D., {Wills} K., {Robinson} T.~G.,
  {Jackson} N., 2003, \mnras, 342, 861

\bibitem[{{Taylor} {et~al}\mbox{.}(2006){Taylor}, {Sanders}, {Fabian}, \&
  {Allen}}]{Taylor06}
{Taylor} G.~B., {Sanders} J.~S., {Fabian} A.~C., {Allen} S.~W., 2006, \mnras,
  365, 705

\bibitem[{{Tchekhovskoy}, {Narayan} \& {McKinney}(2011){Tchekhovskoy},
  {Narayan}, \& {McKinney}}]{Tchekovskoy11}
{Tchekhovskoy} A., {Narayan} R., {McKinney} J.~C., 2011, \mnras, 418, L79

\bibitem[{{Terashima} {et~al}\mbox{.}(2002){Terashima}, {Iyomoto}, {Ho}, \&
  {Ptak}}]{Terashima02}
{Terashima} Y., {Iyomoto} N., {Ho} L.~C., {Ptak} A.~F., 2002, \apjs, 139, 1

\bibitem[{{Terashima} \& {Wilson}(2003)}]{Terashima03}
{Terashima} Y., {Wilson} A.~S., 2003, \apj, 583, 145

\bibitem[{{Tremaine} {et~al}\mbox{.}(2002){Tremaine}, {Gebhardt}, {Bender},
  {Bower}, {Dressler}, {Faber}, {Filippenko}, {Green}, {Grillmair}, {Ho},
  {Kormendy}, {Lauer}, {Magorrian}, {Pinkney}, \& {Richstone}}]{Tremaine02}
{Tremaine} S. {et~al.}, 2002, \apj, 574, 740

\bibitem[{{Vasudevan} \& {Fabian}(2007)}]{Vasudevan07}
{Vasudevan} R.~V., {Fabian} A.~C., 2007, \mnras, 381, 1235

\bibitem[{{Vestergaard} \& {Barthel}(1993)}]{Vestergaard93}
{Vestergaard} M., {Barthel} P.~D., 1993, \aj, 105, 456

\bibitem[{{Wachter}, {Leach} \& {Kellogg}(1979){Wachter}, {Leach}, \&
  {Kellogg}}]{Wachter79}
{Wachter} K., {Leach} R., {Kellogg} E., 1979, \apj, 230, 274

\bibitem[{{Walsh}, {Barth} \& {Sarzi}(2010){Walsh}, {Barth}, \&
  {Sarzi}}]{Walsh10}
{Walsh} J.~L., {Barth} A.~J., {Sarzi} M., 2010, \apj, 721, 762

\bibitem[{{Wu}, {Yuan} \& {Cao}(2007){Wu}, {Yuan}, \& {Cao}}]{Wu07}
{Wu} Q., {Yuan} F., {Cao} X., 2007, \apj, 669, 96

\bibitem[{{Yi} \& {Boughn}(1998)}]{YiBoughn98}
{Yi} I., {Boughn} S.~P., 1998, \apj, 499, 198

\bibitem[{{Yuan} {et~al}\mbox{.}(2002){Yuan}, {Markoff}, {Falcke}, \&
  {Biermann}}]{Yuan02}
{Yuan} F., {Markoff} S., {Falcke} H., {Biermann} P.~L., 2002, \aap, 391, 139

\end{thebibliography}
